\documentclass[a4paper]{article}

\usepackage[pages=all, color=black, scale=1, pages=all, opacity=0, vshift=5mm]{background}

\usepackage[margin=1in]{geometry} % full-width

\usepackage{amsmath}
\usepackage{amsthm}
\usepackage{amssymb}

\usepackage[utf8]{inputenc}
\usepackage{hyperref}
\hypersetup{
	unicode,
	pdfproducer={LaTeX},
	pdfcreator={pdflatex}
}

\usepackage{enumitem} %roman numbers
\usepackage[super]{nth} % for dates
\usepackage{verbatim} % \begin and \end {comment}
\usepackage{hyperref} % to use \href
\usepackage{subcaption}
\usepackage{eurosym}
\usepackage{booktabs} % lines in tables
\usepackage{url}

\usepackage{graphicx, color}
\usepackage{algorithm, algpseudocode} % use algorithm and algorithmicx for typesetting algorithms
\usepackage{mathrsfs} % for \mathscr command

\title{An Agent-Based Model of COVID-19 Diffusion to Plan and Evaluate Intervention Policies}
\author{Gianpiero Pescarmona$^1$ \and Pietro Terna$^1$$^2$ \and Alberto Acquadro$^1$ \and Paolo Pescarmona$^3$ \and Giuseppe Russo$^4$ 
\and Emilio Sulis$^1$ \and Stefano Terna$^5$}

\date{
	$^1$University of Torino, Italy \\ \texttt{\{gianpiero.pescarmona, pietro.terna, alberto.acquadro, emilio.sulis\}@unito.it}\\%
	$^2$ Fondazione Collegio Carlo Alberto, Italy\\
	$^3$University of Groningen, The Netherlands \\ \texttt{p.p.pescarmona@rug.nl}\\
	$^4$Centro Einaudi, Torino, Italy \\ \texttt{russo@centroeinaudi.it}\\
	$^5$\url{tomorrowdata.io}  \\ \texttt{stefano.terna@tomorrowdata.io}\\[2ex]
	August, 2021%\today
}

\begin{document}
	\maketitle
	
	\begin{abstract}
A model of interacting agents, following plausible behavioral rules into a world where the Covid-19 epidemic is affecting the actions of everyone. The model works with (i) infected agents categorized as symptomatic or asymptomatic and (ii) the places of contagion specified in a detailed way. The infection transmission is related to three factors: the characteristics of both the infected person and the susceptible one, plus those of the space in which contact occurs. The model includes the structural data of Piedmont, an Italian region, but we can easily calibrate it for other areas. The micro-based structure of the model allows factual, counterfactual, and conditional simulations to investigate both the spontaneous or controlled development of the epidemic.\newline\indent
The model is generative of complex epidemic dynamics emerging from the consequences of agents' actions and interactions, with high variability in outcomes and stunning realistic reproduction of the successive contagion waves in the reference region. 
There is also an inverse generative side of the model, coming from the idea of using genetic algorithms to construct a meta-agent to optimize the vaccine distribution. This agent takes into account groups' characteristics---by age, fragility, work conditions---to minimize the number of symptomatic people.		
	\end{abstract}

	\tableofcontents
	
%%%%%%%%%%%%%%%%%%%%%%%%%%%%%%%%%%%%%%%%%%%%%%%
%%%%%%%%%%%%%%%%%%%%%%%%%%%%%%%%%%%%%%%%%%%%%%%
\section{A quick introduction to our agent-based epidemic model}
\label{intro}

The starting point is a compartmental model with Susceptible, Infected, and Recovered people (S.I.R.), but adding both a more detailed breakdown of the subjects involved in the contagion process \cite{Scala:2020aa} and a multi-scale framework to account for the interaction at different dimensional, and spatial levels \cite{doi:10.1142/S0218202520500323}. From the virus micro-level, we move to individuals and up to the collective behavior of the population.

Following \cite{rahmandad2008heterogeneity}, we know that the analysis based on the assumption of heterogeneity strongly differs from S.I.R. compartmental structures modeled by differential equations. The authors of this work argue when it is best to use agent-based models and when it would be better to use differential equation models ponder when it is better to use agent-based models and when it would be better to use differential equation models. Differential equation models assume homogeneity and perfect mixing of characteristics within compartments, while agent-based models can capture heterogeneity in agent attributes and the structure of their interactions. We follow the second approach (about agent-based approach, see Section \ref{why}).

\begin{itemize}

\item

Our model takes into consideration: 

\begin{enumerate}[label=\roman*]
\item infected agents categorized as symptomatic or asymptomatic and 
\item the places of contagion specified in a detailed way, thanks to agent-based modeling capabilities. 
\end{enumerate}

 \item
The infection transmission is related to three factors: the infected person's characteristics and those of the susceptible one, plus those of the space in which a contact occurs.

\end{itemize}

Finally, we subscribe the call of \cite{squazzoni2020} to <<cover the full behavioural and social complexity of societies under pandemic crisis>> and we work arguing that <<the study of collective behavior must rise to a ``crisis discipline'' just as medicine, conservation, and climate science have, with a focus on providing actionable insight to policymakers and regulators for the stewardship of social systems>>, as in \cite{Bak-Colemane2025764118}. 

A look at the structure of the whole presentation.
In Section \ref{why}, we discuss models and specifically agent-based models; in Section \ref{biochem}, the molecular support to agents' intrinsic susceptibility construction; in Section \ref{ourModel}, the structure of the model, with the daily sequence of the agents' actions. Section \ref{howWorks} introduces a detailed description of the internal model mechanisms, with: conditional actions in Section \ref{cond}, parameters in Section \ref{par} and agents' interaction in Section \ref{inter}.

A technique for contagion representation is introduced in Section \ref{contag}. Then we explore simulation cases in Section \ref{exploring}, building several batches of runs and comparing extreme situations in Sections \ref{withoutWith}.

Section \ref{actual} reports the actual epidemic data in the reference region. With those data, we verify factual and counterfactual analyses in Section \ref{facCounterfac}. Considering the possibility of calculating infection indicators without delays (Section \ref{indicator}), we experiment with the effect of adopting the control measure with 20 days of anticipation (Section \ref{anticip}). In Section \ref{frag} we verify another counterfactual policy, that of concentrating the efforts uniquely in defense of fragile persons. Section recap \ref{recap} summarizes these results.

The final application of the model is dedicated to a planning exercise on vaccination campaigns (Section \ref{planning}). We introduce an analysis of the vaccine mechanism in the perspective of our model (Section \ref{vaccineAction}), using both planned strategies (Sections \ref{plain}, \ref{wise}) and genetic algorithms (Section \ref{GAquotas}). The GAs goal is to optimize the behavior of a meta-agent, deciding the sequence of the vaccinations.

%%%%%%%%%%%%%%%%%%%%%%%%%%%%%%%%%%%%%%%%%%%%%%%
\subsection{Why models? Why agents? Why another model?}
\label{why}

Why another model, and most of all, why models? With \cite{epstein2008model}:
\begin{quote}
The choice (\ldots) is not whether to build models; it's whether to build explicit ones. In explicit models, assumptions are laid out in detail, so we can study exactly what they entail. On these assumptions, this sort of thing happens. When you alter the assumptions that is what happens. By writing explicit models, you let others replicate your results.
\end{quote}

With even more strength:
\begin{quote} 
I am always amused when these same people challenge me with the question,``Can you validate your model?'' The appropriate retort, of course, is,``Can you validate yours?'' At least I can write mine down so that it can, in principle, be calibrated to data, if that is what you mean by ``validate'' a term I assiduously avoid (good Popperian that I am).
\end{quote}

To reply to ``why agents?'', with \cite{axtell2000agents} we define in short what an agent-based model is:
\begin{quote} 
An agent-based model consists of individual agents, commonly implemented in software as objects. Agent objects have states and rules of behavior. Running such a model simply amounts to instantiating an agent population, letting the agents interact, and monitoring what happens. That is, executing the model---spinning it forward in time---is all that is necessary in order to ``solve'' it.
\end{quote}

More in detail:
\begin{quote} 
There are, ostensibly, several advantages of agent-based computational modeling over conventional mathematical theorizing. First, [\ldots] it is easy to limit agent rationality in agent-based computational models. Second, even if one wishes to use completely rational agents, it is a trivial matter to make agents heterogeneous in agent-based models. One simply instantiates a population having some distribution of initial states, e.g., preferences. That is, there is no need to appeal to representative agents. [\ldots] Finally, in most social processes either physical space or social networks matter. These are difficult to account for mathematically except in highly stylized ways. However, in agent-based models it is usually quite easy to have the agent interactions mediated by space or networks or both.
\end{quote}

In \cite{inverseGen} we have a relevant step ahead, considering \emph{inverse generative social science}:
\begin{quote} 
The agent-based model (ABM) is the principal scientific instrument for understanding how individual behaviors and interactions, the micro-world, generates change and stasis in macroscopic social regularities. So far, agents have been iterated forward to generate such explananda as settlement patterns, scaling laws, epidemic dynamics, and many other phenomena [6]. But these are all examples of the forward problem: we design agents and grow the target phenomenon. The motto of generative social science is: ``If you didn't grow it, you didn't explain it.'' \cite{epstein1999agent} 
But there may be many ways to grow it! How do we find `all' the non-trivial generators? This is inverse generative social science---agent architectures as model outputs not model inputs---and machine learning can enable it.
\end{quote} 

And now, "why another?" As a commitment to our creativity, using our knowledge to understand what is happening. Indeed, with arbitrariness: it is up to others and time to judge.

As any model, also this one is based on assumptions: time will tell whether these were reasonable hypotheses. Modeling the Covid-19 pandemic requires a scenario and the actors. As in a theater play, the author defines the roles of the actors and the environment. The characters are not real, they are prebuilt by the author, and they act according to their peculiar constraints. If the play is successful, it will run for a long time, even centuries. If not, we will rapidly forget it. Shakespeare?s Hamlet is still playing after centuries, even if the characters and the plot are entirely imaginary. The same holds for our simulations: we are the authors, we arbitrarily define the characters, we force them to act again and again in different scenarios. However, in our model, the micro-micro assumptions are not arbitrary but based on scientific hypotheses at the molecular level, the micro agents? behaviors are modeled in an explicit and realistic way. In both plays and simulations, we compress the time: a whole life to two or three hours on the stage. In a few seconds, we run the Covid-19 pandemic spread in a given regional area.

%%%%%%%%%%%%%%%%%%%%%%%%%%%%%%%%%%%%%%%%%%%%%%%
\subsection{The molecular basis of SARS-CoV-2 infection}
\label{biochem}

To fully understand what the word infection means, we have previously to define the scenario where life takes place.

We start with the properties of life on earth's surface \cite{pescarmona2002life}. 
Making a long story short, basically life is a dissipative process fueled by energy supplied by the sun. As the sun has been shining for billions of years the biological systems on earth expanded exponentially in a finite environment, and they became limited in their growth due to the shortage of available atoms/molecules (``nutrients'').
The competition for the limiting nutrients in each local environment (``niche'') will locally drive the selection and will explain the complexity of the interactions of the different organisms in any environment, from the microscopic to the social level.
The ground, the soil and the ponds, are overcrowded with bacteria, algae, molds, insects and so on. They help to keep the ground healthy and ready for cultivation. 
We too are populated by microorganisms, the gut, the mucosae, the skin. But when we feel healthy we do not realize they are there; but sometimes we don't feel well, we are sick. We have a disease, we need a culprit: somebody different from us, a virus, a bacterium, a protozoan that infected us.

In most cases, the same agent is shared by people surrounding us, but most of them are healthy, few are sick.
The coexistence/cooperation between organisms sharing the same ``niche'' is the rule after billions of years of evolution. The disease is the exception.
The asymptomatic infection is the rule, the symptomatic infection is the disease.
The simplest explanation for the rise of the symptoms is the competition of different organisms for a limiting  ``nutrient''.

In the case of the Herpesvirus family and man, the limiting ?nutrient? is Iron.
Virus Ribonucleotide reductase is an enzyme with an affinity for iron higher than human cells. Infected cells survive quite well until iron availability covers the needs of both host and virus. In the case of iron shortage (evaluated as the level of serum ferritin) the infected cells are forced to reduce heme synthesis, necessary for the respiratory chain, and hence ATP synthesis. Less ATP, loss of many cellular functions, symptoms. In our experience, most of the people seropositive to HSV had no symptoms, provided they had serum ferritin levels => 90 ug/dl. The lower the ferritin level, the higher the frequency of the symptoms. The level of ferritin depends on genetic, dietary, environmental factors, explaining the variability of the clinical manifestations \cite{gennero2010iron}.

In the case of HSV, the virus metabolism is well known and studied for tens of years.
In the case of SARS-CoV-2 our experience is in the range of months and the identification of the limiting ``nutrient'' is only speculative.

On the basis of the data collected up to now, Cysteine could be the most relevant.
One of the coauthors here, G.~Pescarmona, with other contributors, has recently developed a software able to easily compare the amino acids (AA) percentage and some selected ratios between couples of them, using Uniprot proteins repository as data source \cite{vernone2013human}. Using this software, it has been possible to compare the AA percentage in different tissues \cite{vernone2019analysis} demonstrating the limiting role of AA local availability on the synthesis of specific proteins.

\begin{figure}[t]
\center
\includegraphics[width=0.7\textwidth]{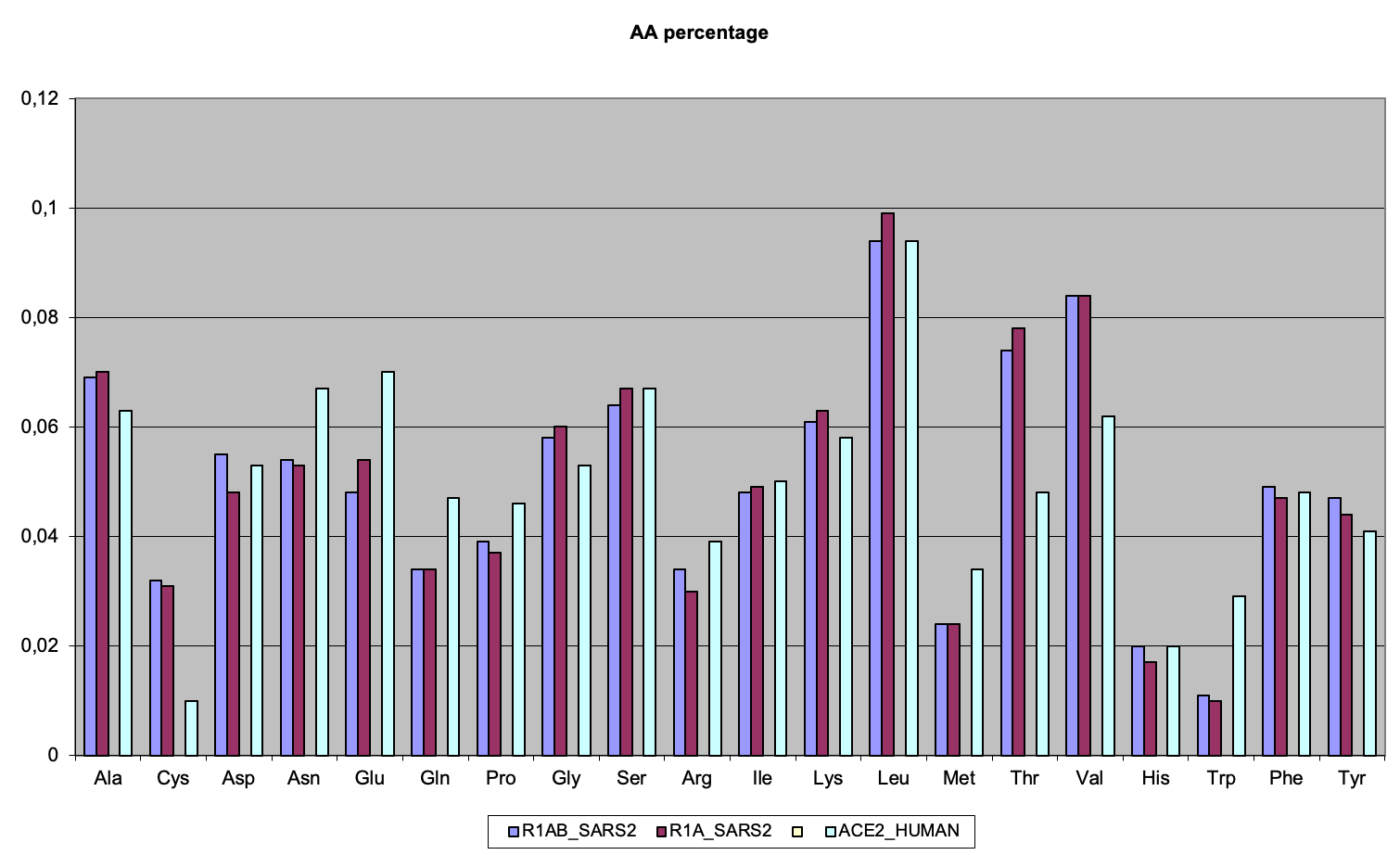}
\caption{Comparison of the Amino Acids percentage in the most representative SARS-Cov-2 proteins and the human ACE2, receptor on the surface of host cells}
\label{ch1}
\end{figure}

From the beginning of the pandemic, it has been clear that ACE2 was the preferred ligand for the Spyke protein and that cells expressing it were the perfect host for the fast synthesis of viral protein \cite{scialo2020ace2}. Our working hypothesis is that the best host cell is one producing a protein with a similar AA percentage.
From Fig. \ref{ch1} we can extract the following info: most AA percentages of the viral proteins are similar, with the exception of Cysteine (lower) and Methionine and Tryptophan (higher) to the ACE2. Higher methionine associates with faster protein synthesis, higher Tryptophan with higher nucleic acid synthesis. A perfect environment for replication of both RNA and virus proteins. Expression of viral proteins with high Cysteine decreases free cysteine and therefore Glutathione (GSH) synthesis, with impaired antioxidant defense and increased ROS activity. Increased ROS activity has been one of the first well-identified mechanisms of viral infection \cite{Shenoy2020} and their scavenging by GSH has been proposed as a preventive/therapeutic approach to the symptomatic disease \cite{Silvagno_2020}.

\begin{figure}[t]
\center
\includegraphics[width=0.7\textwidth]{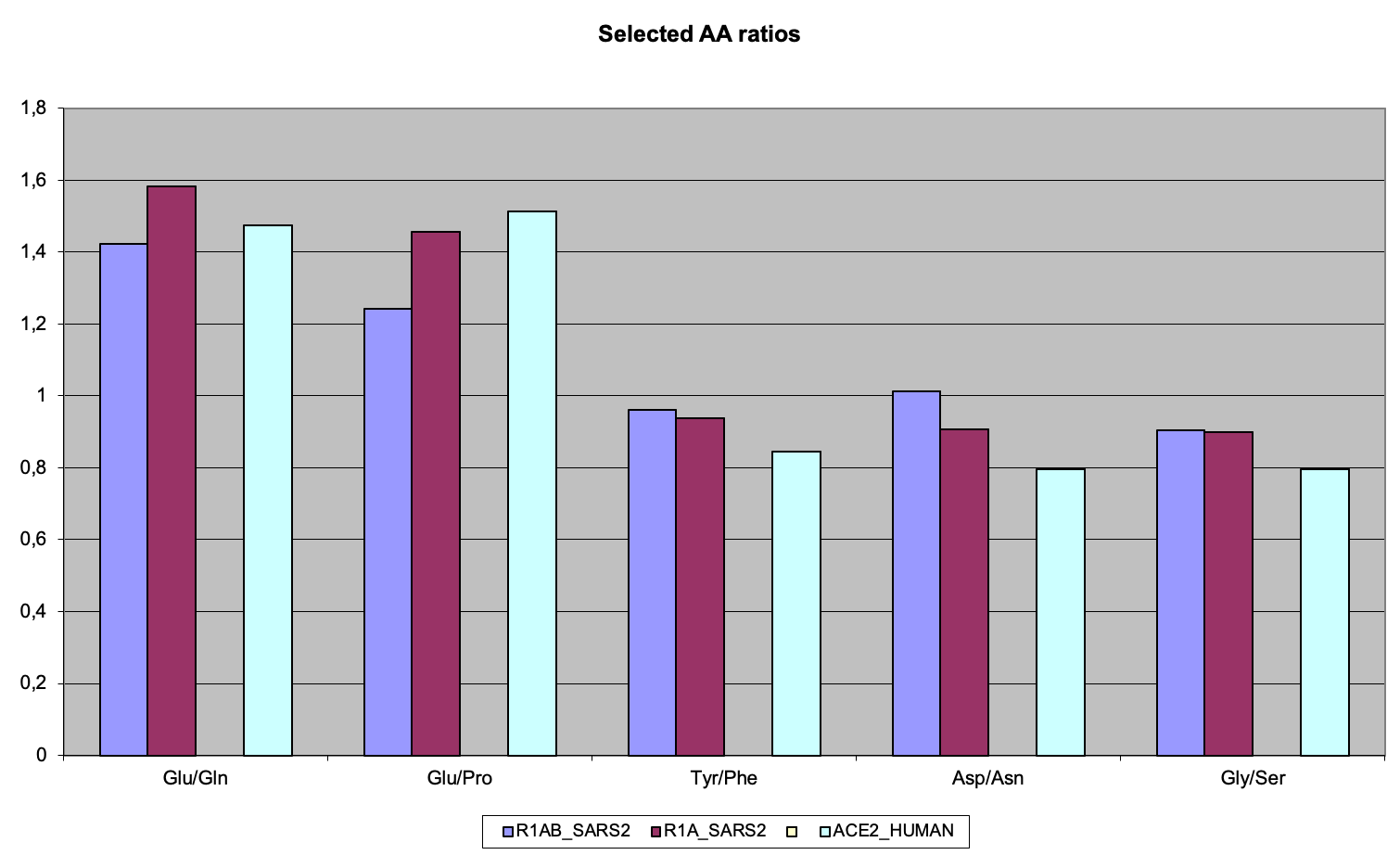}
\caption{Comparison of some ratios between selected Amino Acids in the most representative SARS-Cov-2 proteins and the human ACE2, receptor on the surface of host cells. These ratios supply specific information about the local metabolic condition inside the cell \cite{vernone2019analysis}}.
\label{ch2}
\end{figure}

Also, the ratio between AA couples, in Fig. \ref{ch2}, shows good similarity with the exception of the Spyke protein, as far as the ratios including glutamate are involved, but the almost perfect coincidence between the catalytic proteins of the virus and ACE2 explains the reproductive advantage of entering a cell expressing it.
The interesting information that we get from this approach is that the cysteine deprivation of the naturally infected cells is shared also by cells induced to produce Spyke protein, independently by the vector used.
Moreover, whilst the full virus enters the cells expressing on the outer surface ACE2, and we can identify them and try to imagine the long-term effects of infection, in the case of vaccines the synthetic vectors should allow the entry in any kind of cells.

In conclusion, we can expect oxidative damage (ROS increase and inflammation) in any kind of cell in our body. The extent of the inflammation will vary according to so many variables: age, diet, drugs, previous silent sites of inflammation, to make almost impossible the prediction of the localization and gravity of the side effects.

\subparagraph{The cytokine storm}

The cytokine storm is a synthetic definition of the set of reactions leading to the symptomatic COVID-19 and to death. The core process of the infection is the unbalance between ACE/ACE2. SARS-CoV-2 binds to ACE-2 and sequester it, causing an ACE prevalence and a sharp increase of ROS (\cite{soy2020cytokine,hu2021cytokine}).
All pre-existing processes leading to the prevalence of ACE are pro-inflammatory, those leading to a prevalence of ACE2, are anti-inflammatory.
A low level of the active Vitamin D (1,25-dihydroxy-Vitamin D) leads to an increased expression of ACE. Cortisol has the same, but with an independent mechanism, effect on ACE expression and, additionally, decreases the expression of ACE2.
In all the cases the ROS released by ACE activity are scavenged by GSH and its ancillary enzymes. Downstream of ROS, the inflammatory pathway includes NF-kB, TNF-alpha,  IL-6, PLA2, COX1,  and COX2.

From the clinical point of view the COVID-19 pandemic is affecting differently the world population: in presence of conditions such as aging, diabetes, obesity, and hypertension the virus triggers a lethal cytokine storm and patients die from acute respiratory distress syndrome, whereas in many cases the disease has a mild or even asymptomatic progression \cite{gao2021risk}). The identification of the biochemical patterns underlying the severe disease may allow the identification of fragile people in need of more accurate protection.

\begin{figure}[t]
\center
\includegraphics[width=0.7\textwidth]{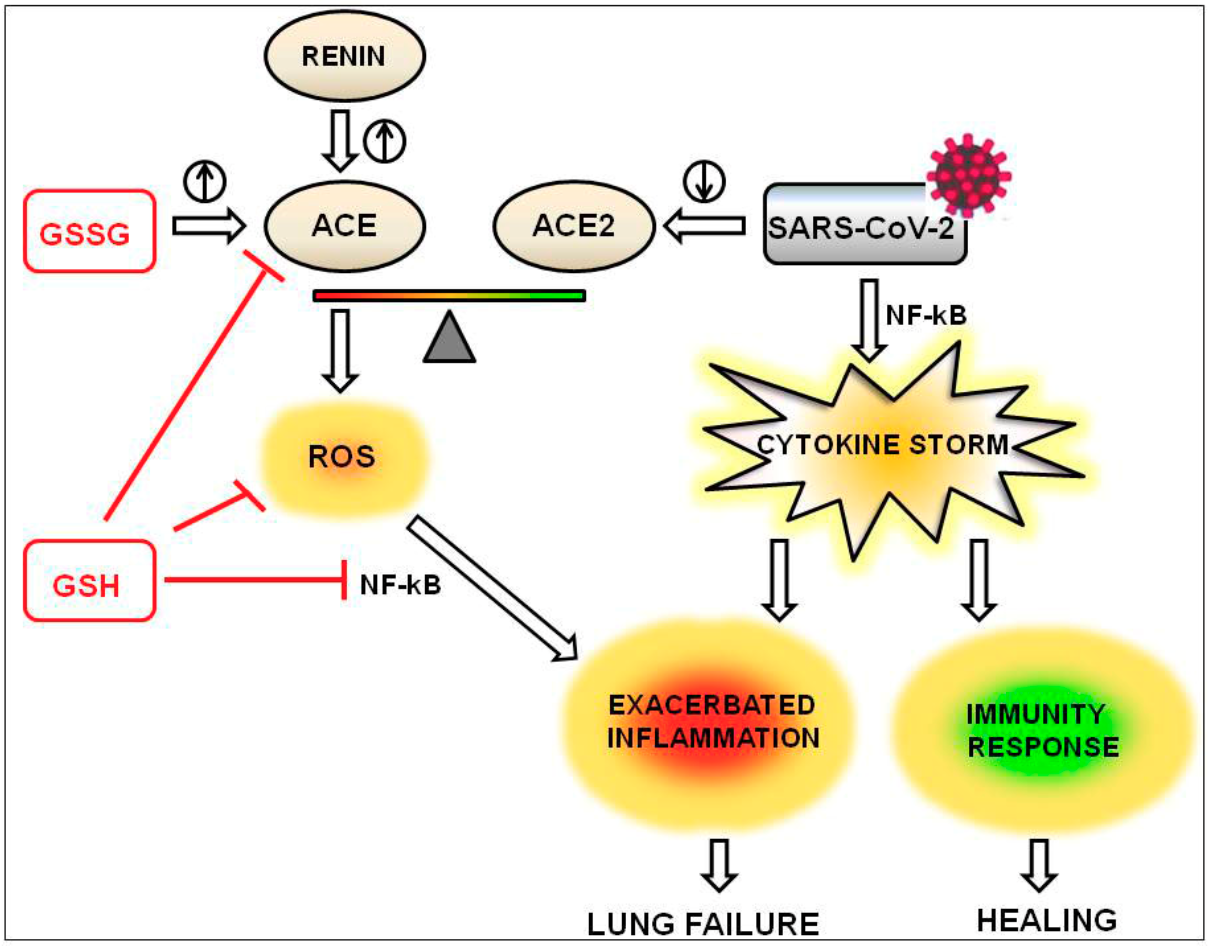}
\caption{All the main agents involved in the inflammatory response during COVID-19 are depicted here, with their relationships (reprinted with permission from \cite{Silvagno_2020})}
\label{ace}
\end{figure}

\begin{table}[t]
\center
\begin{tabular}{llllll}
\toprule
Risk factors~~~ & DHEA~~~ &	Cortisol & GSH~~~~ & Vit~D~~ & BMR~~~~ \\
\midrule
Aging & low & high & low & low & low \\
Diabetes & low & high & low & low & low \\
Hypertension & low & high & low & low & ? \\
Obesity & low & high & low & low & low \\
Diuretics & - & high & - & - & ? \\
Drugs & - & - & low & low & ? \\
Air pollution & - & - & low & - & ? \\
Paracetamol & - & - & low & - & ? \\
Cloroquine & - & - & low & - & ? \\
Glucocorticoids & - & high & - & - & ? \\
Ibuprofen & - & - & - & - & ? \\
Aspirin	\\
\bottomrule
\end{tabular}
\caption{A synopsis of all the metabolic features associated with the clinical conditions favoring a severe COVID-19 development. DHEA: Dehydroepiandrosterone, GSH: Glutathione, Vit D: 25(OH)-Vitamin D, BMR: Basal Metabolic Rate}
\label{riskFactors}
\end{table}

Combining the biochemical determinants listed in Table \ref{riskFactors} within the model described in Fig. \ref{ace} is possible to evaluate the risk for every individual, or class of similar individuals, of developing a severe form of the disease.

DHEA is an adrenal hormone, a precursor of testosterone and estrogens, that activates heme synthesis. Heme is required for plenty of reactions, including the respiratory chain (ATP synthesis) and Vitamin D activation. ATP is required for GSH synthesis, BMR reflects the activity of the respiratory chain and therefore depends again on heme. Heme synthesis requires iron, whose availability depends on diet, correct digestion, and absorption. 
Table \ref{riskFactors} lists also some of the environmental factors that can interfere with the molecules involved in the COVID-19 dependent inflammatory response.
Environmental pollution and drugs abuse in older people are among the factors that can explain the excess mortality in developed countries.
The therapeutic use of paracetamol, chloroquine, and glucocorticoids to prevent severe symptoms looks inappropriate on the basis of their action mechanism.

This set of considerations can be used to tentatively identify and protect fragile people but can be easily modified according to the epidemiological data. Unfortunately, up to now, the prevailing approach has been different, and not so much data about the characteristics of the patients severely ill have been published to allow validation of our criteria for fragility.

%%%%%%%%%%%%%%%%%%%%%%%%%%%%%%%%%%%%%%%%%%%%%%%
\subsection{Our model}
\label{ourModel}

With our model, we move from a macro compartmental vision to a meso and micro-analysis capability. Its main characteristics are:

\begin{itemize}

\item
scalability: we take into account the interactions between virus and molecules inside the host, determining individual susceptibility; the interactions between individuals in more or less restricted contexts; the movement between different environments (home, school, workplace, open spaces, shops); the movements occur in different parts of the daily life, as in \cite{ghorbani2020assocc};
in detail, the scales are: 

\begin{itemize}
\item
	\emph{micro}, with the internal biochemical mechanism involved in reacting to the virus, as in \cite{Silvagno_2020}, from where we derive the critical importance assigned to an individual attribute of intrinsic susceptibility related to the age and previous morbidity episodes; the model indeed incorporates the medical insights and consistent perspectives of one of its co-authors, former full professor of clinical biochemistry, signing also the quoted article; a comment on Lancet \cite{horton2020offline} consistently signals the syndemic character of the current event: <<Two categories of disease are interacting within specific populations---infection with severe acute respiratory syndrome coronavirus 2 (SARS-CoV-2) and an array of non-communicable diseases (NCDs)>>;
\item
	\emph{meso}, with the open and closed contexts where the agents behave, as reported above;
\item	
	\emph{macro}, with the emergent effects of the actions of the agents;
	
\end{itemize}

\item
granularity: at any level, the interactions are partially random and therefore the final results will always reflect the sum of the randomness at the different levels; changing the constraints at different levels and running multiple simulations should allow the identification of the most critical points, where to focus the intervention.

\end{itemize}

Summing up, S.I.s.a.R. (\href{https://terna.to.it/simul/SIsaR.html}{https://terna.to.it/simul/SIsaR.html}) is an agent-based model designed to reproduce the diffusion of the COVID-19 using agent-based modeling in NetLogo \cite{NetLogo}. We have Susceptible, Infected, symptomatic, asymptomatic, and Recovered people: hence the name S.I.s.a.R. The model works on the structural data of Piedmont, an Italian region, but we can quite easily calibrate it for other areas. It reproduces the events following a realistic calendar (national or local government decisions, as in Section \ref{par}), via its script interpreter. At the above address, it is also possible to run the code online without installation. Into the \emph{Info} sheet of the model, we have more than 20 pages of Supporting Information about both the structure and the calibration of the model.

The micro-based structure of the model allows factual, counterfactual, and conditional simulations. Examples of counterfactual situations are those considering:

\begin{enumerate}[label=\roman*]
\item different timing in the adoption of the non-pharmaceutical containment measures;
\item an alternative strategy, focusing exclusively on the defense of fragile people.
\end{enumerate}

The model generates complex epidemic dynamics, emerging from the consequences of agents' actions and interactions, with high variability in outcomes, and with a stunning realistic reproduction of the contagion waves that occurred in the reference region. 

We take charge of the variability of the epidemic paths within the simulation, running batches of executions with 10,000 occurrences for each experiment.

Following \cite{inverseGen}, the AI and inverse generative side of the model comes from constructing a meta-agent optimizing the vaccine distribution among people groups---characterized by age, fragility, work conditions---to minimize the number of symptomatic people (as deceased persons come from there).

We can characterize the action of the planner both:
\begin{enumerate}[label=\roman*]
\item introducing ex-ante rules following ``plain'' or ``wise'' strategies that we imagine as observers or
\item evolving those strategies via the application of a genetic algorithm, where the genome is a matrix of vaccination quotas by people groups, with their time range of adoption. 
\end{enumerate}

%%%%%%%%%%%%%%%%%%%%%%%%%%%%%%%%%%%%%%%%%%%%%%%
\section{How S.I.s.a.R. works}
\label{howWorks}

We have two initial infected individuals in a population of 4,350 individuals, on a scale of 1:1000 with Piedmont. The size of the initial infected group is out of scale: it is the smallest number ensuring the epidemic's activation in a substantial number of cases. Initial infected people bypass the incubation period. For plausibility reasons, we never choose initial infected people among persons in nursing homes or hospitals. The presence of agents in close spaces---such as classrooms, factories, homes, hospitals, nursing homes---is set with realistic numbers, out of scale: e.g., a classroom contains 25 students, a home two persons, large factories up to 150 employees, small ones up to 15, etc.; the movements occur in different parts of the daily life, as in \cite{ghorbani2020assocc}.

\begin{figure}[t]
\center
\fbox{\includegraphics[width=0.9\textwidth,,angle=0]{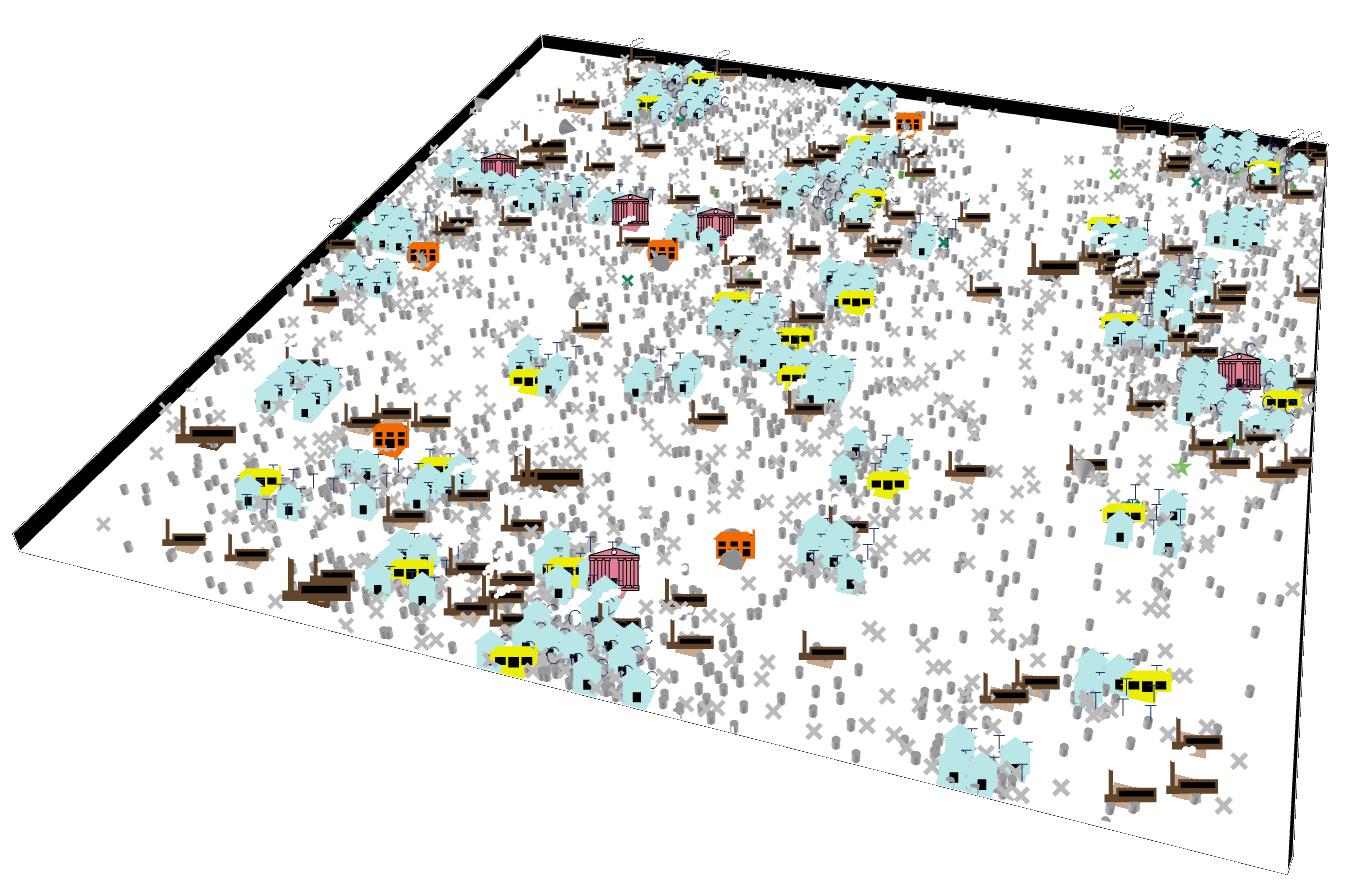}}
\caption{A live 3D picture of the model world}
\label{3D}
\end{figure}

In Fig. \ref{3D} we have a 3D representation of the model world, with one of the possible random maps that the simulation generates. Persons are in gray, houses in cyan, nursing homes in orange, hospitals in pink, schools in yellow, factories (with shops and offices) in brown. Persons have a cylinder as shape, if regular or robust (young); a capital X if fragile; temporary their colors can be: red, if symptomatic; violet, if asymptomatic; turquoise, if symptomatic recovered; green, if asymptomatic recovered.

Doing the batches of repetitions of the simulation, we use random maps to have a neutral effect of the structure of the space. 

We can set: 
\begin{itemize}
\item min and max duration of the individual infection;

\item the length of the incubation interval;

\item the critical distance, i.e., the radius of the possibility of infection in open air, with a given probability;

\item the corrections of that probability, due to the personal characteristics of both active and the passive agents; 

\begin{itemize}
\item active agents can be symptomatic or asymptomatic, with different spreading characteristics (see \ref{D} in Section \ref{par});

\item passive agents, as receivers, can be robust (young), regular, fragile, and extra fragile.
\end{itemize} 

\end{itemize} 

We have two main types of contagion: (a) within a radius, for people moving around, temporary in a house/factory/nursing home/hospital; (b) in a given space (room or apartment) for people resident in their home or in a hospital or in a nursing home or being in school or in a working environment.

People in hospitals and nursing homes can be infected in ways (a) and (b). Instead, while people are at school, they can only receive the disease from people in the same classroom, where only teachers and students are present, so this is a third infection mechanism (c). In all cases, the personal characteristics of the recipients are decisive.

We remark that workplaces are open to all persons, as clients, vendors, suppliers, external workers can go there. In contrast, schools are reserved for students and school operators.

All agents have their home, inside a city, or a town. The agents also have usual places (UPs) where they act and interact, moving around. These positions can be interpreted as free time elective places. When we activate the schools, students and teachers have both UPs and schools; healthcare operators have both UPs and hospitals or nursing homes; finally, workers have both UPs and working places. In each day (or tick of the model), we simulated full sequences of actions.

\begin{figure}[H]
\center
\fbox{\includegraphics[width=1.3\textwidth,,angle=90]{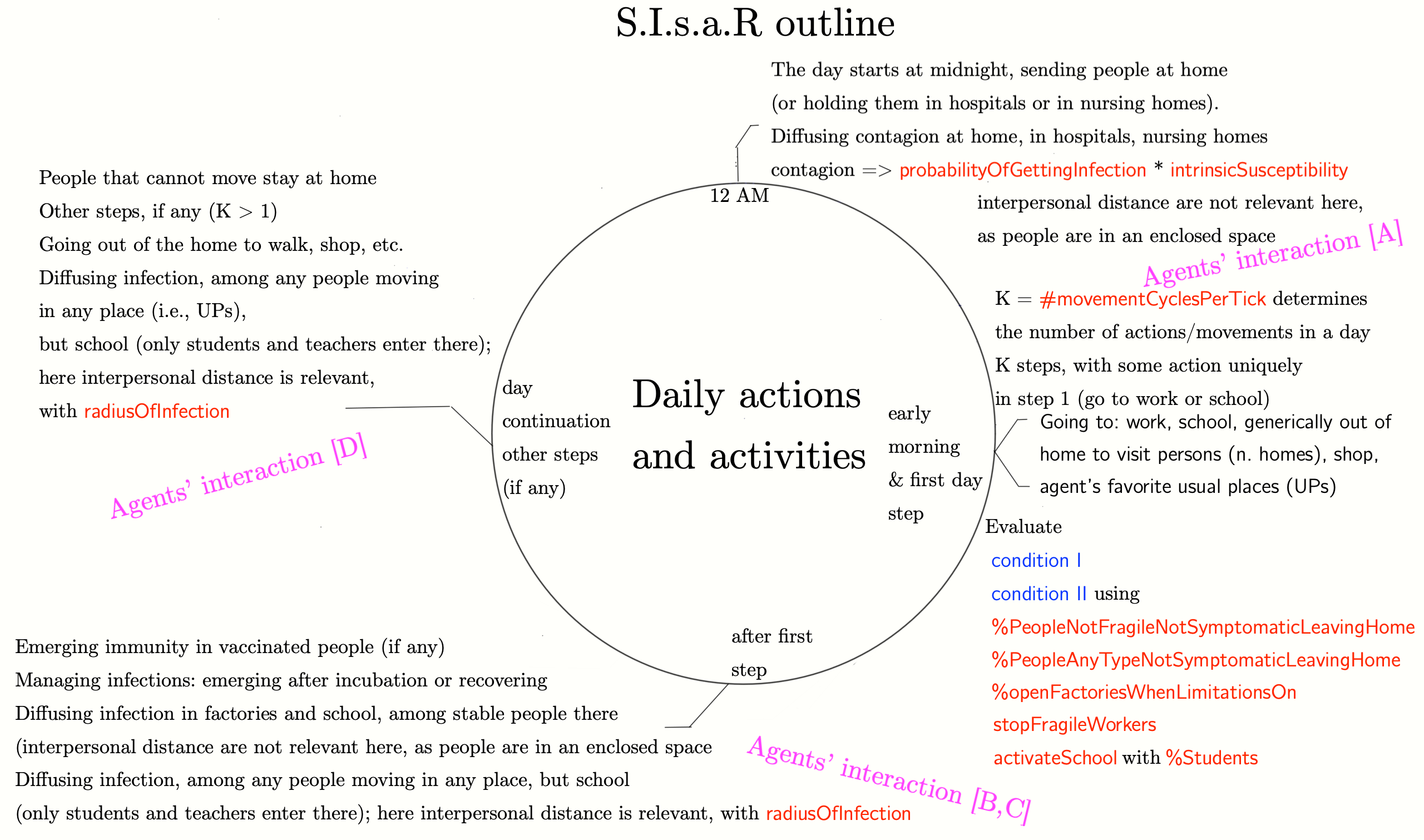}}
\caption{A day in the simulation, with $N$ repetitions where $N$ is the duration of a given outbreak; look at: Section \ref{cond} for the rules of the conditional actions; Section \ref{par} for the parameter definitions; and Section \ref{inter} for details on the agent interactions}
\label{outline}
\end{figure}

Fig. \ref{outline} describes what happens during every \emph{day} in our simulated world, with the daily sequences of actions.

%%%%%%%%%%%%%%%%%%%%%%%%%%%%%%%%%%%%%%
\subsection{Conditional actions}
\label{cond}

Agents' movements in space, to go to work, school, and other UPs are subject to two interrelated general conditions.

\begin{enumerate}[label=\Roman*]

\item Symptomatic persons are at home or in a hospital or a nursing home and do not move. 

\item People not constrained by \emph{condition I} can move if (primary rule) there are no general limitations (e.g., lockdown) \emph{OR} if one of the following sub-conditions applies:

\begin{enumerate}
\item agents who are hospital healthcare operators or nursing home healthcare operators;

\item all people, according to the probability of moving of the whole non-symptomatic agents (\ref{par}, \ref{p3});

\item regular people, according to the probability of moving of the regular non-symptomatic agents
(\ref{par}, \ref{p4});

\item workers, if all the factories are open or it is open their own workplace
(\ref{par}, \ref{p5});

\item teachers, if the schools are open
(\ref{par}, \ref{p6});

\item students, if the schools are open, but with a possible quota limitation
(\ref{par}, \ref{p7}).

\end{enumerate}

\end{enumerate}

%%%%%%%%%%%%%%%%%%%%%%%%%%%%%%%%%%%%%%
\subsection{Parameter definition}
\label{par}

We define the parameters of Fig. \ref{outline}, also with their short names used in program scripts, in round brackets. The values of the parameters are reported in detail in Appendix~1~-~Parameter~values (Section \ref{app1}).

\begin{enumerate}[label=\roman*]

\item \label{p1} \emph{probabilityOfGettingInfection} (\verb|prob|) is the base probability of getting infected, to be multiplied by the \emph{intrinsicSusceptibility} factor (\ref{p2}); it is activated if the subject is within a circle of radius (\ref{p8}) with an infected person; values at (\ref{app1}, \ref{pp1});

\item \label{D} \emph{D\%}, without the short name, is the percent increasing or decreasing factor of the contagion spread of an asymptomatic subject, compared to that of a symptomatic one, value at (\ref{app1}, \ref{pD});

\item \label{p2} the \emph{intrinsicSusceptibility} in defined in Eq. \ref{intrinsic} 
\begin{equation}
intrinsicSusceptibility = intrinsicSusceptibilityFactor^{groupFragility}
\label{intrinsic}
\end{equation}
with \emph{intrinsicSusceptibilityFactor} set to 5, and $groupFragility$ exponent set to:

\begin{description}
\item [1] for extra-fragile persons,
\item [0] for fragile persons,
\item [-1] for regular persons,
\item [-2] young people from 0 to 24 years old;
\end{description}

\item \label{p3} \emph{\%PeopleAnyTypeNotSymptomaticLeavingHome} (\verb|%PeopleAny|)
determines, in a probabilistic way, the number of people of any kind going around in case of limitations/lockdown; the limitations operate only if the lockdown is on (into our simulated world, from day 20); values at (\ref{app1}, \ref{pp3}); 

\item \label{p4} \emph{\%PeopleNotFragileNotSymptomaticLeavingHome} (\verb|%PeopleNot|)
determines, in a probabilistic way, the number of regular people going around in case of limitations/lockdown;
as above, the limitations operate only if the lockdown is on (into our simulated world, from day 20); values at (\ref{app1}, \ref{pp4});

we try to reproduce the uncertainty of the decisions in the real world into the model via frequent changes of the parameters \ref{p3} and \ref{p4};

NB, the parameters \ref{p3} and \ref{p4} produce independent effects, as in the following examples: (a) the activation of \emph{\%PeopleAny at 31, 0} and, simultaneously, of \emph{\%PeopleNot at 31, 80}, means that people had to stay home on that day, but people specifically not fragile could go out in 80\% of the cases; (b) \emph{\%PeopleAny at 339, 80} and, simultaneously, \emph{\%PeopleNot at 339, 100} means that fragile and not fragile persons cannot always go around, but only in the 80\% of the cases; instead, considering uniquely non-fragile persons they are free to go out; the construction is an attempt to reproduce a fuzzy situation;

in future versions of the model, we will define the quotas straightforwardly:
\begin{itemize}
\item \verb|%FragilePeopleNotSymptomaticLeavingHome|;
\item \verb|%NotFragilePeopleNotSymptomaticLeavingHome|;
\end{itemize}

\item \label{p5} \emph{\%openFactoriesWhenLimitationsOn} (\verb|%Fac|) 
determines, in a probabilistic way, the factories (small and large industries, commercial surfaces, private and government offices) that
are open when limitations are on; if the factory of a worker is open, the subject can go to work, not considering the restrictions (but uniquely in the first step of activity of each day); values at (\ref{app1}, \ref{pp5}); 

\item \label{p6} \emph{stopFragileWorkers} (\verb|sFW|) is \emph{off} (set to 0) by default; if \emph{on} (set to 1), fragile workers (i.e., people fragile due to prior illnesses) can move out of their homes following the \ref{p3} and \ref{p4} parameters, but cannot go to work; in the \emph{off} case, workers (fragile or regular) can go to their factory (if open) also when limitations are on; values at (\ref{app1}, \ref{pp6});

alternatively, we also have the \verb|fragileWorkersAtHome| parameter; if \emph{on} (set to 1) the total of the workers is unchanged, but the workers are all regular; we can activate this counterfactual operation uniquely at the beginning of the simulation;

\item \label{p7} when \emph{activateSchools} (\verb|aSch|) is \emph{on} (set to 1), teachers and students go to school avoiding restrictions (but uniquely in the first step of activity of each day); \emph{\%Students} (\verb|%St|) sets the quota of the students moving to school; the residual part is following the lessons from home; values at (\ref{app1}, \ref{pp7});

\item \label{p8} following \emph{radiusOfInfection} (\verb|radius|), the effect of the contagion---outside enclosed spaces, or there, but for temporary presences---is possible within that distance; values at (\ref{app1}, \ref{pp8});

\item \label{p9} \emph{asymptomaticRegularInfected\%} and \emph{asymptomaticFragileInfected\%}, are the parameters determining the percentage of asymptomatic persons after a contagion for non fragile (all cases) or fragile people;
they are without short names, as they come directly from the model interface; we can see the interface online, activating the model at \href{https://terna.to.it/simul/SIsaR.html}{https://terna.to.it/simul/SIsaR.html}; values at (\ref{app1}, \ref{pp9}).

\end{enumerate}

%%%%%%%%%%%%%%%%%%%%%%%%%%%%%%%%%%%%%%
\subsection{Agents' interaction}
\label{inter}

We underline that our simulation tool is not based on micro-simulation sequences, calculating the contagion agent by agent, on the base of their characteristics and ex-ante probabilities. It implements a true agent-based simulation, with the agents acting and, most of all, interacting. The effect is that of generating continuously contagion situations.

Each run creates a population with expected characteristics, but also with random specifications, to assure the heterogeneity in agents. The daily choices of the agents are partially randomized, to reproduce real-life variability.

Contagions arise from agents' interactions, in four time phases, as specified in Fig. \ref{outline}:

\begin{enumerate}[label=\Alph*]

\item - in houses (at night), hospitals, nursing homes;

\item - in schools and workplaces in general, among people stable there;

\item - in the places above (excluding schools) by people temporary there and in open spaces (UPs above);

\item - interactions mainly in open spaces (UPs above).

\end{enumerate}

%%%%%%%%%%%%%%%%%%%%%%%%%%%%%%%%%%%%%%%%%%%%%%%%%%%%%%%%%
\section{Contagion representation}
\label{contag}

We introduce a tool analyzing the contagions' sequences in simulated epidemics and identifying the places where they occur.

 \begin{itemize}
 \item
We represent each infected agent as a horizontal segment (from the starting date to the final date of the infection) with vertical connections to other agents receiving the disease from it.

 \item
We represent the new infected agents via further segments at an upper level. 

 \item
We display multiple information using three elements.
 \begin{itemize}
 \item Colors in horizontal segments (areas of the infections): black for unknown places, gray for open spaces, cyan for houses, orange for nursing homes, pink for hospitals, yellow for schools, brown for factories, with shops and offices.
 \item Vertical connecting segments keep the same color of the horizontal generating one. 
 \item Line thickness; proportional to fragility.
 \item Styles: dotted lines for incubation, dashed lines for asymptomatic subjects, solid lines of symptomatic ones.
 \end{itemize}
 \item
This graphical presentation enables understanding at a glance how an epidemic episode is developing. In this way, it is easier to reason about countermeasures and, thus, to develop intervention policies. 
 \end{itemize}

At \href{https://github.com/terna/contagionSequence}{https://github.com/terna/contagionSequence} we have the program \emph{sequentialRecords.ipynb}, generating these sequences.

\begin{figure}[t]
\center
\includegraphics[width=0.9\textwidth]{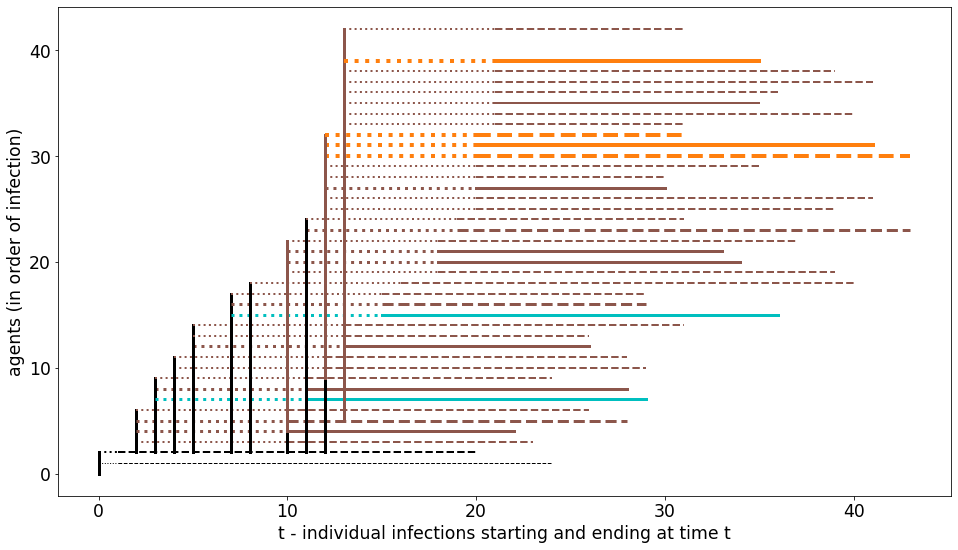}% with control case 473323 474697 in SIsaR_0.9.4.1 experiments 2 seeds with control-table_10000.csv, file withControl_473323_474697.csv
\caption{A case with containment measures, first 40 infections: workplaces (brown) and nursing homes (orange) interweaving}
\label{shortExample}
\end{figure}

Fig. \ref{shortExample} is useful as an example. 
We start with two agents from the outside, with black as the color code (unknown place). The first one is young---as reported by the thickness of the segment, with the infection starting at day 0 and finishing at day 22---and asymptomatic (dashed line); it infects no one. The second one---regular, as reported by the thickness of the segment, with the infection starting at day 0 and finishing at day 15---is asymptomatic (dashed line) and infects four agents on day 2. All the four infected agents receive the infection at work (brown color) and turn to be asymptomatic after the days of incubation (dotted line); the first and the fourth are regular agents; the second and the third are fragile ones.

Continuing the analysis: on day 3, the second agent infects three other agents (at home, at work, at work) [\dots]; on day 13, agent number five infects seven regular agents at work and an extra-fragile one in a nursing home (orange color), etc.

If a vertical segment changes its color, we have an agent in an upper layer infecting someone on the same day of the infection transmitted by an agent in a lower row, so we lose some graphical information.

\begin{figure}[t]
\center
\includegraphics[width=0.48\textwidth]{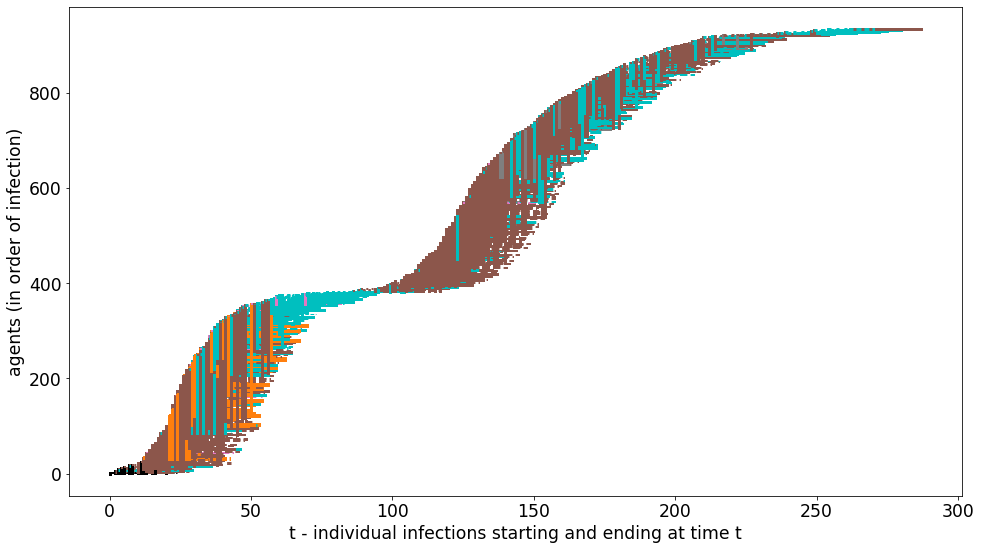}% with control case 473323 474697 in SIsaR_0.9.4.1 experiments 2 seeds with control-table_10000.csv, file withControl_473323_474697.csv
\caption{A case with containment measures, the whole epidemics: workplaces (brown) and nursing homes (orange) and then houses (cyan), with a bridge connecting two waves}
\label{workplacesNursingHomesFull}
\end{figure}

In Fig. \ref{workplacesNursingHomesFull} we see the example of an epidemic with non-pharmaceutical containment measures in adoption: a first wave shows an interlaced effect of contagions at home, in nursing homes, and at work. After a phase in which contagions develop mainly at home, a skinny bridge connects the first wave to a second one, which restarts from workplaces. The thickness of the \emph{snake} of the contagions measures the stock on infects agents on a given date; the slope reports the speediness of the epidemic development; the upper vertical coordinate reports the cumulative number of infected people.

In Appendix~2~-~A gallery of contagion sequences (Section \ref{app2}), we have several examples of contagion sequences.

%%%%%%%%%%%%%%%%%%%%%%%%%%%%%%%%%%%%%%%%%%%%%%%%%%%%%%%%%
\section{Exploring scenarios with simulation batches}
\label{exploring}

The sequences described in Section \ref{contag} offer suggest possible interventions, but are single cases. To explore systematically the introduction of factual, counterfactual, and prospective actions, we need to analyze batches of simulations. In this perspective, each simulation run---whose length coincides with the disappearance of symptomatic or asymptomatic contagion cases---is a datum in a set of different duration and contagion outcomes. To compare the consequences of each batch's basic assumptions, we need to represent compactly the results emerging from simulation repetitions . 

We use blocks of ten thousand repetitions. Besides summarizing the results with the usual statistical indicators, we adopt the technique of the heat-maps. With \cite{steinmann2020don}, our goal is that of making comparative analyzes, not forecasts. This consideration is consistent with the enormous standard deviation values that are intrinsic to the specific reality. 

At \href{https://github.com/terna/readSIsaR\_BatchResults}{https://github.com/terna/readSIsaR\_BatchResults} we have the codes producing the maps of the batches. A heat-map is a double histogram: in our application, it displays each simulated epidemic's duration in the $x$ axis and the total number of the symptomatic, asymptomatic, and deceased agents in the $y$ axis (on a scale of 1:1000). Each cell contains the number of epidemics with $x$ duration and $y$ outcome. Besides the number, a logarithmic color scale improves the readability of the maps.

%%%%%%%%%%%%%%%%%%%%%%%%%%%%%%%%%%%%%%%%%%%%%%%%%%%%%%%%%
\subsection{Epidemics without and with control measures}
\label{withoutWith}

\begin{figure}[t]

 \begin{subfigure}{0.48\textwidth}
 \centering
 \fbox{\includegraphics[width=0.95\textwidth]{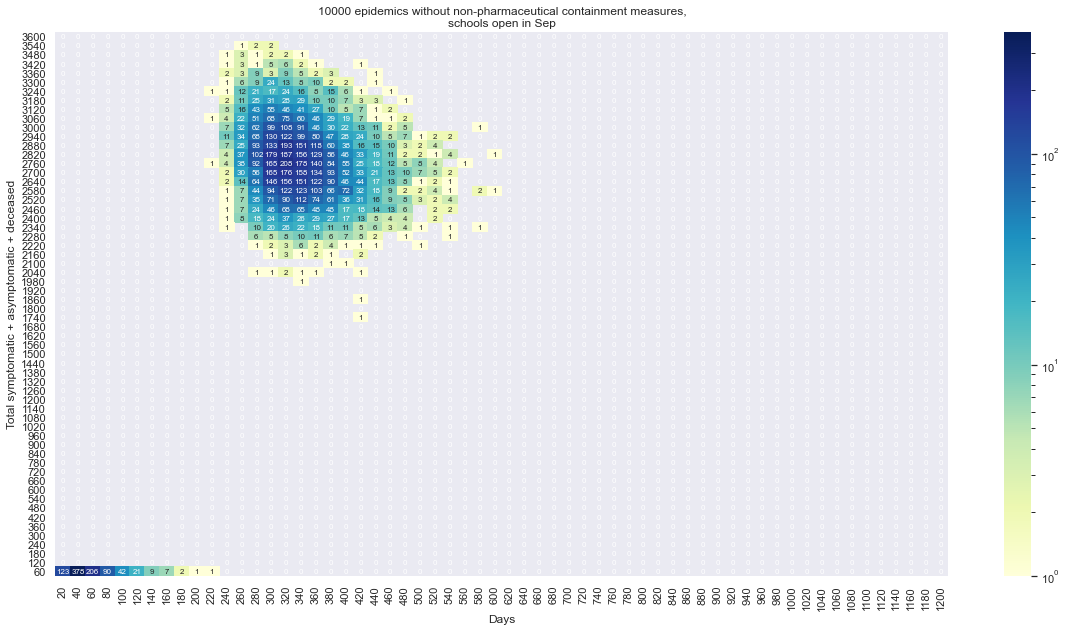}}
 \subcaption{Outbreaks without non-pharmaceutical containment measures}
 \label{10kNoControl}
 \end{subfigure}
 \hfill
 \begin{subfigure}{0.48\textwidth}
 \centering
 \fbox{\includegraphics[width=0.95\textwidth]{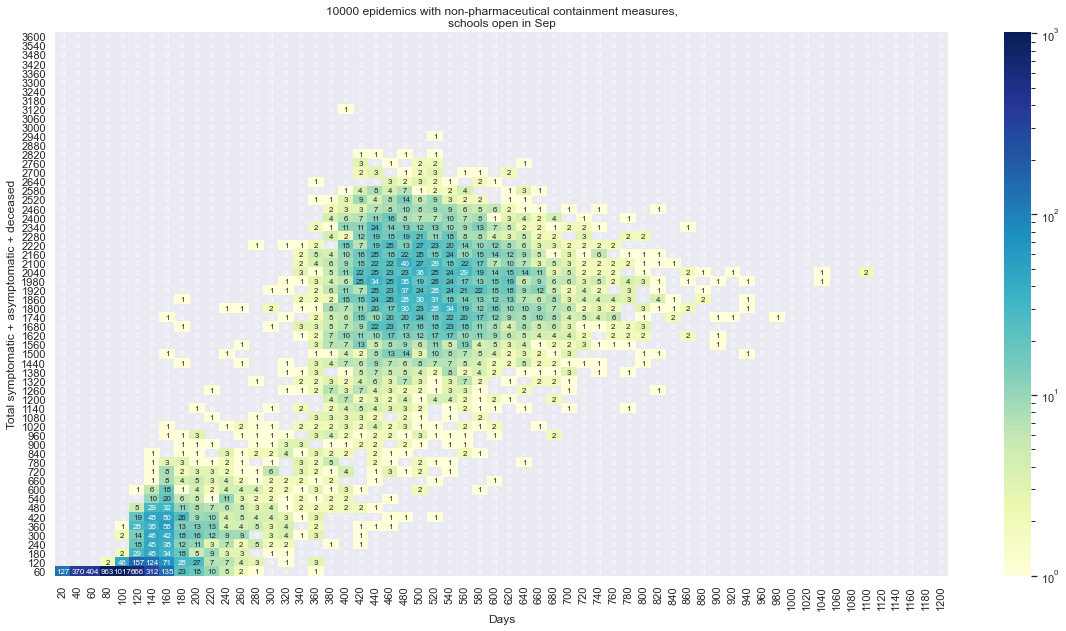}}
 \subcaption{Outbreaks with non-pharmaceutical containment measures}
 \label{10kBasicC}
 \end{subfigure}
 \caption{Starting our analyses: 10,000 epidemics in Piedmont}
 \label{withoutWithFig}
\end{figure}

As a starting point, we compare the situations represented in Figs. \ref{10kNoControl} and \ref{10kBasicC}. In Fig. \ref{10kNoControl}, the heat-map reports the distribution in duration and infection causation of 10,000 simulated outbreaks left to spread without any control; coherently, with the school always open. 
The results in Table \ref{noCTab} are scary. The concentration of the cases in the heat-map shows that, except a few instances spontaneously concluding in a short period (left bottom corner), produces a heavy \emph{cloud} of cases lasting one year or one year and a half, hitting (as symptomatic, asymptomatic and deceased) from 2,000 to 3,500 persons on a total of 4,350 in the region (scale of 1:1000).

In Fig. \ref{10kBasicC} and the related Table \ref{basicCTab}, we report a similar simulation batch of 10,000 runs of the model, but with the adoption of the basic non-pharmaceutical containment measures, registered in the values of the parameters in Appendix~1~-~Parameter~values (Section \ref{app1}). A calendar is at \href{https://terna.to.it/simul/calendario092.pdf}{https://terna.to.it/simul/calendario092.pdf}, and the model---version 0.9.6--- is updated until April 2021. The results are dramatically different, showing the efficacy of the containment measures.

\begin{table}[t]
\center
\begin{tabular}{lrrr}
\toprule
(000) & symptomatic & totalInfected\&Deceased & duration \\
\midrule
mean & 969.46 & 2500.45 & 303.10 \\
std & 308.80 & 802.88 & 93.50 \\
\bottomrule
\end{tabular}
\caption{Mean values and standard deviations in Fig. \ref{10kNoControl} cases}
\label{noCTab}
\end{table}

\begin{table}[t]
\center
\begin{tabular}{lrrr}
\toprule
(000) & symptomatic & totalInfected\&Deceased & duration \\
\midrule
mean & 344.22 & 851.64 & 277.93 \\
std & 368.49 & 916.41 & 213.48 \\
\bottomrule
\end{tabular}
\caption{Mean values and standard deviations in Fig. \ref{10kBasicC} cases}
\label{basicCTab}
\end{table}

%%%%%%%%%%%%%%%%%%%%%%%%%%%%%%%%%%%%%%%%%%%%%%%%%%%%%%%%%
\subsection{Actual data}
\label{actual}

\begin{figure}[t]

 \begin{subfigure}{0.48\textwidth}
 \centering
 \fbox{\includegraphics[width=0.95\textwidth]{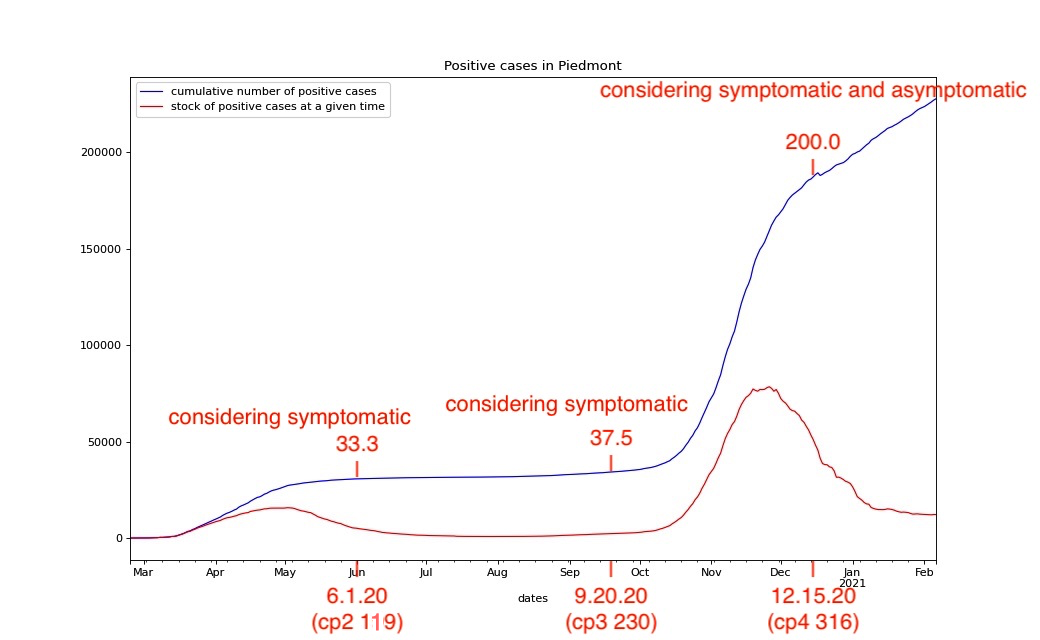}}
 \subcaption{Critical points in epidemic dynamic in Summer and Fall 2020 in Piedmont}
 \label{actualA}
 \end{subfigure}
 \hfill
 \begin{subfigure}{0.48\textwidth}
 \centering
 \fbox{\includegraphics[width=0.95\textwidth]{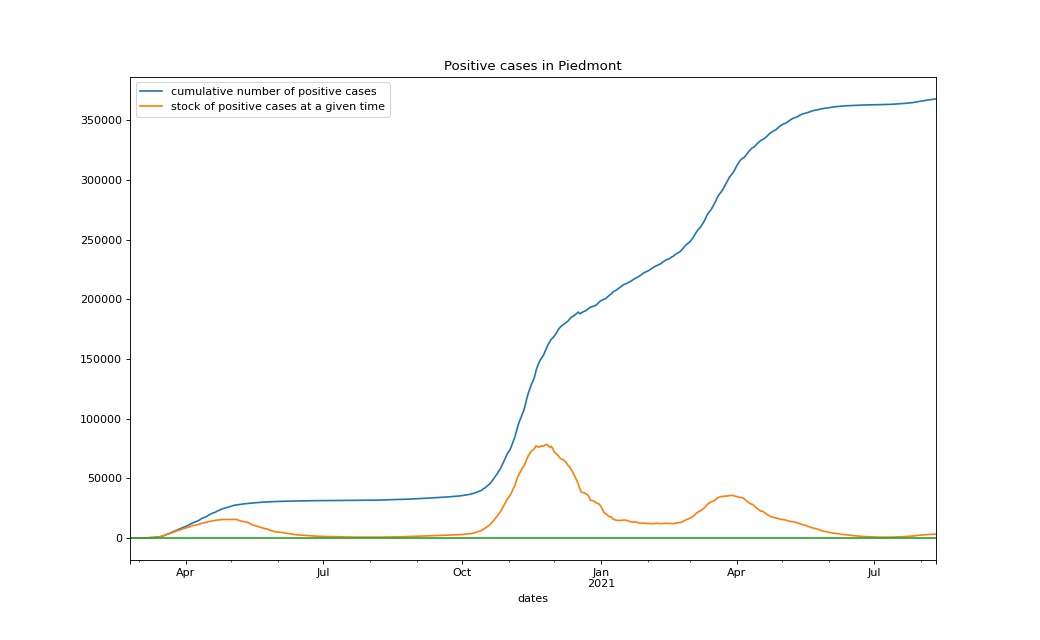}}
 \subcaption{Data in Piedmont until July 2021, showing three waves}
 \label{actualB}
 \end{subfigure}
 \caption{Actual data}
 \label{actualData}
\end{figure}

The critical points for our simulation experiments in Piedmont are Summer and Fall 2020 in Fig. \ref{actualA}, where we have the time series of the first part of Piedmont's actual epidemic. The blue line represents the cumulative number of infected persons. Initially, only symptomatic cases were accounted for, but after the 2020 Summer, with more generalized tests, also asymptomatic patients are included:

\begin{itemize}

\item from \href{http://www.protezionecivile.it/web/guest/department}{http://www.protezionecivile.it/web/guest/department}, the Italian Civil Protection Department web site, we find at \href{https://github.com/pcm-dpc/COVID-19}{https://github.com/pcm-dpc/COVID-1}, i.e., the repository of regional data; 

\item we observe data about symptomatic infected people in the first wave, but from October 2020, data are mixed: in the above \emph{git} repository, in October and November, we had ``Positive cases emerged from clinical activity'', unfortunately then reported as ``No longer populated'' (from the end of November, our observation) and ``Positive cases emerging from surveys and tests, planned at national or regional level'', again ``No longer populated'' (from the end of November, our observation); 

\item as a consequence, the subdivision between symptomatic and asymptomatic cases is impossible after that date.
\end{itemize}

Considering the dynamic of the data in Fig. \ref{actualA}, we search within the simulation batch for cases with both:

\begin{enumerate}[label=\roman*]
\item numbers of infected persons quite similar at \verb|cp2| and at \verb|cp3|; besides, numbers not too different from those of the figure; 
(with \emph{cp}, we indicate the internal check points of the simulation program; in Fig. \ref{actualA} we also report the number of days from the beginning of the epidemic for each check point); 

\item the number of infected persons at \verb|cp4| has to be significantly greater than those at the previous check point.

\end{enumerate}

In a lot of cases, epidemics satisfying condition (i) fail to match condition (ii); both the situations happen only in less than the 1.5\% of the instances in a batch of ten thousand epidemic. We can guess that the second wave registered in Piedmont after the Summer ``pause'' is due to new infected agents coming from outside and restarting the contagion process.

Other critical points in our analysis are the day on which the vaccination campaign starts, 373 of the simulation (Feb. \nth{12}, 2021), and the day of the effectiveness of the initial vaccinations, 40 days later, day 413 (Mar. \nth{22}, 2021). At those dates, within the simulations, we can find either the presence of many infected agents or of few ones, as effectively was the situation in Piedmont.

NB, we concluded model calculations in April 2021. In Fig. \ref{actualB}, the time series covering the whole period.

%%%%%%%%%%%%%%%%%%%%%%%%%%%%%%%%%%%%%%%%%%%%%%%%%%%%%%%%%
\section{Factual and counterfactual analyses}
\label{facCounterfac}

\begin{figure}[H]

 \begin{subfigure}{0.48\textwidth}
 \centering
 \fbox{\includegraphics[width=0.95\textwidth]{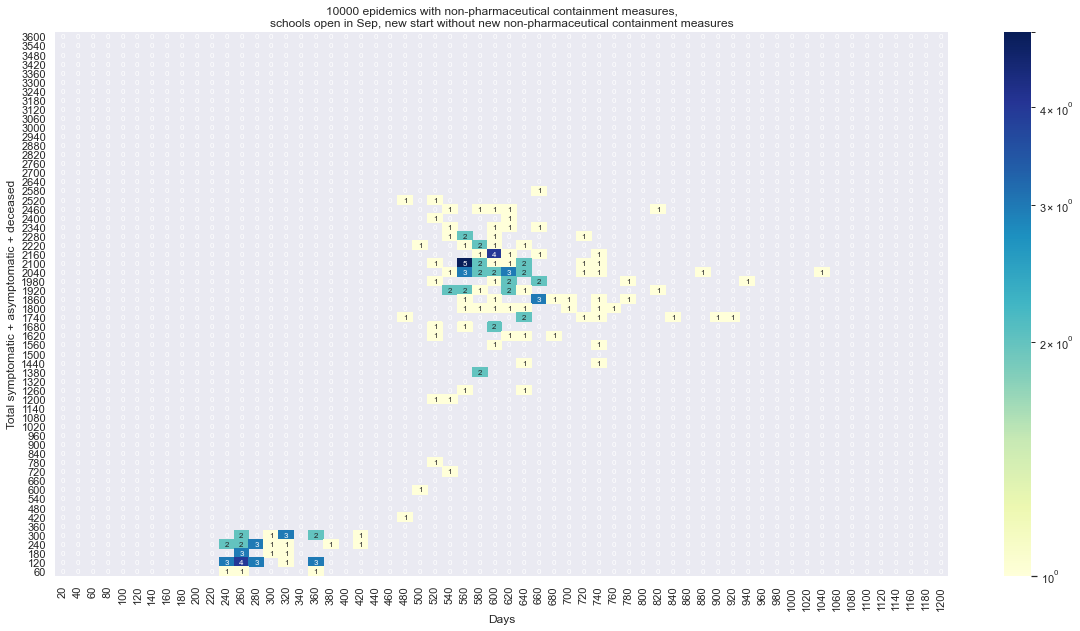}}
 \subcaption{First wave with non-pharmaceutical containment measures, spontaneous second wave, without specific measures}
 \label{spontaneousSec}
 \end{subfigure}
 \hfill
 \begin{subfigure}{0.48\textwidth}
 \centering
 \fbox{\includegraphics[width=0.95\textwidth]{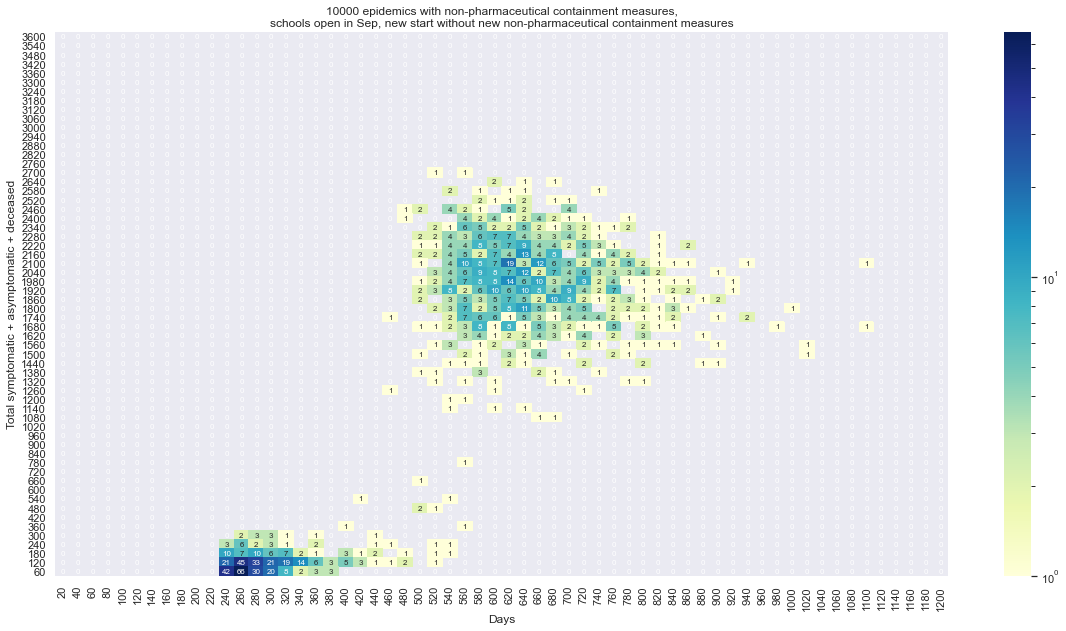}}
 \subcaption{First wave with non-pharmaceutical containment measures, forcing the second wave, without specific measures}
 \label{secondFromOut}
 \end{subfigure}
 \begin{subfigure}{0.48\textwidth}
 \centering
 \fbox{\includegraphics[width=0.95\textwidth]{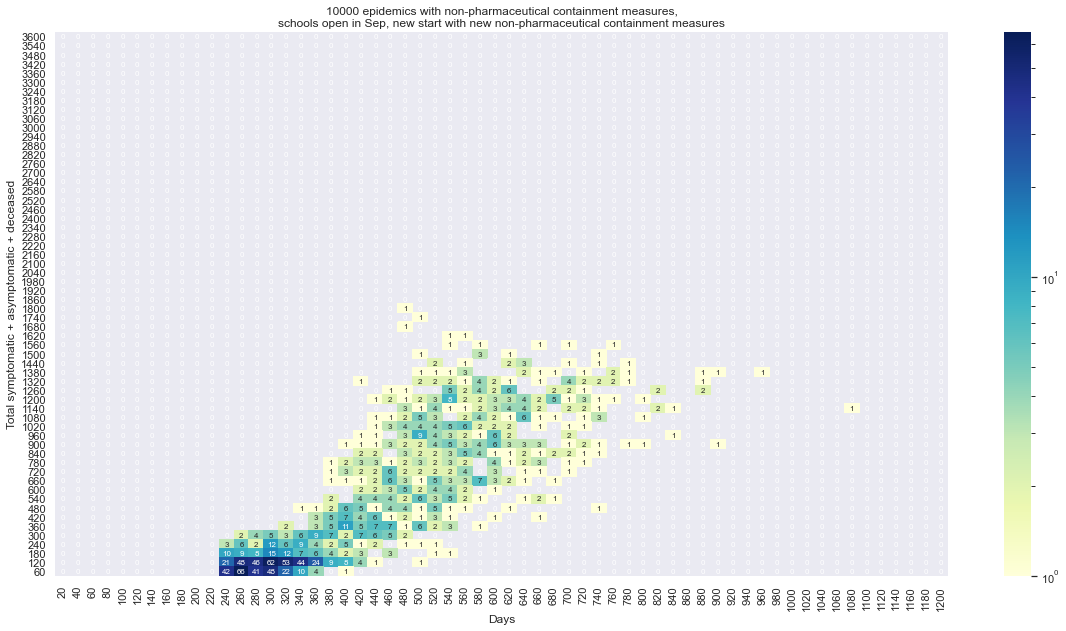}}
 \subcaption{First wave with non-pharmaceutical containment measures, forcing the second wave, \emph{with new specific non-ph. containment measures}}
 \label{secondFromOutWith}
 \end{subfigure}
 \hfill
 \begin{subfigure}{0.48\textwidth}
 \centering
 \fbox{\includegraphics[width=0.95\textwidth]{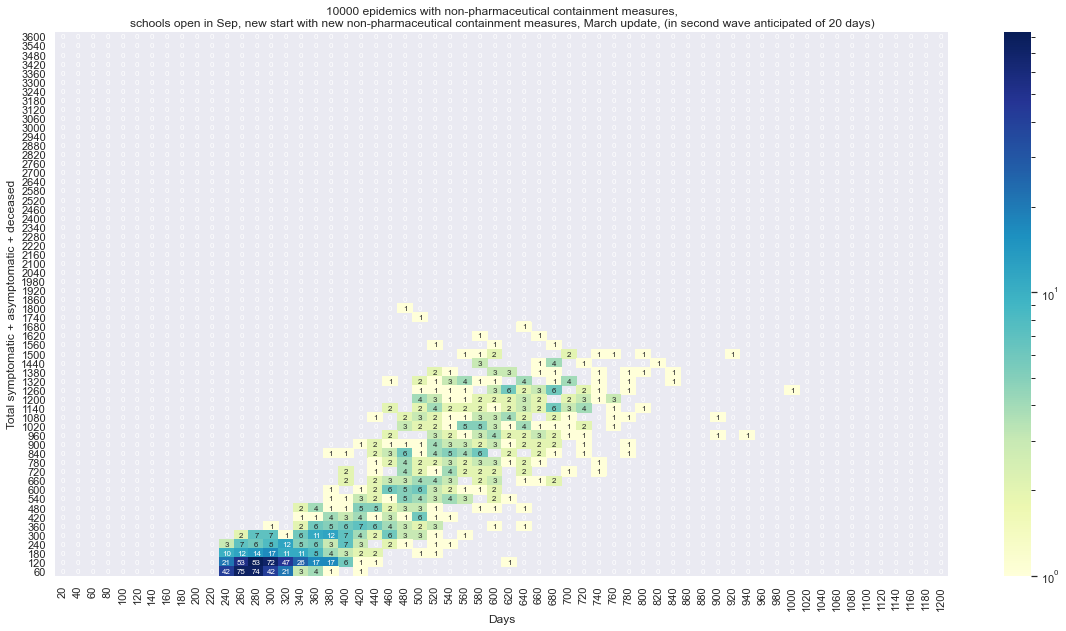}}
 \subcaption{First wave with non-pharmaceutical cont. meas,, forcing the second w., \emph{with new specific non-ph. cont. meas., acting 20 days in advance}}
 \label{secondFromOut-20}
 \end{subfigure}
 \center
 \begin{subfigure}{0.48\textwidth}
 \centering
 \fbox{\includegraphics[width=0.95\textwidth]{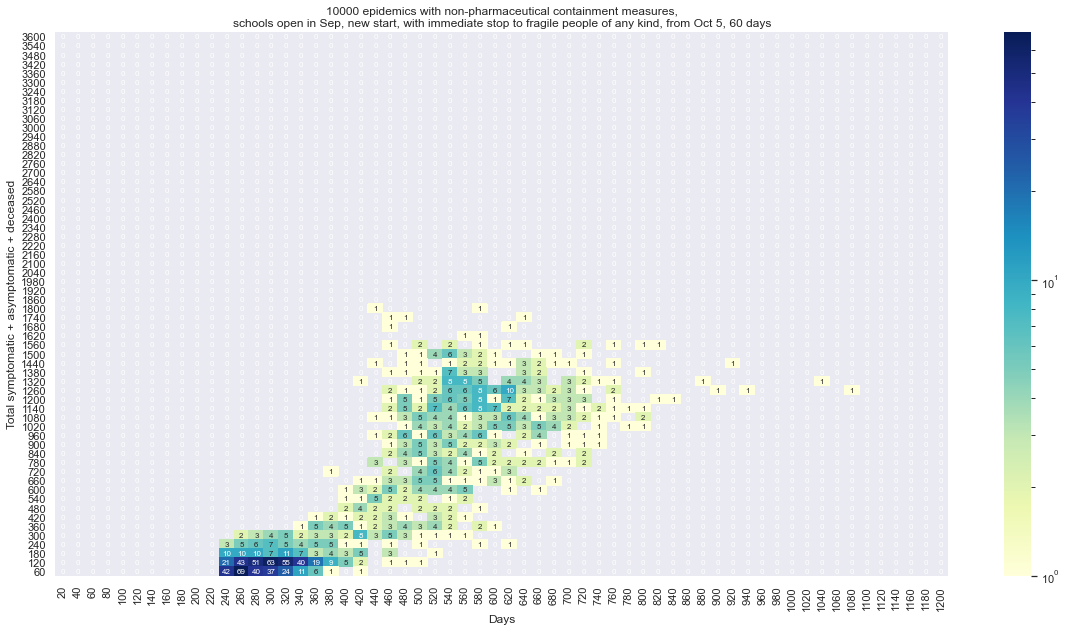}}
 \subcaption{First wave with non-ph. cont. meas., forcing the sec. wave; \emph{in sec. wave, uniquely stopping fragile people, including fragile workers}}
 \label{secondFromOutNoFrag}
 \end{subfigure}
 \caption{Heat-maps of the factual and counterfactual analyses}
 \label{casesHeatM}
\end{figure}

In Fig. \ref{casesHeatM} we collect the heat-maps of the experiments:

\begin{itemize}

\item observing the emergence of spontaneous second waves, in the absence of specific control measures (Sections \ref{spont});

\item causing the emergence of the second wave through infections from outside, again in the absence of specific control measures (Section \ref{secondWithout});

\item causing the emergence of the second wave through infections from outside, in the presence of specific control measures (Section \ref{secondWith});

\item reproducing the case of Section \ref{secondWith}, anticipating by twenty days the start and end of all control measures (Section \ref{anticip});

\item reproducing the case of Section \ref{secondWith}, limiting the control measures to fragile workers and other fragile people (Section \ref{frag}).

\end{itemize}

%%%%%%%%%%%%%%%%%%%%%%%%%%%%%%%%%%%%%%%%%%%%%%%%%%%%%%%%%
\subsection{Spontaneous second wave, without specific containment measures}
\label{spont}
% selectResults10kStableSeedsCPoints_basicControlB_schoolOpenSeptChangingWorld_plusHMlog,ipynb
% using 10kCtrl1.csv from SIsaR_0.9.5.4.1trials10kCtrl1.nlogo

In an initial plain batch of runs of the Piedmont model, we count only 140 cases of epidemics with both the absence of new contagions in Summer 2020 and their explosion in Fall, as in Fig. \ref{actualA}.

The steps are:
\begin{itemize}

\item we select, first of all, the 170 cases of epidemics that. on June \nth{1} have, a number of symptomatic agents in the (10, 70] interval (with mean: 37.9) and, on September \nth{20}, a number of symptomatic agents in the (20, 90] interval (with mean: 60.4);

\item due to the lack of data described in Section \ref{actual}, to compare December \nth{15} and September \nth{20} situations, we use symptomatic plus asymptomatic agents' count;

\item we observe the existence of 140 outbreaks with the required characteristics; the December mean of the infected agents is 648.7, sensibly larger than the actual value: $\approx$200.0. 

\end{itemize}

We overestimate the reality being the long-lasting simulated outbreaks, the larger ones, and, most of all, having no containment measures operating in the simulations.

Fig. \ref{spontaneousSec} and Table \ref{selSpontWave2Tab} show the outbreaks with similar cumulative numbers before and after the Summer 2020 ``pause'' (170 cases), with the second wave (140 cases) in the absence of containment measures.

140 out of 10,000, i.e., 1.4\%, is a very light spontaneous ratio for the second wave occurred in the Fall. The transition to the third wave, that we see in Fig. \ref{actualB}, is easy to explain, as the second wave never completely ended.

\begin{table}[t]
\center
\tiny
\begin{tabular}{lrrrrrrrrrrrrr}
\toprule
(1000) & Jun~1 & & Sep~20 & & Dec~15 & & Feb~1 & & May~1 & & Dec~15\\
&2020 &&2020 &&2020&&2021&&2021&&2020\\
&&&&&&&&&&& to~end \\
cum.~v. & sym. & all & sympt. & totalInf. & sympt. & totalInf. & sympt. & totalInf. & sympt. & totalInf. & sympt. & totalInf. & days\\
\midrule
count & 170.0 & 170.0 & 170.0 & 170.0 & 140.0 & 140.0 & 131.0 & 131.0 & 128.0 & 128.0 & 140.0 & 140.0 & 140.0 \\
mean & 37.9 & 100.2 & 60.4 & 159.3 & 248.4 & 648.7 & 432.2 & 1109.5 & 656.3 & 1655.5 & 701.1 & 1757.9 & 594.2 \\
std & 16.4 & 61.0 & 19.6 & 71.7 & 167.4 & 424.3 & 220.4 & 538.4 & 215.4 & 513.3 & 246.4 & 599.7 & 118.9 \\
\bottomrule
\end{tabular}
\caption{Spontaneous second wave, without specific measures}
\label{selSpontWave2Tab}
\end{table}

%%%%%%%%%%%%%%%%%%%%%%%%%%%%%%%%%%%%%%%%%%%%%%%%%%%%%%%%%
\subsection{Second wave, new infections from outside, without specific containment measures}
\label{secondWithout}
% selectResults10kStableSeedsCPoints_basicControlB_schoolOpenSeptChangingWorldNewStart_plusHMlog.ipynb
% using 10kCtrl1NStart.csv from SIsaR_0.9.5.4.1trials10kCtrl1NStart.nlogo

To generate a framework consistent with the presence of a second wave after a period of substantial inactivity of the epidemic, we introduced two cases of infected persons coming back from outside after Summer vacancies, conventionally on September \nth{1}, 2020.

As above, the steps are:
\begin{itemize}

\item we select, first of all, the 1407 cases of epidemics that. on June \nth{1} have, a number of symptomatic agents in the (10, 70] interval (with mean: 35.6) and, on September \nth{20}, a number of symptomatic agents in the (20, 90] interval (with mean: 40.0);

\item due to the lack of data described in Section \ref{actual}, to compare December \nth{15} and September \nth{20} situations, we use symptomatic plus asymptomatic agents' count;

\item we observe the existence of 1044 outbreaks with the required characteristics; the December mean of the infected agents is 462.1, again sensibly larger than the actual value: $\approx$200.0. 

\end{itemize}

We overestimate the reality being the simulations run without the adoption of containment measures.

Both Fig. \ref{secondFromOut} and Table \ref{selForceWave2Tab} show the outbreaks with similar cumulative numbers before and after the Summer 2020 ``pause'' (1407 cases), with the second wave of 1044 cases. In the absence of containment measures, we have a heavy cloud as that of Fig. \ref{10kNoControl}, with infected people of any kind in a range approximately of 1,500 to 2,800 realizations, with an equivalence, to the Piedmont scale, to 1,5-2,8 millions of subjects.

The number of cases is now sufficient to evaluate the effects of factual (Section \ref{secondWith} and counterfactual (Sections \ref{anticip} and \ref{frag}) simulation experiments.

\begin{table}[t]
\center
\tiny
\begin{tabular}{lrrrrrrrrrrrrr}
\toprule
(1000) & Jun~1 & & Sep~20 & & Dec~15 & & Feb~1 & & May~1 & & Dec~15\\
&2020 &&2020 &&2020&&2021&&2021&&2020\\
&&&&&&&&&&& to~end \\
cum.~v. & sym. & all & sympt. & totalInf. & sympt. & totalInf. & sympt. & totalInf. & sympt. & totalInf. & sympt. & totalInf. & days\\
\midrule
count & 1407.0 & 1407.0 & 1407.0 & 1407.0 & 1044.0 & 1044.0 & 1005.0 & 1005.0 & 980.0 & 980.0 & 1044.0 & 1044.0 & 1044.0 \\
mean & 35.6 & 72.7 & 40.0 & 84.1 & 180.4 & 462.1 & 354.1 & 900.4 & 623.8 & 1563.3 & 726.6 & 1810.9 & 620.9 \\
std & 14.1 & 42.6 & 16.7 & 52.8 & 134.6 & 354.6 & 213.8 & 535.4 & 217.9 & 527.0 & 221.9 & 544.0 & 110.8 \\
\bottomrule
\end{tabular}
\caption{Second wave, new infections from outside, without specific measures}
\label{selForceWave2Tab}
\end{table}

%3%%%%%%%%%%%%%%%%%%%%%%%%%%%%%%%%%%%%%%%%%%%%%%%%%%%%%%%%
\subsection{Second wave, new infections from outside, with new specific containment measures}
\label{secondWith}
% selectResults10kStableSeedsCPoints_basicControlB_schoolOpenSeptOctMarControlChangingWorldNewStart_plusHMlog.ipynb
% using 10kCtrl1NStartCtrl2M.csv from SIsaR_0.9.5.4.1trials10kCtrl1NStartCtrl2M.nlogo

Repeating the third step above:
\begin{itemize}

\item we observe the existence of 874 outbreaks with the required characteristics; the December mean of the infected agents is 340.6, closer to the actual value ($\approx$200.0) due to the introduction into the simulation of specific control measures for the second wave.. 

\end{itemize}

We always overestimate the reality because the surviving epidemics are the larger ones.

In Fig. \ref{secondFromOutWith} we see that the heavy cloud of the previous figure dissolved, and in Table \ref{selForceWave2Contr2Tab} the numbers in italic emphasize the positive effects of the containment interventions on the cases of epidemic continuation (which have also dropped in quantity).

\begin{table}[t]
\center
\tiny
\begin{tabular}{lrrrrrrrrrrrrr}
\toprule
(1000) & Jun~1 & & Sep~20 & & Dec~15 & & Feb~1 & & May~1 & & Dec~15\\
&2020 &&2020 &&2020&&2021&&2021&&2020\\
&&&&&&&&&&& to~end \\
cum.~v. & sym. & all & sympt. & totalInf. & sympt. & totalInf. & sympt. & totalInf. & sympt. & totalInf. & sympt. & totalInf. & days\\
\midrule
count & 1407.0 & 1407.0 & 1407.0 & 1407.0 & 874.0 & 874.0 & 719.0 & 719.0 & 523.0 & 523.0 & 874.0 & 874.0 & 874.0 \\
mean & 35.6 & 72.7 & 40.0 & 84.1 & \emph{130.0} & \emph{340}.6 & \emph{194.4} & \emph{512.8} & \emph{295.7} & \emph{791.2} & 252.7 & 666.4 & 494.1 \\
std & 14.1 & 42.6 & 16.7 & 52.8 & 83.9 & 232.6 & 104.1 & 276.9 & 119.1 & 300.6 & 156.8 & 416.4 & 122.7 \\
\bottomrule
\end{tabular}
\caption{Second wave, new infections from outside, with new specific measures}
\label{selForceWave2Contr2Tab}
\end{table}

%%%%%%%%%%%%%%%%%%%%%%%%%%%%%%%%%%%%%%%%%%%%%%%%%%%%%%%%%
\subsection{Calculating the reproduction number without delays}
\label{indicator}
The reproduction number $R_t$ \cite{BettencourtRibeiro2008, CorietAl2013} \begin{quote}is the average number of secondary cases of disease caused by a single infected individual over his or her infectious period\end{quote}
and is defined as follows:
\begin{equation}
\large{R_t = \frac{I_t}{\sum_{s=1}^t w_s I_{t-s}}}
\label{RtDef}
\end{equation}
where:
\begin{itemize}
\item
$I_t$ is the number of new infected individuals at time $t$
\item 
$w_s=\Gamma(s; \alpha,\beta)$ is the infectivity profile, usually approximated with the serial interval distribution \cite{CorietAl2013}; it shapes the infectious period of each individual by weighting the infected individuals so that when their period is over, they do not count any more in the sum; it is usually assumed to be a the Gamma distribution \cite{CorietAl2013}
\begin{itemize}
\item 
there is great uncertainty on the parameters of the Gamma distribution, which have been fitted to different values on different national data-sets (\cite{RoyalSocSAGESPI-M}, table 1 page 25)
\item following the Istituto Superiore di Sanit\`{a} (ISS), italian $R_t$ estimates are based on the parameters fitted in \cite{CeredaetAl2020}, namely $\alpha = 1.87$ and $\beta = 0.28$
\end{itemize}
\end{itemize}

\begin{figure}[t]
\center
\includegraphics[width=0.7\textwidth]{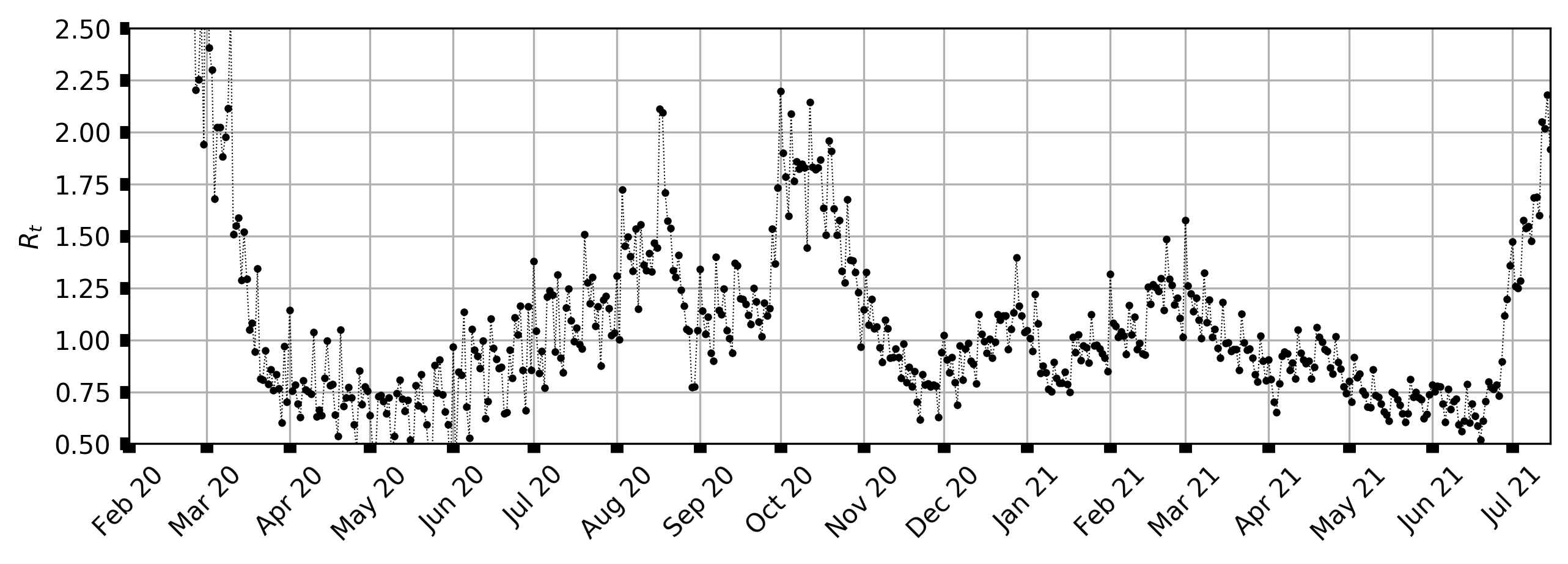}
\caption{Naive $R_t$ calculated on raw infected cases by symptoms onset date, data-set by ISS}
\label{naive_Rt_symptoms_onset}
\end{figure}

While Eq. \ref{RtDef} could in principle be applied naively to any time series of the new infected cases, it usually leads to noisy results caused by the noise contained in the original raw series, as can be seen in Fig. \ref{naive_Rt_symptoms_onset}. Moreover, despite the noisy content of the original signal, the naive approach does not give any clue on the confidence interval of the result, which is fundamental if the reproduction number has to be used to take decisions about the restrictions.

The most widely adopted approach to extract statistics about the $R_t$ estimate, and hence its confidence interval, is to apply Bayesian statistical inference, assuming a prior distribution for the serial interval and a posterior for the reproduction number \cite{CorietAl2013}. 

While Bayesian inference allows us to compute any kind of statistics on the estimate, it still fails  dealing with the noise in the original signal, leading again to spiky estimates of $R_t$.

\begin{figure}[t]
\center
\includegraphics[width=0.7\textwidth]{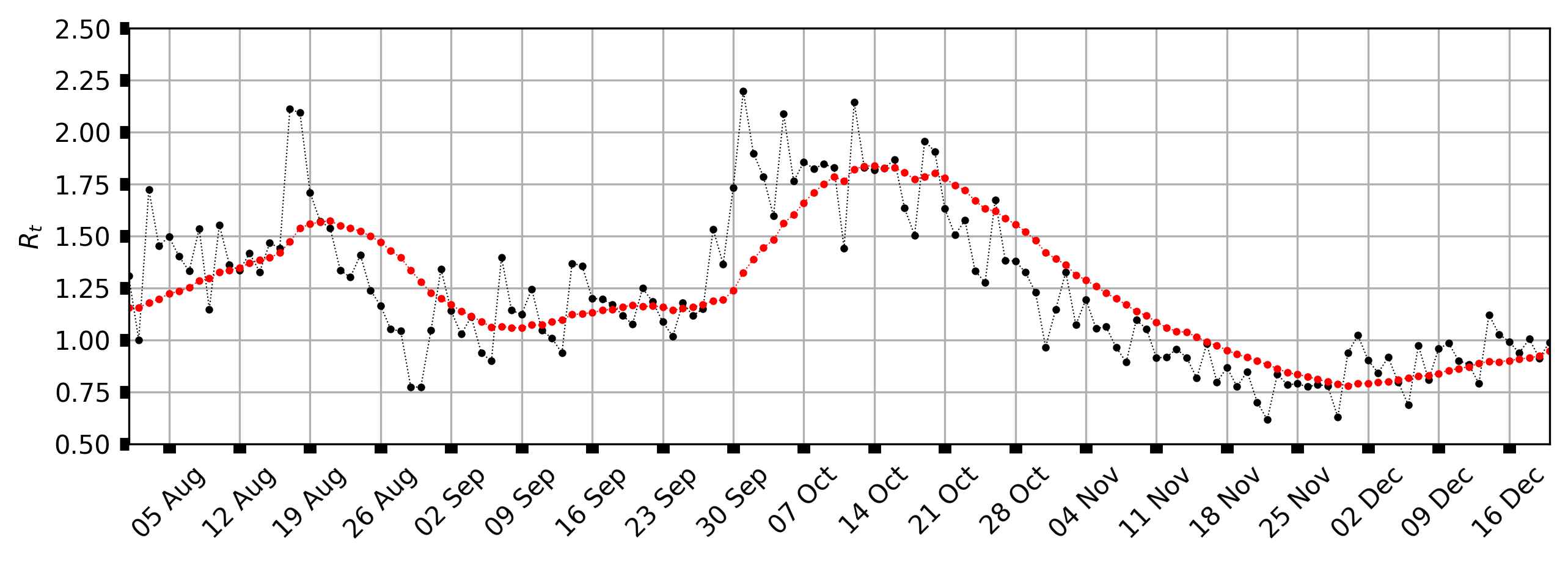}
\caption{In black $R_t$ calculated on raw infected cases, in red the average $R_t$ calculated over a window of 14 days; both series are by symptoms onset date, data-set by ISS}
\label{naive_Rt_symptoms_onset_window_14}
\end{figure}

The standard solution to smooth out the noise is to assume that the transmissibility is constant over a time window (e.g. a week): we can then estimate the average $R_t$ over the time window \cite{CorietAl2013}, by computing the total number of new infected cases over a window $\tau$ instead of those of each single day: $\hat{I}_{t,\tau}=\sum_{s=t-\tau}^{t} I_s$ and replacing it to $I_t$ in Eq. \ref{RtDef}; note that this is equivalent to compute $R_t$ on the average of $I_t$ over the window, as Eq. \ref{RtDef} is invariant under constant scaling of $I_t$. 

The result is smoothed, but it turns out that it is delayed by the size of the windows. Fig. \ref{naive_Rt_symptoms_onset_window_14} shows $R_t$ calculated over a 14 days rolling windows (14 days is the window size officially adopted in Italy); it is clearly visible that the average $R_t$ is systematically delayed: maximizing the cross correlation of the signals confirms a measure of the delay of 14 days.

\subparagraph{Official data-sets}
Data used in all the computations refers to the following sources:
\begin{itemize}
    \item data-set by ISS: count of new infected individuals by symptoms onset date, at \url{https://github.com/tomorrowdata/COVID-19/blob/main/data/sources/ISS/covid_19-iss_2021-07-30T22:34:44\%2B00:00.inizio_sintomi.csv} downloaded on Jul 30
    \item data-set by Protezione Civile: count of new infected individuals by notification date, at \url{https://github.com/pcm-dpc/COVID-19}{https://github.com/pcm-dpc/COVID-19} dowloaded on Jul 31
\end{itemize}

\subsubsection{Tikhonov regularization to smooth the original signal}

As an alternative solution to averages, we adopt Tikhonov regularization to the original signal, which does not introduce delays.

$I_t$ is smoothed by fitting a series to represent the derivative of $I_t$ and then integrating it back to the original signal, which then results in a smoothed one.

We search for the differential signal $\vec{\omega}$ such that:
$$\vec{I} = \vec{X}\cdot \vec{\omega}$$
where $\vec{I}$ denotes the array of elements $I_t$ and $\vec{X}$ is the matrix representing the integration operator:
$$\vec{X} = \begin{bmatrix} 
1 & 0 & ... & 0 \\
1 & 1 & ... & 0 \\
... & ... & ...  & ... \\
1 & 1 & ... & 1 \\
\end{bmatrix}
$$

$\vec{\omega}$ is obtained by minimizing the following cost function:
\begin{equation}
F(\vec{\omega}) = \left\|\vec{I} - \vec{X}\cdot\vec{\omega}\right\|^2 +\alpha^2 \left\|\vec{\Gamma}\cdot\vec{\omega}\right\|^2
\label{TickCostFunction}
\end{equation}

Hence the derivative $\vec{\omega}$ is fitted using a Ridge regression with a generalized Tikhonov regularization factor:
\begin{itemize}
    \item 
    $\vec{\Gamma}$: the Tikhonov regularization matrix, chosen to be the second derivative operator;
    \item
    $\alpha$: the regularization factor.
\end{itemize}

The regularization factor penalizes the spikes in the second derivative, forcing the derivative to be a smoothed signal. Once the derivative is fitted, the original signal is reconstructed by applying again the integral matrix to the differentiated smoothed signal; denoting the smoothed signal by $\Bar{\vec{I}}$:
$$\Bar{\vec{I}} = \vec{X}\cdot\vec{\omega}$$

The parameter $\alpha$ can be obtained by searching the maximal smoothness constrained to the desired degree of information still available in the signal. It can be shown empirically that $\alpha = 100$ represents a reasonable trade-off.

Once we have $\Bar{I}_t$ it can be fed into Eq. \ref{RtDef} to obtain the reproduction number computed on the smoothed signal, which we denote by $\Bar{R}_t$.

\begin{figure}[t]
\center
\includegraphics[width=0.7\textwidth]{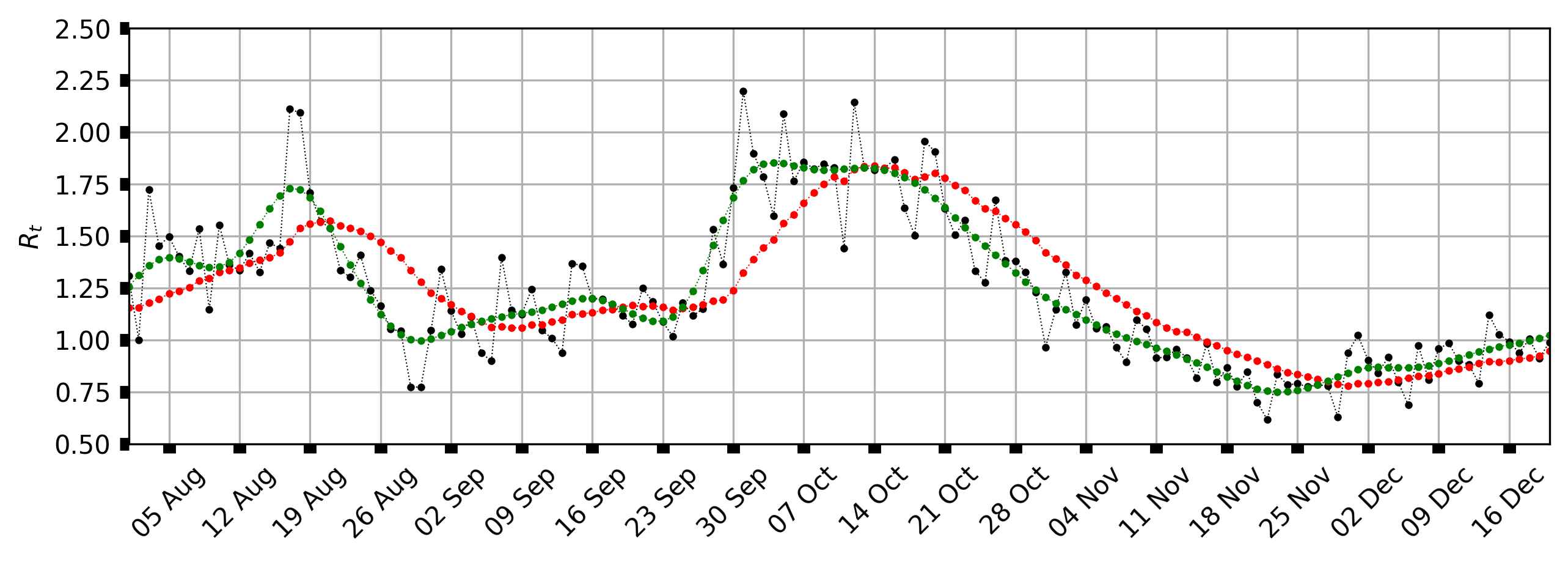}
\caption{In black $R_{t}$ calculated on raw infected cases, in red the average $R_t$ calculated over a window of 14 days; in green $\Bar{R}_{t}$ calculated on the signal smoothed with Tikhonov regularization; each series is by symptoms onset date, data-set by ISS}
\label{naive_Rt_symptoms_onset_smoothed}
\end{figure}

Fig. \ref{naive_Rt_symptoms_onset_smoothed} shows in green the result of calculating $\Bar{R}_t$ on the signal smoothed by minimizing Eq. \ref{TickCostFunction}. It is clearly visible that the green line anticipates the red one: maximizing the cross correlation of the original noisy $R_t$ wrt $\Bar{R}_t$ confirms a measure of the delay of 0 days.

\subsubsection{Do not wait for the symptoms onset date}
\label{notification_date}

Delays are not as important in literature, where we usually look at historical data, as they are in policy making, where we do need near real-time data.

\begin{figure}[t]
\center
\includegraphics[width=0.7\textwidth]{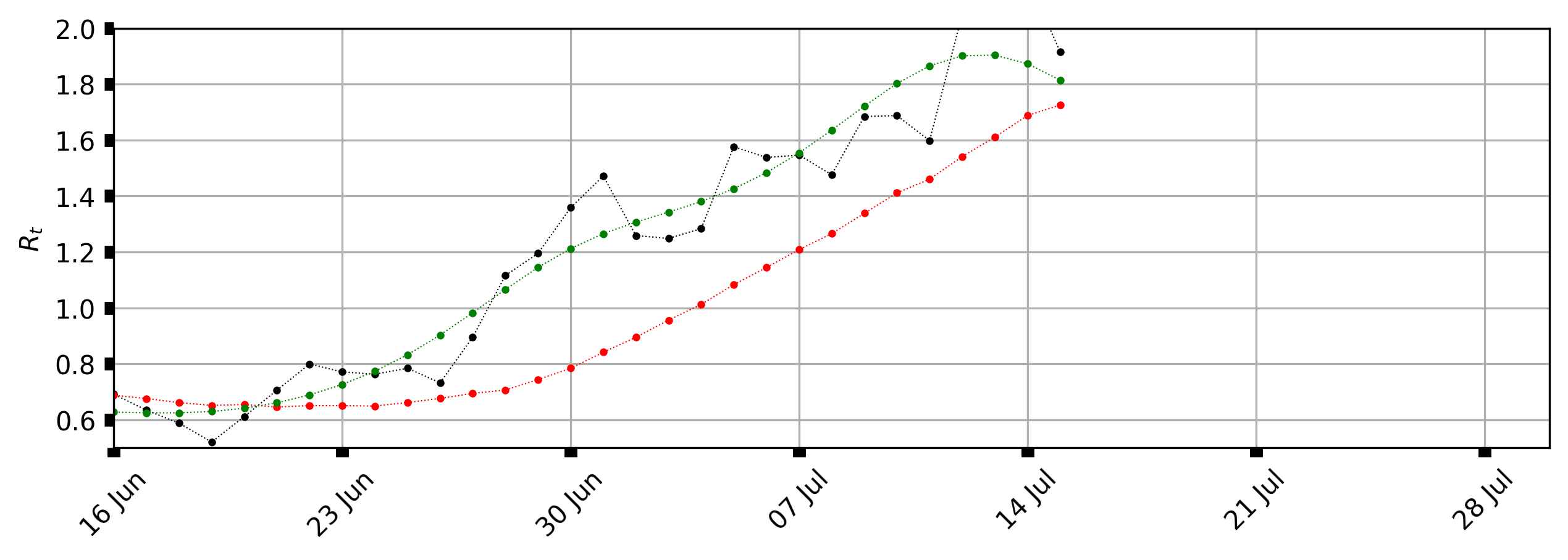}
\caption{Zoom of Fig. \ref{naive_Rt_symptoms_onset_smoothed} to the most recent data available and consolidated, data-set by ISS}
\label{naive_Rt_symptoms_onset_smoothed_zoom}
\end{figure}

Fig. \ref{naive_Rt_symptoms_onset_smoothed_zoom} shows the zoom of the series in Fig. \ref{naive_Rt_symptoms_onset_smoothed} to the "present" days (which is Jul 30 at the time of writing). The last available "consolidated" count of new infected cases dates back to Jul 15, as the full process of data collection must be completed if we want to know the symptoms onset date. This problem, known as right censoring, is true for every country, with delays which vary depending on the particular data collection process. Moreover, it is well known in Italy that the collection process greatly depend on the pressure that the epidemic is producing on the Health System.

Instead of using the distribution of new infected cases by symptoms onset date, we propose to adopt the smoothed distribution by notification date as the input for Eq. \ref{RtDef}, to obtain $\hat{\Bar{R}}_{t}$. The difference is that as soon as a case is detected, it is notified. The advantage that the series is consolidated by definition, without the need of past revisions, comes with the following drawbacks:
\begin{enumerate}
    \item there is a certain amount of delay from the symptoms onset date to the notification date;
    \item the series accounts for more noise, as it makes no distinctions between symptomatic cases and asymptomatic cases.
\end{enumerate}

\begin{figure}[t]
\center
\includegraphics[width=0.7\textwidth]{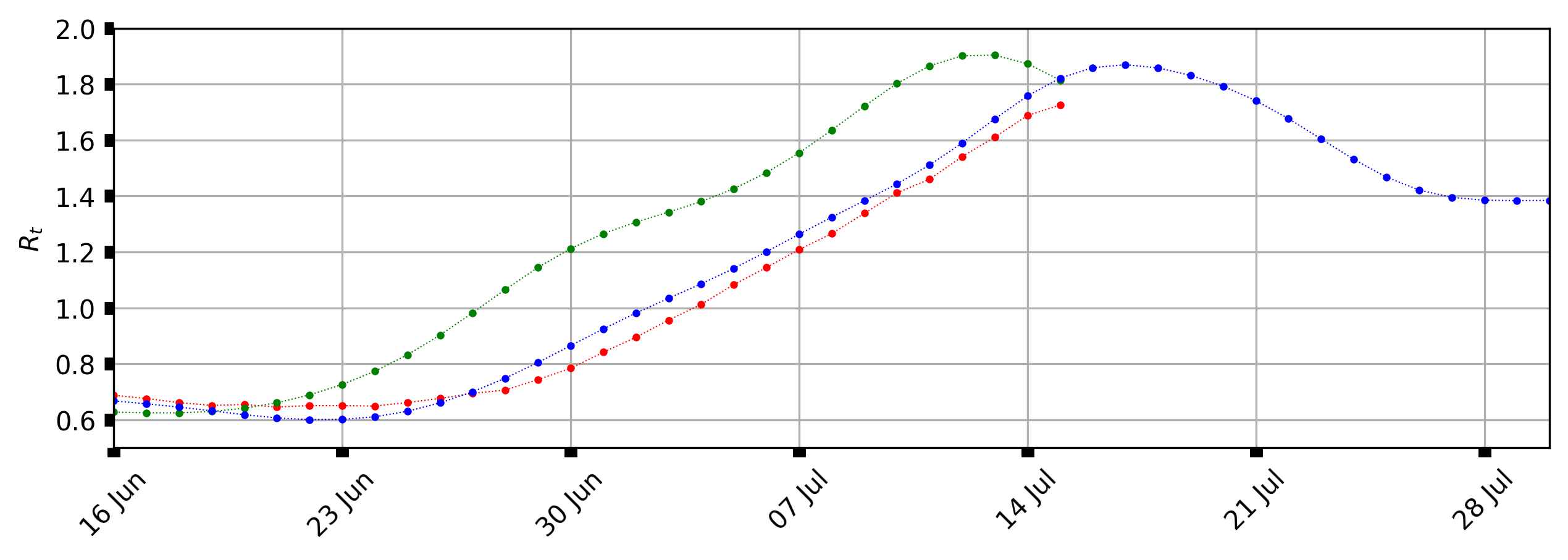}
\caption{In red the average $R_t$ calculated over a window of 14 days on the symptoms onset distribution, data-set by ISS; in green $\Bar{R}_{t}$ calculated on the symptoms onset distribution smoothed with Tikhonov regularization, data-set by ISS; in blue $\hat{\Bar{R}}_{t}$ calculated on the notification date distribution smoothed, data-set by Protezione Civile}
\label{naive_Rt_symptoms_onset_vs_notification}
\end{figure}

Fig. \ref{naive_Rt_symptoms_onset_vs_notification} shows the comparison of three $R_t$ calculations: average $R_t$ calculated on a 14 day windows (in red), $\Bar{R}_t$ calculated on smoothed cases by symptoms onset date (in green) and $\hat{\Bar{R}}_{t}$ calculated on smoothed cases by notification date (in blue). The blue line exhibits a delay wrt the green one, but it is still anticipating the red line. Maximizing the cross correlations provides the following measures of the relative delays:
\begin{itemize}
    \item $\Bar{R}_t$ anticipates $\hat{\Bar{R}}_{t}$ by 8 days, but the last available value of $\Bar{R}_t$ dates back 15 days prior to the present;
    \item $\hat{\Bar{R}}_{t}$ anticipates $R_t$ (calculated on a 14 day windows) by 6 days.
\end{itemize}

Hence, we can conclude that, thanks to the smoothing procedure without delays (via Tikhonov regularization), we can replace the distribution of new cases by symptoms onset date with the distribution by notification date, obtaining the following advantages:
\begin{enumerate}
    \item earn 6 days of anticipation with respect to the averaged $R_t$;
    \item being able to compute the reproduction number up to the present, without having to wait for varying consolidation times in the data collection processes.
\end{enumerate}

\subsubsection{Residuals}

As the original raw series are noisy and uncertain, we want a method to extract the noise and use it to calculate confidence intervals on the estimated $R_t$, in a way such that confidence intervals can directly reflect the uncertainty in the effective measuring process. This is much more relevant as we plan to estimate the reproduction number on the series of new infected cases by notification date, which includes both symptomatic and asymptomatic cases, with the latter exhibiting high noise.

The noise can be measured by the relative residuals of the signal with respect to its smoothed version, $\epsilon_t = (I_t - \Bar{I}_t)/\Bar{I}_t$.

\begin{figure}[t]
\center
\includegraphics[width=0.7\textwidth]{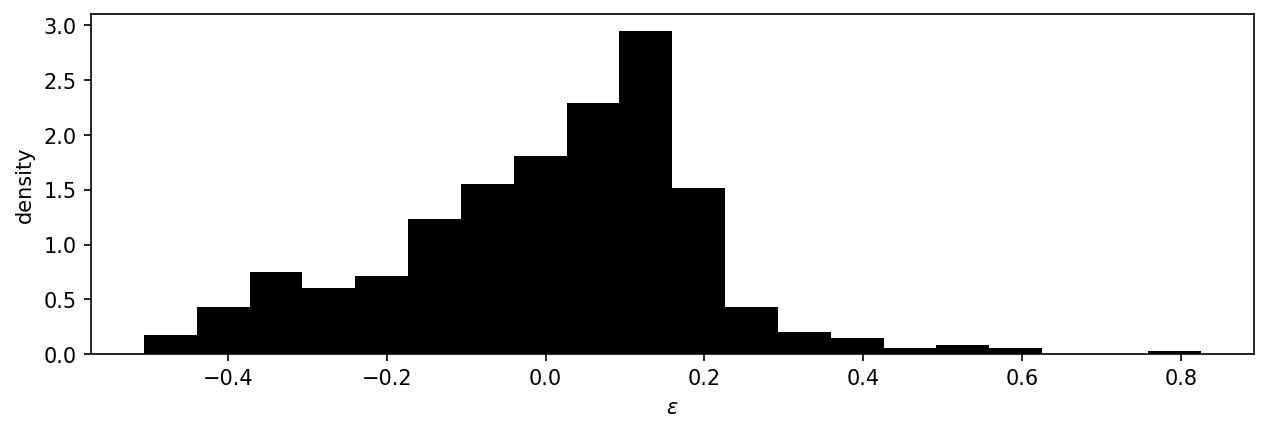}
\caption{distribution of $\epsilon_t = (I_t - \Bar{I}_t)/\Bar{I}_t$ calculated on the series of new infected cases by notification date, data-set by Protezione Civile}
\label{epsilon_tikhonov_distribution}
\end{figure}

Fig. \ref{epsilon_tikhonov_distribution} shows that the distribution of $\epsilon_t$ calculated on the series of new infected cases by notification date \textbf{is unbalanced}.

It turns out that the unbalancing is directly related to the weekly seasonality which affects the series (the seasonality can be seen in Fig. \ref{naive_Rt_symptoms_onset} or \ref{naive_Rt_symptoms_onset_window_14}). The reason is that the smoothing obtained by Eq. \ref{TickCostFunction} is not able to capture the seasonality. 

\subsubsection{Deseasoning via singular value decomposition}
\label{RSVD}

Standard techniques to deal with seasonality, like SARIMA (Seasonal Autoregressive Integrated Moving Average), rely on moving averages. 

To avoid the delays introduced by moving averages, we instead adopt 
Regularized Singular Value Decomposition (RSVD) proposed by Lin, Huang and Mcelroy in \cite{LinHuangMcelroy}. RSVD allows to detect the seasonal component of the signal by casting the signal vector into a matrix whose columns are the seasons and the rows are the repetitive periods of a complete series of seasons. Singular Value Decomposition is then applied to the matrix so that singular values represent the seasonal component of the signal. Each seasonal component is regularized via Tikhonov regularization, following the hypothesis that each seasonal component must change smoothly, period after period. The Tikhonov regularization parameter is fitted via "leave one out cross validation".

The advantage of this method with respect to the SARIMA approach is that \emph{we do not need to take moving averages}, and we don't need to tune any meta-parameter of the model.

The python porting of the original R code is available in the supplementary material at \href{https://github.com/tomorrowdata/COVID-19}{https://github.com/tomorrowdata/COVID-19}, within the library \texttt{covid19\_pytoolbox}. The following features have been added to the original work:
\begin{itemize}
    \item take the logarithm of the seasonal series, to remove exponential trends;
    \item differentiate the signal to a desired degree, to remove non-stationary trends in the original data, with an augmented Dickey?Fuller (ADF) test to check if any non-stationary component is present;
    \item apply Tikhonov regularization to the deseasoned signal to obtain the trend.
\end{itemize}

Denoting by $\Tilde{I}_t$ the trend of the raw signal $I_t$ after removing the seasonality, we obtain the following decomposition of the original series:
\begin{equation}
    I_t = \Tilde{I}_t + S_t + \Tilde{E}_t
\end{equation}
where $S_t$ is the seasonal component and $\Tilde{E}_t$ is the residual after deseasoning.

\begin{figure}[t]
\center
\includegraphics[width=0.7\textwidth]{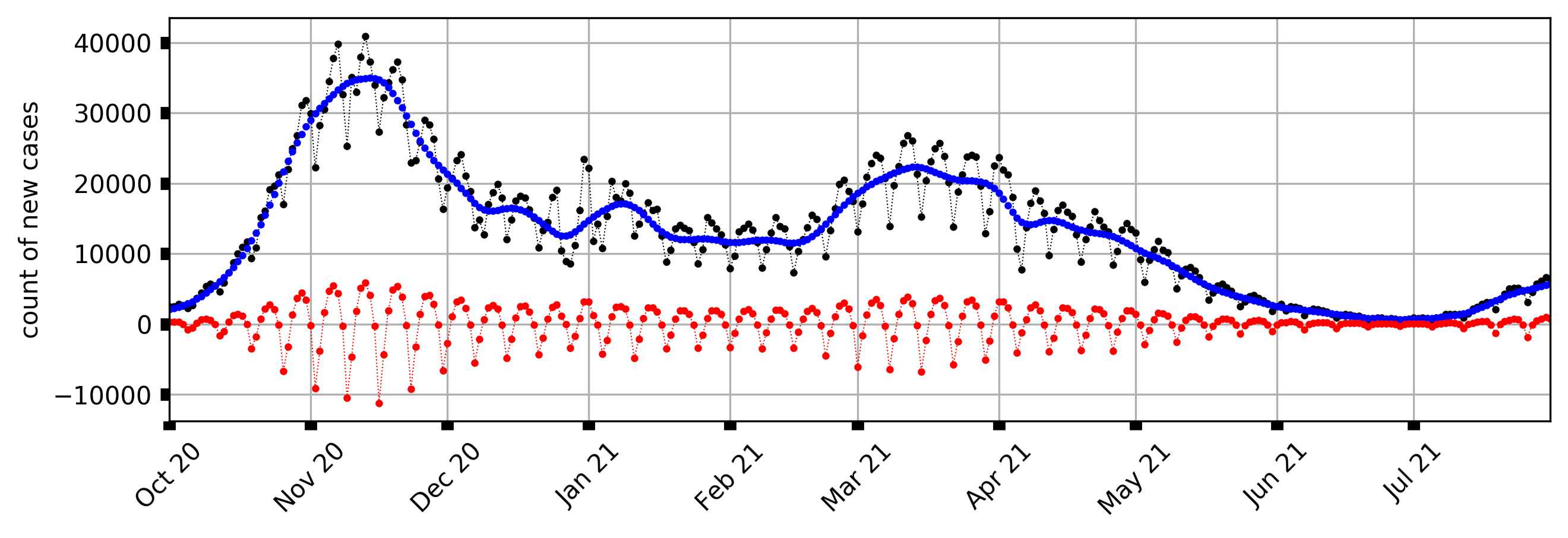}
\caption{Raw series $I_t$ of new infected cases by notification date (in black), its trend $\Tilde{I}_t$ after removing the seasonal component (in blue), the seasonal components $S_t$ (in red); data-set Protezione Civile}
\label{new_cases_deseason}
\end{figure}

Fig. \ref{new_cases_deseason} shows the result of applying RSVD to the series of new infected cases by notification date. RSVD has been applied to the logarithm of the second difference of the original series, with the ADF test confirming the removal of any non-stationary component. The smoothness of the seasonal components can be noted clearly.

\subsubsection{Residuals of the deseasoned series}

\begin{figure}[t]
\center
\includegraphics[width=0.7\textwidth]{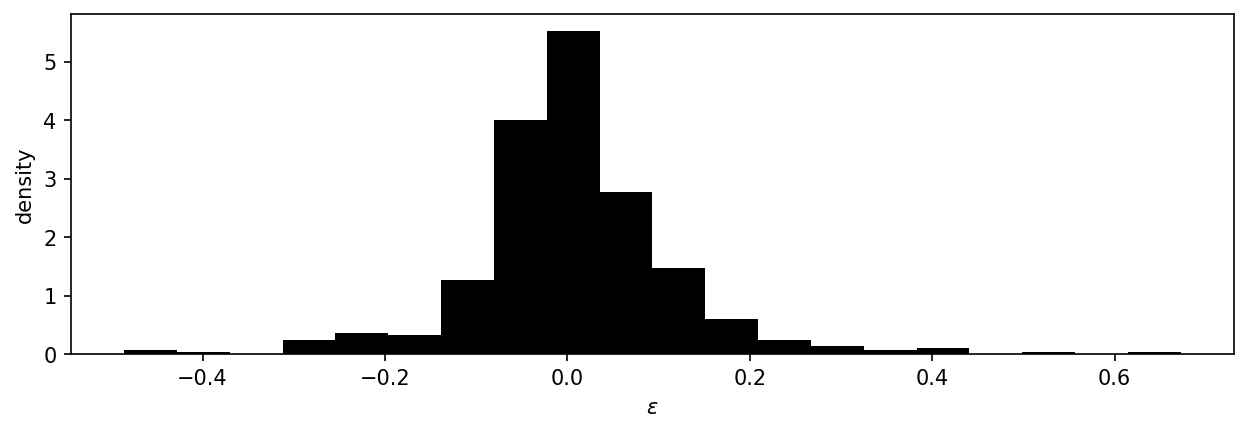}
\caption{distribution of $\Tilde{\epsilon}_t = \Tilde{E}_t/\Tilde{I}_t$ calculated on the series of new infected cases by notification date, data-set by Protezione Civile}
\label{epsilon_RSVD_distribution}
\end{figure}

Now that we have removed the seasonality, we can look at the distribution of the residuals again. Fig. \ref{epsilon_RSVD_distribution} shows the distribution of the relative residuals after removing the seasonal component: removing the seasonality produced a much more balanced, almost gaussian, distribution, if compared to Fig. \ref{epsilon_tikhonov_distribution}.

\subsubsection{Putting it all together with Markov chain Monte Carlo}

We start from the series $I_t$ of new infected cases by notification date, as explained in \ref{notification_date}. We then apply RSVD to obtain the deseasoned smoothed trend $\Tilde{I}_t$ of the series and the respective relative residuals $\Tilde{\epsilon}_t = \Tilde{E}_t/\Tilde{I}_t$, as explained in \ref{RSVD}. 

With those ingredients, we can setup Markov chain Monte Carlo simulations to sample multiple chains of $R_t$ values, as follows, denoting by $C(\cdot)$ the chains obtained via sampling:
\begin{enumerate}
    \item $C(R_t)$ chains are sampled from a prior normal distribution, with $\mu=1.3$ and $\sigma=10$; a Gaussian process could be used instead, but it is less computationally efficient; the length of the chains is the same as the length of $I_t$;
    \item $C(\Tilde{\epsilon}_t)$ chains are sampled from a prior normal distribution, with $\mu=\sum_{s=t-7}^t{\Tilde{\epsilon}_s}/7$ and $\sigma=\sqrt{\sum_{s=t-7}^{t}(\Tilde{\epsilon}_s-\mu)^2/7}$; the length of the chains is the same as the length of $I_t$;
    \item $C(\Tilde{I}_t)$, the chains of new cases with random noise, are obtained as $\Tilde{I}_t+\Tilde{I}_t \cdot C(\Tilde{\epsilon}_t)$;
    \begin{itemize}
        \item \textbf{note}: this is where the original noise of the series is transferred to the simulation, so that the confidence interval will account for uncertainties in the original series;
    \end{itemize}
    \item the estimated count $T_t$ of new cases in each day of the chain is calculated from Eq. \ref{RtDef} as $C(T_t) = C(R_t) \cdot \sum_{s=1}^t w_s C(\Tilde{I}_{t-s})$;
    \item finally, a posterior Poisson distribution is tested via Monte Carlo, between the estimated cases, $T_t$, and the expected ones, $\Tilde{I}_t$.
\end{enumerate}

We sample 4 chains with 1000 iterations each discarded for tuning, and 500 iterations each kept for sampling. The final data-set contains 2000 samples from which day by day statistics, like the confidence interval, can be calculated.

\begin{figure}[t]
\center
\includegraphics[width=0.7\textwidth]{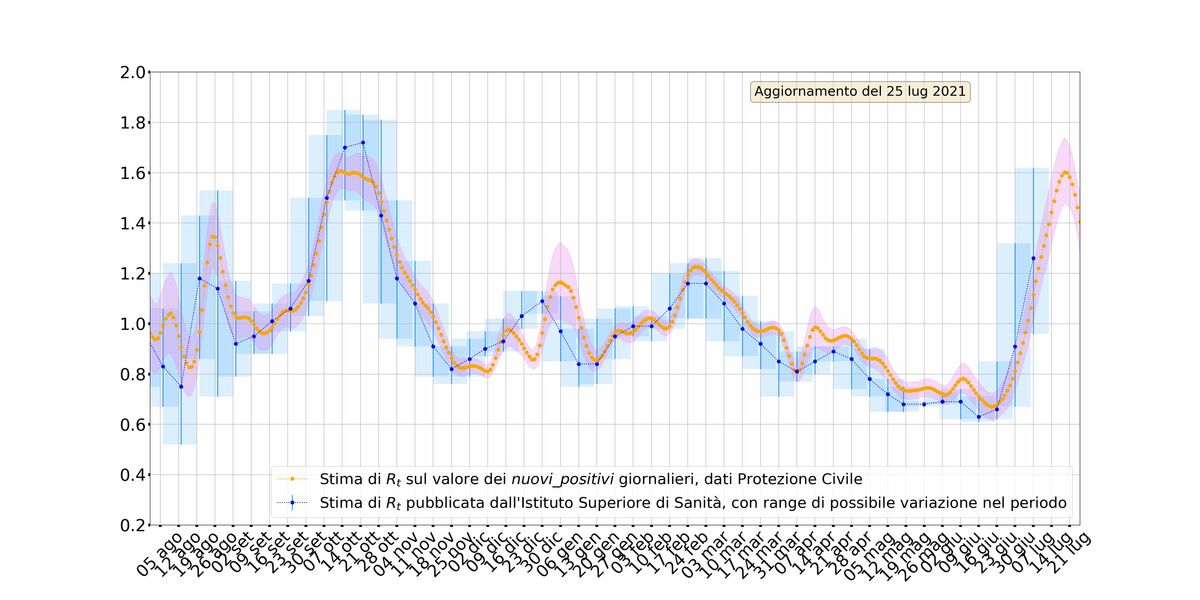}
\caption{In blue, the $R_t$ values as reported by the Istituto Superiore di Sanit\`{a} and in red the anticipated calculation published regularly, from the end of November 2020, at \href{https://mondoeconomico.eu}{https://mondoeconomico.eu} by Stefano Terna}
\label{Rt}
\end{figure}

Fig. \ref{Rt} shows the result, where the confidence interval (in violet) succeeds in representing periods of higher uncertainty in the data.

%%%%%%%%%%%%%%%%%%%%%%%%%%%%%%%%%%%%%%%%%%%%%%%%%%%%%%%%%
\subsection{Second wave, new infections from outside, introducing 20 days in advance the new specific containment measures}
\label{anticip}
%selectResults10kStableSeedsCPoints_basicControlB_schoolOpenSeptOctMar-20ControlChangingWorldNewStart_plusHMlog.ipynb
% using 10kCtrl1NStartCtrl2M-20.csv from SIsaR_0.9.5.4.1trials10kCtrl1NStartCtrl2M-20.nlogo

The counterfactual situation described in this section---inspired by Section \ref{indicator}---is related to the start and end dates of the actions of containment, both occurring 20 days in advance, with a natural barrier set on October \nth{5}, 2020. Before that date, no one could plan to start new control measures.

As in the last two sections, we have 1407 cases of epidemics alive at the critical dates of June \nth{1} and September \nth{20}, after a Summer interval characterized by a quiet phase.
Considering December \nth{15} and September \nth{20} situations, the second wave epidemics are 769, again decreasing because the anticipated actions have eliminated some other cases. The December mean of the infected agents is 294.2, still higher than the actual value ($\approx$200.0). We always overestimate the mean of the epidemic effects, being the surviving epidemics the larger ones.

Comparing Fig. \ref{secondFromOut-20} and Fig. \ref{secondFromOutWith} the difference is not evident; instead, the italic figures, and most of all, the red bold ones---in Table \ref{selForceWave2Contr2M-20Tab}---report clearly the comparative advantage of this counterfactual experiment with respect to the values of Table \ref{selForceWave2Contr2Tab}.

\begin{table}[t]
\center
\tiny
\begin{tabular}{lrrrrrrrrrrrrr}
\toprule
(1000) & Jun~1 & & Sep~20 & & Dec~15 & & Feb~1 & & May~1 & & Dec~15\\
&2020 &&2020 &&2020&&2021&&2021&&2020\\
&&&&&&&&&&& to~end \\
cum.~v. & sym. & all & sympt. & totalInf. & sympt. & totalInf. & sympt. & totalInf. & sympt. & totalInf. & sympt. & totalInf. & days\\
\midrule
count & 1407.0 & 1407.0 & 1407.0 & 1407.0 & 769.0 & 769.0 & 637.0 & 637.0 & 471.0 & 471.0 & 769.0 & 769.0 & 769.0 \\
mean & 35.6 & 72.7 & 40.0 & 84.1 & \textbf{{\color{red}112.2}} & \textbf{{\color{red} 294.2}} & \emph{172.0} & \emph{467.9} & \emph{276.5} & \emph{748.6} & 248.9 & 663.4 & 499.3 \\
std & 14.1 & 42.6 & 16.7 & 52.8 & 66.8 & 188.4 & 91.5 & 251.3 & 112.9 & 286.9 & 158.0 & 417.5 & 124.1 \\
\bottomrule
\end{tabular}
\caption{Second wave, new infections from outside, with new specific measure anticipation of -20 days}
\label{selForceWave2Contr2M-20Tab}
\end{table}

%%%%%%%%%%%%%%%%%%%%%%%%%%%%%%%%%%%%%%%%%%%%%%%%%%%%%%%%%
\subsection{Second wave, new infections from outside, with a unique intervention measure: stopping fragile people for 60 days}
\label{frag}
%selectResults10kStableSeedsCPoints_basicControlB_schoolOpenSeptNoFragOCT05-60dControlChangingWorldNewStart_plusHMlog.ipynb
% using 10kCtrl1NStartNoFragOCT05-60d.csv from SIsaR_0.9.5.4.1trials10kCtrl1NStartNoFragOCT05-60d.nlogo

The second counterfactual experiment is based on an immediate stop to the circulation of fragile persons and specifically of fragile workers, plus isolating nursing homes and hospitals. Schools are always open in this experiment. The decision is activated on October \nth{5}, 2020, when the second wave was becoming evident. 
In \cite{phillips2021coronavirus} we have important consideration suggesting the importance of taking into account fragility in a long-term fighting perspective against this kind of epidemics.

As in the last three sections, we have 1407 cases of epidemics alive at the critical dates of June \nth{1} and September \nth{20}, after a Summer interval characterized by a quiet phase.
Considering December \nth{15} and September \nth{20} situations, the second wave epidemics are 886, lightly above the values of Section \ref{secondWith} and Section \ref{anticip}, but without locking the economy and the society as a whole. The December mean of the infected agents is 326.3, higher than the actual value ($\approx$200.0) for the explained overestimation bias.

Comparing Fig. \ref{secondFromOutNoFrag} and Fig. \ref{secondFromOutWith} the difference is not evident; instead, the italic figures, and most of all, the violet bold ones---in Table \ref{selForceWave2NoFragTab}---signal the close proximity of the effects of this counterfactual experiment with those of Table \ref{selForceWave2Contr2Tab}.

\begin{table}[H]
\center
\tiny
\begin{tabular}{lrrrrrrrrrrrrr}
\toprule
(1000) & Jun~1 & & Sep~20 & & Dec~15 & & Feb~1 & & May~1 & & Dec~15\\
&2020 &&2020 &&2020&&2021&&2021&&2020\\
&&&&&&&&&&& to~end \\
cum.~v. & sym. & all & sympt. & totalInf. & sympt. & totalInf. & sympt. & totalInf. & sympt. & totalInf. & sympt. & totalInf. & days\\
\midrule
count & 1407.0 & 1407.0 & 1407.0 & 1407.0 & 886.0 & 886.0 & 761.0 & 761.0 & 637.0 & 637.0 & 886.0 & 886.0 & 886.0 \\
mean & 35.6 & 72.7 & 40.0 & 84.1 & \textbf{{\color{violet}128.1}} & \textbf{{\color{violet}326.3}} & \emph{211.0} & \emph{555.1} & \emph{323.3} & \emph{862.1} & 301.1 & 792.3 & 515.5 \\
std & 14.1 & 42.6 & 16.7 & 52.8 & 89.6 & 234.2 & 118.1 & 306.7 & 126.4 & 315.9 & 170.7 & 450.2 & 116.9 \\
\bottomrule
\end{tabular}
\caption{Second wave, new infections from outs., stop fragile people. 60 days from Oct. 5}
\label{selForceWave2NoFragTab}
\end{table}

%%%%%%%%%%%%%%%%%%%%%%%%%%%%%%%%%%%%%%%%%%%%%%%%%%%%%%%%%
\subsection{To recap}
\label{recap}

Table \ref{keyResultsT} reports the different cases synthetically and, most of all allows an easy comparative interpretation of the actual and counterfactual situations.

\begin{table}[t]
\center
\footnotesize
\begin{tabular}{lrrrrrr}
\toprule
Scenarios & & Dec~15,~20 & & Dec~15,~20\\
& & & & to~end~~~~~~ \\
 & & sympt. & totalInf. & sympt. & totalInf. & days \\ 

\midrule

no containment & count & 140.0 & 140.0 & 140.0 & 140.0 & 140.0 \\
in spontaneous & mean & 248.4 & 648.7 & 701.1 & 1757.9 & 594.2 \\
second wave & std & 167.4 & 424.3 & 246.4 & 599.7 & 118.9 \\

\bottomrule
no containment & count & 1044.0 & 1044.0 & 1044.0 & 1044.0 & 1044.0 \\
in forced & mean & 180.4 & 462.1 & 726.6 & 1810.9 & 620.9 \\
second wave & std & 134.6 & 354.6 & 221.9 & 544.0 & 110.8 \\

\bottomrule
basic containment & count & 874.0 & 874.0 & 874.0 & 874.0 & 874.0 \\
in forced & mean & \emph{130.0} & \emph{340}.6 & 252.7 & 666.4 & 494.1 \\
second wave & std & 83.9 & 232.6 & 156.8 & 416.4 & 122.7 \\
 
\bottomrule
-20 days cont. & count & 769.0 & 769.0 & 769.0 & 769.0 & 769.0 \\
in forced & mean & \textbf{{\color{red}112.2}} & \textbf{{\color{red} 294.2}} & 248.9 & 663.4 & 499.3 \\
second wave & std & 66.8 & 188.4 & 158.0 & 417.5 & 124.1 \\

\bottomrule
frag. subj. \& \\ 
workers control & count & 886.0 & 886.0 & 886.0 & 886.0 & 886.0 \\
in forced & mean & \textbf{{\color{violet}128.1}} & \textbf{{\color{violet}326.3}} & 301.1 & 792.3 & 515.5 \\
second wave & std & 89.6 & 234.2 & 170.7 & 450.2 & 116.9 \\

\bottomrule
\end{tabular}
\caption{Report of the key results, with count, mean, and std}
\label{keyResultsT}
\end{table}

%%%%%%%%%%%%%%%%%%%%%%%%%%%%%%%%%%%%%%%%%%%%%%%%%%%%%%%%%
\section{Economic analysis of the of interventions}
\label{CBanalysis}

The pandemic has an impact on the general economy. First, we take into account the additional health expenditure, which in Piedmont has risen from \EUR{8,880} million to \EUR{9,200} million, with an increase in pressure on GDP of 0.2\%. It is an increment that cannot be generalized. In other regions and States, health expenditure has even decreased, due to the lower demand for diagnostic and treatment services, precisely because of the pandemic and the precautionary reduced access to health services. Apart from the additional health expenditure, the main impact to be considered is the loss in production induced by the contagion containment measures, i.e., the so-called ``lockdown'' of the economy and the associated mobility bans. 

The impact assessment of production stoppages and mobility bans can be measured by applying an Input-Output model. The main quality of Input-Output models is the possibility of determining the total effect of changes in output in all sectors of the economy due to a unit change in final demand in a given sector. This is achieved by applying a matrix of multipliers, i.e., Leontief's inverse matrix, to a sectoral vector of demand changes. The inverse matrix makes it possible to calculate the sum of the direct impact of the stopped productions, sector by sector, and the indirect impact, due to the infinite feedback on the purchases of the affected sectors from the first drop in demand received. However, the standard representation is not complete. The literature tends to extend these effects to consider the feedback not only by the purchases of the impacted sectors but also by the drop in the final demand of households affected by the unexpected change in income through their marginal propensity to consume. This third effect is the so-called induced impact.

The matrix of direct, indirect and induced impacts of the Piedmont economy in Table \ref{PiedmontMatrix} has been originally estimated by one of the authors.

\begin{table}[H]
\centering
\begin{tabular}{lrrrrrr}
\toprule
& & & & total & \\
& final /total & & & induced & production &  added value \\
& demand & direct impact & indirect impact & impact & multiplier & multiplier \\
\midrule
agriculture & 0.53	& 1.40 & 0.50 & 1.30	 & 3.20 & 1.40\\
manufacturing & 0.33 & 1.80 & 1.20 & 1.60 & 4.50 & 1.60\\
construction & 0.38 & 1.70 & 0.90 & 1.60 & 3.20 & 1.60\\
distribution & 0.53 & 1.40 & 0.50 & 1.30 & 3.10 & 1.60\\
services & 0.50 & 1.50 & 0.50 & 1.40 & 3.40 & 1.50\\
\bottomrule
\end{tabular}
\caption{Multipliers of direct, indirect and induced impact, and overall impact as well added value (GDP) multipliers per 1 euro of final demand change, related to each of the five sectors on the rows}
\label{PiedmontMatrix}
\end{table}

As we can see, the economic effects of lockdowns can be very different depending on whether they selectively affect one sector (normally the sectors most affected are the last two, distribution and services), or whether all sectors are affected. The manufacturing sector, which is strongly linked with other sectors, has a total, direct, indirect, and induced multiplication coefficient of 4.5 times the initial reduction in final demand. Therefore, to calculate the impacts, we started from three different assumptions, or scenarios, which we have called A, B and C.

[A] The restrictions affected all economic activities that could be stopped, safeguarding only those businesses that were essential. This meant stopping approximately half of the regional production system. Schooling was only permitted with distance learning. This case occurred in the period from 9 March 2020 to the end of April 2020.

[B] Only businesses in sectors whose activities were rated with a high risk of contagion were stopped: these activities included non-food retail trade, the tourism restaurant and hotel sector, the sport, recreation and entertainment sector, the cultural sector, and, of course, the whole education system, that was served by distance learning. The transportation sector was legally active but still impacted by an almost obligatory drop in demand. This case actually occurred at different times during 2020 and 2021 and significantly from October 2020 until spring 2021, with a break of a few weeks during the winter.

[C] Purely theoretical and not put in place, it was considered to stop only the fragile workers, leaving intact the education and all the activities stopped in case B. In this case, fragile workers are estimated to be 14\% of the total, based on a national projection of the total number of 5.6 million fragile people under 65 in Italy. To make the calculation of the impact realistic, we assume that all fragile workers received sickness compensation equal to the lost wage that impacted on the overall tax loss, increasing it; we also assume that half of the production of fragile workers could still be produced with overtime or temporary work by other workers.

The results of the simulations are reported in Table \ref{ABC} and are expressed in points per thousand of Piedmont's GDP.

\begin{table}[t]
\centering
\begin{tabular}{lrrrr}
\toprule
& Scenario A & Scenario B & Scenario C \\
\midrule
\emph{daily impacts} \\
total production & -4.86 & -1.23 & -0.69  \\
added value & -2.12 & -0.55 & -0.30 \\
taxes & -0.91 & -0.24 & -0.35 \\
\\
\emph{monthly impacts} \\
total production &-145.7 & -36.9 & -20.7  \\
added value & -63.7 & -16.6 & -9.1 \\
taxes & -27.4 & -7.1 & -10.6 \\
\bottomrule  
\end{tabular}
\caption{Economic impacts of the pandemic with three hypotheses of non-pharmaceutical containment measures applied; values are expressed in GDP points/1000}
\label{ABC}
\end{table}

In Table \ref{ABC}, each day of closure of productive activities (leaving only the essential ones and schools closed, i.e., open for distance learning) to counter the contagion and allow access to hospitals produces a loss of income (added value) equal to 2.1 per thousand of GDP and a worsening of the fiscal budget by 0.9 per thousand of GDP. One month of total closures, therefore, would cost an income loss of 6.4 percent and a worsening of the fiscal balance of 2.7 percent (Scenario A, actually implemented in Italy from 9 March 2020 to April).

Conversely, limiting closures to only distribution activities (non-food), as well as to school (open as distance learning), sports, culture and leisure, tourism, and restaurant services (as in the light lockdown established in October 2020 and subsequent months, Scenario B) would have produced a daily income loss of 0.55 per thousand of GDP (equivalent to 1.6 percent per month) and a fiscal loss of 0.24 per thousand per day and 0.7 percent of GDP per month. 

The solution of protecting at home (and paying) only fragile workers, leaving all schools and productive activities open, would reduce the loss of income to 0.3 per thousand per day (0.91 percent per month). Although this solution (Scenario C, never actually implemented) is more convenient concerning the overall income loss, even 1/7 of that of scenario A and ? of that of scenario B, it costs slightly more in fiscal terms than scenario B (-0.35 per thousand per day, instead of -0.24). However, it would seem preferable because it is the only option of the three that would allow the regular operation of the schools. According to the reliable Invalsi tests performed in 2021, the percentage of pupils in Italian schools who have not reached the minimum learning standards has increased by 10 percentage points based on the total number of pupils. If we were to put this loss of human capital on an economic balance sheet, we would have to consider the full cost of an additional year of schooling for 10 percent of the school population, both in terms of the cost of additional education plus the income lost for a 1-year delay in subsequent employment. A raw estimate of this cost would appear to be 1.3\% of GDP, of which 0.58\% for the additional cost of education and 0.75\% for the income lost by postponing entry into employment by 1-year. 

Following Table \ref{ABC2}, in a C scenario, the cost of pandemic restrictions, for 3 months (hypothesis), would be 0.2\% of annual GDP for increased health expenditures + 2.7\% of direct, indirect and induced value added (GDP) losses, plus 3.19\% of GDP of public budget deterioration, while there would be no human capital losses. The total losses in scenario C would be 6.1 percent of annual GDP for three full months of restrictions. In scenarios B and A the total loss would have been much higher and specifically 8.6\% and 28.8\% of the pre-Covid GDP, respectively. It is also worth noting the distribution of losses by row. In scenario C, the losses in value-added, and thus the recession damage to the economy to be recovered, would be minimal, and the losses due to insufficient human capital formation would be zero. Nevertheless, the policies adopted have preferred the adoption of scenarios A and B.

\begin{table}[t]
\centering
\begin{tabular}{lrrrr}
\toprule
& Scenario A & Scenario B & Scenario C \\
& three months & three months & three months \\
\midrule
more health expenditure & -2.0 & -2.0 & -2.0  \\
added value or GDP loss & --191.0 & -49.8 & -27.2 \\
tax loss & -82.1 & -21.4 & -31.9 \\
human capital loss &--13.4 & -13.4 & 0.0 \\
total loss (GDP/1000) & -288.6 & -86.6 & -61.1 \\
\bottomrule  
\end{tabular}
\caption{Total losses simulating the 3 scenarios A, B, C, from activity and mobility restrictions, in GDP points/1000}
\label{ABC2}
\end{table}

%%%%%%%%%%%%%%%%%%%%%%%%%%%%%%%%%%%%%%%%%%%%%%%%%%%%%%%%%
\section{Planning vaccination campaigns}
\label{planning}

\subsection{Some notes on vaccines}
\label{vaccineAction}

Vaccines are biological products made from killed or attenuated microorganisms, from viruses or from some of their components (antigens), or from substances they produce made safe by chemical (e.g., formaldehyde) or heat treatment, while maintaining their immunogenic properties (\href{https://www.who.int/vaccines}{https://www.who.int/vaccines}); today, vaccines can be composed of proteins obtained by recombinant DNA techniques using genetic engineering approaches.

They usually contain, in addition to the antigenic fraction, sterile water (or a saline-based physiological solution), adjuvants, preservatives and stabilisers. Adjuvants are included in the vaccine in order to enhance the immune system response; preservatives are added to prevent contamination of the preparate by bacteria; stabilisers are introduced to increase the shelf life of the product and to maintain the properties of the vaccine during storage.

\subparagraph{How vaccines work: a step back in the 18th century}
\label{hints}

Although early forms of empirical immunization appear to have been present in different cultures (India and China; \cite{Boylston2012}), the creation of the first vaccine (for smallpox immunization) dates back to 1798 by Edward Jenner, an English physician. Jenner had noticed that milkmaids who became infected with cowpox (Vaccinia Virus), a virus that causes similar symptoms to human smallpox (Variola Virus or Smallpox virus) but not fatal, did not subsequently develop the disease \cite{jenner1802}. This suggested that Cowpox inoculation could protect against Smallpox. Jenner decided to test his theory by inoculating an eight-year-old boy, the son of his gardener (sic!), James Phipps, with material taken from the cowpox lesions of a local milkmaid. As expected, James developed few local lesions and a modest fever. Two months later Jenner inoculated James with variolous matter from a case of human smallpox,  without a sensible effect was produced: Jenner had proved that the boy had been immunized. By definition, all subsequent immunizations would be called vaccinations as in 1881 Louis Pasteur proposed it as a general term for the new protective inoculations, in honor of Jenner.
 
Once inoculated, vaccines (all of them), mimicking the first contact between man and pathogen, are able to stimulate an immunological response (humoral and cellular) as if this occurred through a natural contagion, although not leading to disease and without giving the associated complications. The rationale behind this phenomenon is immunological memory: the body/immune system that has already experienced a pathogenic microorganism, treasures the experience by responding rapidly to the same microorganism (the absence of immunological memory is the reason why Covid-19 emerged as a problem for humans). For some vaccines it is necessary to make recalls at a distance of time. Normally our body reacts to an unwanted host, but it can take up to two weeks to produce a sufficient amount of antibodies versus the pathogen. In the absence of vaccination, in this interval of time a pathogen can create damage to the body and even lead to death. 
 
\subparagraph{Types of vaccines}
\label{types}

A long way has been covered since 1798, and technologies have steadily improved to arrive at hi-tech vaccines such as those we are using today to fight Covid. The types of vaccine that exist today are:

\begin{itemize}
\item live attenuated vaccines (e.g., measles and tuberculosis): these are produced from infectious agents that have been rendered non-pathogenic;
\item inactivated vaccines (e.g., poxvirus): these are produced using infectious agents killed by heat or chemicals;
\item purified antigen vaccines (e.g., anti-meningococcal): these are produced by purifying specific components (bacterial or viral);
\item anatoxin vaccines (e.g., tetanus): these are produced using molecules from the infectious agent, which are not capable of causing the disease on their own, but which can stimulate/activate the immune defenses of the vaccinee;
\item recombinant protein vaccines (e.g., hepatitis B): these are produced using recombinant DNA technology, which involves inserting genetic material coding for the antigen (a protein/peptide) into microorganisms capable of producing the antigen specifically, allowing it to be purified;
\item recombinant mRNA vaccines (e.g., Pfizer/BioNTec and Moderna): these are produced using an mRNA coding for a target gene encapsulated nanoparticle made with lipid bilayers; this information is able to drive the synthesis of an antigenic protein, through the cell machinery, and the triggering of the immunitary system;
\item recombinant viral vector vaccines (e.g., Astrazeneca/Oxford): these are produced using an DNA coding for a target gene carried within a defective adenovirus (from human or from chimpanzee), able to vehicle the gene within the cell nucleus. This information is able to drive the synthesis of an antigenic protein, through the cell machinery, and the triggering of the immunitary system. 
\end{itemize}

The latter types of vaccine (mRNA and Adenoviral vector-based) were today adopted mainly because they can be manufactured very quickly, and being their production process highly standardized. As a matter of fact, they are the fastest way to create a vaccine in the middle of a pandemic.

\subparagraph{mRNA and adenovirus-based vaccines}
\label{based} 

Let us take a step back. The CoV-SARS-2019 virus has on its surface a protein called Spike (S protein), that it uses to enter a human cell via binding to the ACE2 receptor (Fig. \ref{ace}). The S-protein has therefore been chosen as the specific target to produce a vaccine since it is exposed in large quantities on the surface of the virus.

mRNA-based and adenovirus-based vaccine for Covid target the S-protein through the production of a RNA messengers (mRNAs), the classical molecule that routinely instructs all the cells what to build. Once the S-protein is produced within the body and presented the the immune system, it is considered an antigen, and the body starts producing antibodies against it. The same thing can be done by using a pre-made protein and injecting it, but its production, testing and approval is longer (years to decades) and more expensive.

In a mRNA-based vaccine (e.g., Pfizer/Moderna) the mRNA coding for the Cov-Sars-2019 spike (S protein)  is encapsulated in lipid nanoparticles. This preparate is then injected (usually in the deltoid muscle). After that, the nanoparticles fuse with the cell membranes and mRNA is  released into the cell cytoplasm, without entering in the nucleus nor getting incorporated into the genomic DNA. In a adenovirus-based vaccine (e.g., AstraZeneca) the gene coding for the spike protein is inglobated as DNA in a defective Chimpanzee adenovirus which is not able to proliferate, alone.  This virus once injected latch on the host cell and released DNA (carrying the spike protein gene) in the cell cytoplasm. DNA then migrates to the nucleus where it is transcribed into mRNA, which will migrate to the cytoplasm.

In both the vaccines, at this particular stage the mRNA coding for the Spike uses the cellular machinery (e.g., ribosomes) for being translated into protein, imitating virus-infection-like humoral immunity and cellular immunity (\cite{monslow2020}. Both mRNA-based and adenovirus-based vaccines are able to increase the host?s anti-virus effects by increasing T cells? antigen reactiveness \cite{wang2021}. Normally, are these white blood cells, as the first defenses of the body, to "detect" the presence of the pathogen and to organize a protection by generating specific antibodies to combat it through B-lymphocytes, as particular blood cells deputed to antibody production \cite{ratajczak2018} These antibodies cover the virus and prevent it from attacking our body. 

The immune system memory can be compared to a human's memory. Once it encounters an unwanted visitor, it will remember it and will be able to recognize it in the future. This process typically takes a few weeks for the body to produce the antibodies, but these cells will be there to guard the body for a long time.

%%%%%%%%%%%%%%%%%%%%%%%%%%%%%%%%%%%%%%%%%%%%%%%%%%%%%%%%%
\subsection{Planning a vaccination campaign using genetic algorithms, with non-pharmaceutical containment measures in action}
\label{planGA}

We compare the effect of choosing the vaccination quotas via genetic algorithms (GAs) with two predetermined strategies. Our model considers three hypotheses: vaccinated people still spread the contagion; they do not spread the contagion; they do it in the 50\% of the case. We show here only the results of the first case, the worst (as we write, the Delta variant is spreading, with vaccinated people transmitting the infection). 

The parameters of the GAs side of the model are contained in a special file, as described in the Info sheet of the model; at  \href{https://terna.to.it/simul/SIsaR.html}{https://terna.to.it/simul/SIsaR.html} start the model and look at the Info paragraph named \emph{Using Genetic Algorithms}.

Important dates: 
\begin{itemize}
\item in the internal calendar of the model, day 373 is February \nth{12}, 2021; it is the starting point of the vaccinations in Piedmont; 

\item the effectiveness of the initial vaccinations, 40 days later, starts on day 413 (March \nth{22}, 2021).
\end{itemize}

A technical detail: we simulate the vaccination campaigns with the GAs using the BehaviorSearch program, \href{https://www.behaviorsearch.org}{https://www.behaviorsearch.org}, strictly related to NetLogo.

\subsubsection{Vaccination groups}
\label{vgroups}

We take into consideration seven groups, in order of decreasing fragility, also considering the exposure to contagions:

\begin{enumerate}
\item [\emph{g1}]
	extra fragile people with three components;
	\begin{itemize}
		\item due to intrinsic characteristics: people in living in nursing homes;
		\item due to risk exposure:
		\begin{itemize}
			\item nursing homes operators;
			\item healthcare operators;
 		\end{itemize} 
 	\end{itemize} 
\item [\emph{g2}]
	teachers;
\item [\emph{g3}]
	workers with medical fragility;
\item [\emph{g4}]
	regular workers;
\item [\emph{g5}]
	fragile people without special characteristics;
\item [\emph{g6}]
	regular people, not young, not worker, and not teacher;
\item [\emph{g7}]
	young people excluding special activity cases (a limited number in \emph{g1}).
\end{enumerate}

%%%%%%%%%%%%%%%%%%%%%%%%%%%%%%%%%%%%%%%%%%%%%%%%%%%%%%%%%
\subsection{A specific realistic case}
\label{specific}

\begin{figure}[t]
\center
\includegraphics[scale=0.2]{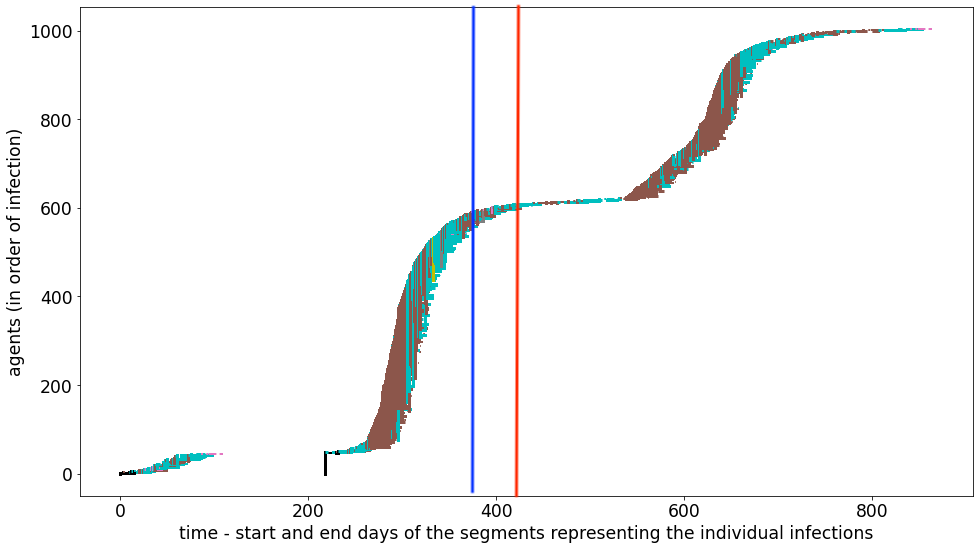}
\caption{Crucial dates: blue line for the starting point of the vaccination campaign and red line for the start of the effectiveness of the initial vaccinations; all the situations without vaccination}
\label{specificCaseGA}
\end{figure}

\begin{figure}[t]
\center
\fbox{\includegraphics[width=0.6\textwidth]{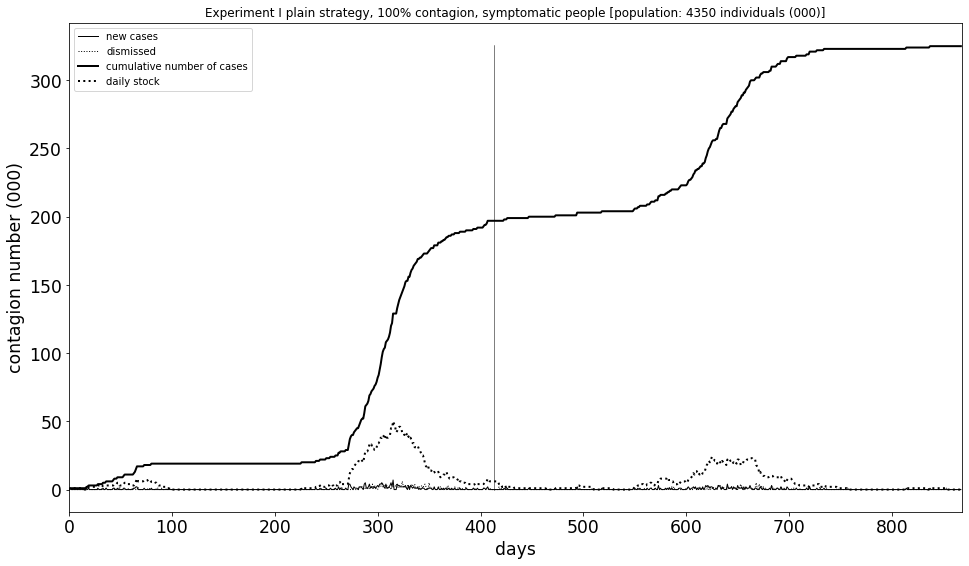}}
\caption{Base symptomatic series; the vertical line at day 413 is not relevant here} 
\label{symptomaticSeries}
\end{figure}

The description of the vaccination effects on an outbreak is quite lengthy. Considering the collection at  \href{https://terna.to.it/simul/GAresultPresentation.pdf}{https://terna.to.it/simul/GAresultPresentation.pdf}, we report here a unique case: the experiment I reported there, maintaining the reference to I in the titles of the figures.
Considering the adoption of the government non-pharmaceutical measures, we search---in the batch of the 10,000 outbreaks of Section \ref{secondWith}---for realizations of sequences similar to the actual events that occurred in Piedmont. As we see in Fig. \ref{specificCaseGA} and the related Fig. \ref{symptomaticSeries}, the artificial case that we adopt for the GAs exploration has the following critical characteristics:

\begin{itemize}
\item [(i)] numbers of infected persons quite similar at \verb|cp2| and at \verb|cp3| in Fig. \ref{actualA}; besides, numbers not too different from those of the same figure;

\item [(ii)] number of infected persons at \verb|cp4| significantly greater than those at the previous checkpoint.
\end{itemize}
% calculation made using SIsaR_0.9.6.nlogo with 66507108 94261776 44424105

In Fig. \ref{symptomaticSeries}, without vaccinations, we have the first wave in Spring 2020, a larger one in Fall 2020, a limited one between the end of 2020 and the beginning of 2021; then, a relatively quiet interval and successively, just while we write these notes, some restarting signals; finally, a fourth wave. Currently, it is in the future, relative to both the time of writing and the time when the calculations were completed (see NB at the end of Section \ref{actual}). Very realistic with Piedmont's actual situation, the limited thickness of the \emph{snake} of Fig. \ref{specificCaseGA}, when vaccinations start and when their effectiveness develops. The hole in the series identifies a period of quasi-extinct epidemics. Then it restarts with the arrival of infected persons from outside.

Here and in the following sections, we analyze the count of symptomatic persons, being the goal of our simulated vaccination campaign exactly that of decreasing the number of symptomatic people, as deceased persons come from there.

% using contagionSeriesByGroups.ipynb on Experiment_I_base.csv 

%%%%%%%%%%%%%%%%%%%%%%%%%%%%%%%%%%%%%%%%%%%%%%%%%%%%%%%%%
\subsection{Vaccination quotas, \emph{plain} strategy}
\label{plain}

The vaccination plans are related to the first dose; the second dose is supposed to be automatically scheduled, with an independent supply. The vaccinated person starts to benefit from immunity 40 days after the first dose.

Considering a \emph{plain} option as that adopted in Table \ref{quotaPlainTab} with, in each day, the quantities of doses of the first column, we will primarily vaccinate the left column groups to move gradually to people of the other columns, as those on the left have already received the vaccine. The order is (\emph{g1}) extra fragile people, (\emph{g2}) teachers, (\emph{g3}) fragile workers, (\emph{g4}) regular workers, (\emph{g5}) fragile people, (\emph{g6}) regular people, (\emph{g7}) young people. In Table \ref{susceptible} we have numbers both of persons in each category at the beginning of this experiment (and in the following ones) and when the vaccination campaign starts.

\begin{table}[t]
\centering
\begin{tabular}{ccccccccc}
\toprule
\begin{tabular}[c]{@{}c@{}}From \\ day\end{tabular} & \begin{tabular}[c]{@{}c@{}}Q. of \\ vaccines \\ (000)\end{tabular} & \emph{g1} & \emph{g2} & \emph{g3} & \emph{g4} & \emph{g5} & \emph{g6} & \emph{g7} \\
\midrule
373 & 5 & 0.1 & 0.1 & 0.1 & 0.1 & 0.1 & 0.1 & 0.1 \\
433 & 10 & 0.1 & 0.1 & 0.1 & 0.1 & 0.1 & 0.1 & 0.1 \\
493 & 10 & 0.1 & 0.1 & 0.1 & 0.1 & 0.1 & 0.1 & 0.1 \\
553 & 10 & 0.1 & 0.1 & 0.1 & 0.1 & 0.1 & 0.1 & 0.1 \\
613 & 20 & 0.1 & 0.1 & 0.1 & 0.1 & 0.1 & 0.1 & 0.1 \\
738 & end \\ 
\bottomrule 
\end{tabular}
\caption{From the day of the first column, considering the quantity of the second column (000), the vaccination of each group follows the quotas of the related columns}
\label{quotaPlainTab}
\end{table}

\begin{table}[t]
\centering
\begin{tabular}{lccccccc}
\toprule
(000) & \emph{g1} & \emph{g2} & \emph{g3} & \emph{g4} & \emph{g5} & \emph{g6} & \emph{g7} \\
\midrule
Susc. at t = 0 & 133 & 84 & 240 & 1560 & 1179 & 254 & 900 \\
Susc. when\\vacc. starts & 124 & 81 & 162 & 1234 & 1032 & 245 & 891 \\
\bottomrule 
\end{tabular}
\caption{Susceptible persons at the beginning of the simulation and when the vaccination campaign starts, day 373, Feb. \nth{12}, 2021}
\label{susceptible}
\end{table}

Some of the coefficients in Table \ref{quotaPlainTab}, and all the successive similar ones, are not used in two situations:

\begin{enumerate}[label=\roman*]
\item \label{quo1} when the persons of a group are fully vaccinated, the quotas in the rows below that day are not relevant;
\item \label{quo2} when the people in the columns to the left of a given column completely absorb the available doses of vaccine on that day (the quotas in that column have unimportant values).
\end{enumerate}

We anticipate that the GAs procedure does not optimize the coefficients of cases \ref{quo1} and \ref{quo2}.

The series that we introduce hereafter are significant from day 413, March \nth{22}, when the initial vaccinations' effectiveness begins, after 40 days from initial vaccinations.

In Fig. \ref{seqPlain} we have the effects of the vaccination plan as numbers of vaccinated persons by groups. In Fig. \ref{timeSeriesPlain} we have the most important outcome: the no vaccination test-bed is that of Fig. \ref{symptomaticSeries}. We note the waves after the vertical line---when vaccinations start to operate---are lower than in the test plot, but anyway, those further waves are there.

%%%%%%%%%%%%%%%%%%%%%%%%%%%%%%%%%%%%%%%%%%%%%%%%%%%%%%%%%
\subsection{Vaccination quotas, \emph{wise} strategy}
\label{wise}

Considering now a \emph{wise} option, as an attempt to mimic the actual (and complex) vaccine distribution in the region, we use the quotas of Table \ref{quotaWiseTab}, with the exact mechanism of the previous section.
We primarily vaccinate the left column groups to move gradually to other columns, but postponing group \emph{g4} (regular workers), \emph{g6} (regular people), and \emph{g7} (young people). In Table \ref{susceptible} we have numbers both of persons in each category at the beginning of this experiment (and in the following ones) and when the vaccination campaign starts. The considerations sub \ref{quo1} and \ref{quo2} in Section \ref{plain} apply also here.

\begin{table}[t]
\centering
\begin{tabular}{ccccccccc}
\toprule
\begin{tabular}[c]{@{}c@{}}From \\ day\end{tabular} & \begin{tabular}[c]{@{}c@{}}Q. of \\ vaccines \\ (000)\end{tabular} & \emph{g1} & \emph{g2} & \emph{g3} & \emph{g4} & \emph{g5} & \emph{g6} & \emph{g7} \\
\midrule
373 & 5 & 0.1 & 0.1 & 0.1 & 0.0 & 0.1 & 0.0 & 0.0 \\
433 & 10 & 0.1 & 0.1 & 0.1 & 0.0 & 0.1 & 0.0 & 0.0 \\
493 & 10 & 0.1 & 0.1 & 0.1 & 0.1 & 0.1 & 0.1 & 0.1 \\
553 & 10 & 0.1 & 0.1 & 0.1 & 0.1 & 0.1 & 0.1 & 0.1 \\
613 & 20 & 0.1 & 0.1 & 0.1 & 0.1 & 0.1 & 0.1 & 0.1 \\
738 & end \\ 
\bottomrule 
\end{tabular}
\caption{From the day of the first column, considering the quantity of the second column (000), the vaccination of each group follows the quotas of the related columns}
\label{quotaWiseTab}
\end{table}

In Fig. \ref{seqWise} we have the effects of the vaccination plan as numbers of vaccinated persons by groups. In Fig. \ref{timeSeriesWise} we have the experiment outcome: the no vaccination test-bed is that of Fig. \ref{symptomaticSeries}. We note the waves after the vertical line---when vaccinations start to operate---are lower  than in the test plot, but we have significant further waves in this case too.

%%%%%%%%%%%%%%%%%%%%%%%%%%%%%%%%%%%%%%%%%%%%%%%%%%%%%%%%%
\subsection{GAs quotas in the experiment, with vaccinated people spreading the infection}
\label{GAquotas}

Finally, this whole section's objective is to use GAs to evolve populations of models by choosing ``genetically'' the parameters to decide daily vaccination. Initially, on a random basis and successively considering them as a genetic chromosome of each model, re-productively crossed with those of other models. The search is for the best fitness related to the goal of reducing the number of symptomatic persons. \cite{miller1998active}, also quoted at \href{https://www.behaviorsearch.org}{https://www.behaviorsearch.org}, is a helpful introduction to the methodology; the sources of the GAs used here are at \href{https://github.com/terna/GAs}{https://github.com/terna/GAs}. The GAs action, determining the vaccination quotas, optimizes the behavior of a \emph{deciding} meta-agent, in a sort of \emph{inverse generative social science perspective} \cite{vu2019toward}.

With the GAs option, we use the quotas of Table \ref{quotaGATab}, with the exact mechanism of the previous section. The considerations sub \ref{quo1} and \ref{quo2} in Section \ref{plain} also apply here. We underline that the GAs procedure does not optimize the coefficients of those two cases, because they do not affect the fitness related to the goal of minimizing the number of symptomatic subjects.  

In Table \ref{susceptible} we have numbers both of persons in each category at the beginning of the experiment and when the vaccination campaign starts.

\begin{figure}[H]

 \begin{subfigure}{0.4\textwidth}
 \centering
 \fbox{\includegraphics[width=0.95\textwidth]{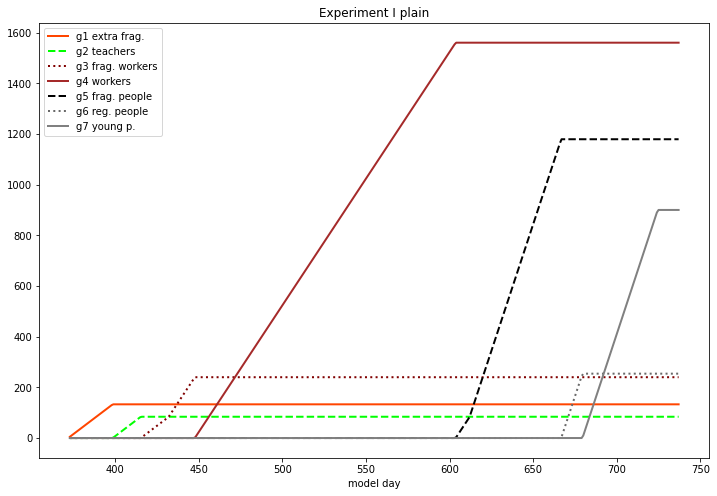}} % experiment_I_vaccination_plots.ipynb
 \subcaption{\emph{Plain} vaccination sequence; on the $y$ axis the number of vaccinated subjects of each group (if vaccination is complete, the line is horizontal)}
 \label{seqPlain}
 \end{subfigure}
 \hfill
 \begin{subfigure}{0.52\textwidth}
 \centering
 \fbox{\includegraphics[width=0.95\textwidth]{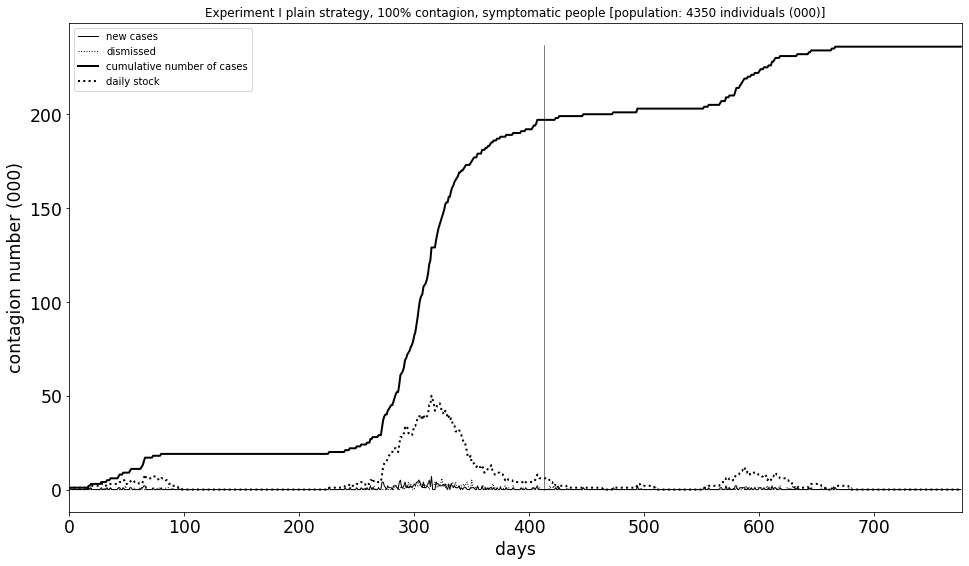}} % using contagionSeriesByGroups.ipynb on Experiment_I_plain_1.csv 
 \subcaption{\emph{Plain} vaccination symptomatic series; the vertical line is at day 413, when the effectiveness of first vaccination starts}
 \label{timeSeriesPlain}
 \end{subfigure}

\begin{subfigure}{0.4\textwidth}
 \centering
 \fbox{\includegraphics[width=0.95\textwidth]{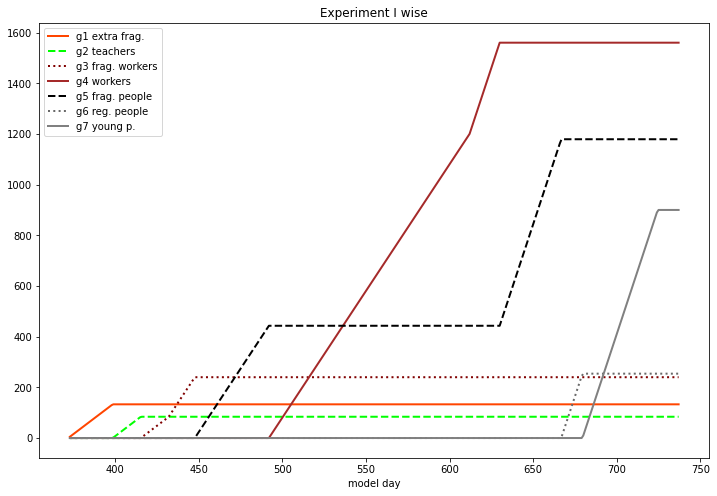}} % experiment_I_vaccination_plots.ipynb
 \subcaption{\emph{Wise} vaccination sequence; on the $y$ axis the number of vaccinated subjects of each group (if vaccination is complete, the line is horizontal)}
 \label{seqWise}
 \end{subfigure}
 \hfill
 \begin{subfigure}{0.52\textwidth}
 \centering
 \fbox{\includegraphics[width=0.95\textwidth]{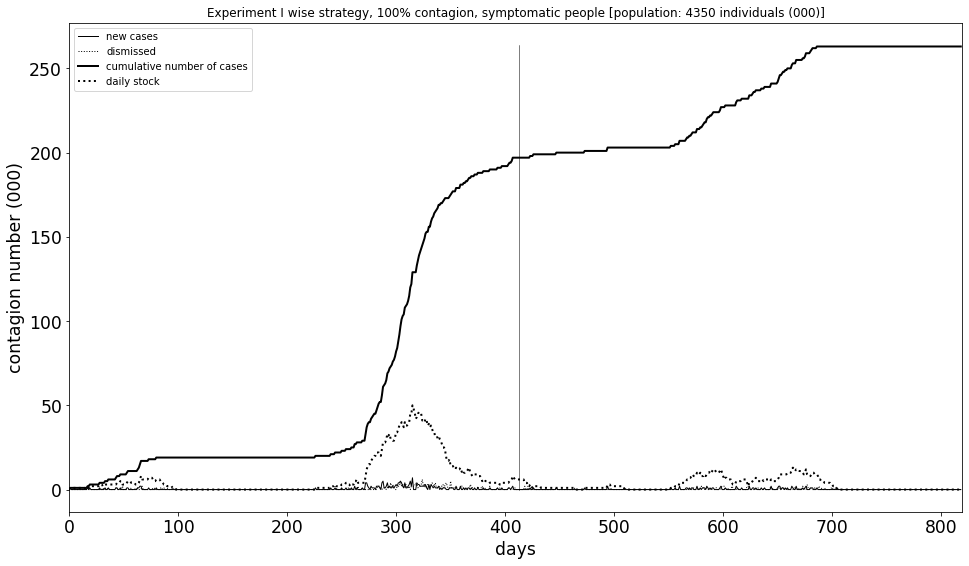}} % using contagionSeriesByGroups.ipynb on Experiment_I_wise_1.csv
 \subcaption{\emph{Wise} vaccination symptomatic series; the vertical line is at day 413, when the effectiveness of first vaccination starts} 
 \label{timeSeriesWise}
 \end{subfigure}

\begin{subfigure}{0.4\textwidth}
 \centering
 \fbox{\includegraphics[width=0.95\textwidth]{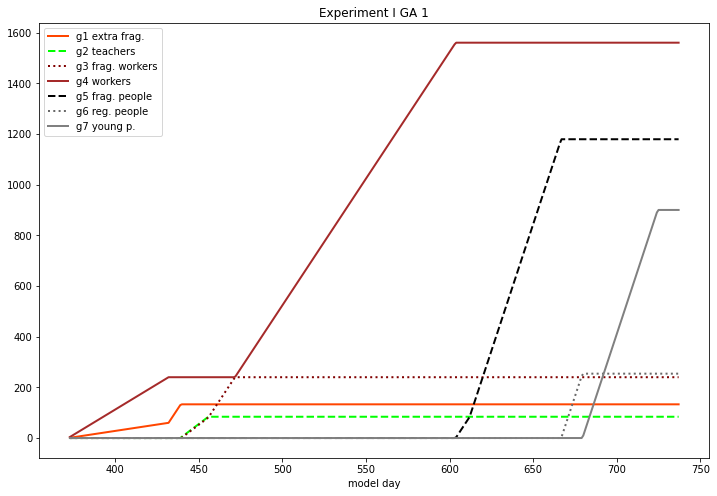}} % experiment_I_vaccination_plots.ipynb
 \subcaption{GA vaccination sequence; on the $y$ axis the number of vaccinated subjects of each group (if vaccination is complete, the line is horizontal)} 
 \label{seqGA}
 \end{subfigure}
 \hfill
 \begin{subfigure}{0.52\textwidth}
 \centering
 \fbox{\includegraphics[width=0.95\textwidth]{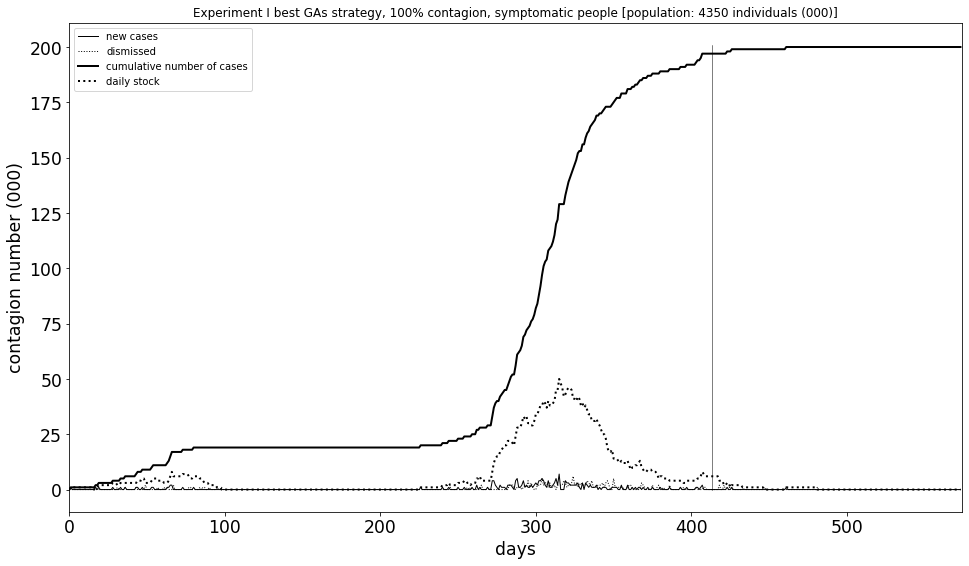}} % using contagionSeriesByGroups.ipynb on Experiment_I_1bestGA.csv 
 \subcaption{GAs vaccination symptomatic series; the vertical line is at day 413, when the effectiveness of first vaccination starts}
 \label{timeSeriesGA}
 \end{subfigure}

 \caption{Vaccination sequences and time series}
 \label{vaccinCases}
\end{figure}

In Fig. \ref{seqGA} we have the effects of the vaccination plan as numbers of vaccinated persons by groups. 
The main attention of the GAs is initially related to the groups: \emph{g4} (workers), \emph{g1} (extra-fragile persons), \emph{g3} (fragile workers), \emph{g2}  (teachers). Then \emph{g5} (fragile people), finally \emph{g6} (regular people), and \emph{g7} (young people).
The priority is for highly circulating persons (workers and teachers), then for fragile persons.

In Fig. \ref{timeSeriesGA} we have the crucial result of this experiment: the no vaccination test-bed is always that of Fig. \ref{symptomaticSeries}. With GAs' choices, the waves after the vertical line---when vaccinations start to operate---disappear, and the whole outbreak is a lot shorter.

\begin{table}[t]
\centering
\begin{tabular}{ccccccccc}
\toprule
\begin{tabular}[c]{@{}c@{}}From \\ day\end{tabular} & \begin{tabular}[c]{@{}c@{}}Q. of \\ vaccines \\ (000)\end{tabular} & \emph{g1} & \emph{g2} & \emph{g3} & \emph{g4} & \emph{g5} & \emph{g6} & \emph{g7} \\
\midrule
373 & 5 & 0.01 & 0 & 0 & 0.79 & 0.18 & 0.38 & 0.19 \\
433 & 10 & 0.94 & 0.06 & 0.32 & 0.54 & 0.19 & 0.83 & 0.5 \\
493 & 10 & 0.97 & 0.97 & 0.74 & 0.79 & 0.2 & 0.14 & 0.52 \\
553 & 10 & 0.98 & 0.83 & 0.02 & 0.39 & 0.99 & 0.04 & 0.48 \\
613 & 20 & 0.52 & 0.01 & 0.83 & 0.6 & 1 & 0.27 & 0.9 \\
738 & end \\ 
\bottomrule 
\end{tabular}
\caption{GAs best strategy with \emph{vaccinated people still spreading the infection}: from the day of the first column, considering the quantity of the second column, the vaccination of each group follows the quotas of the related columns}
\label{quotaGATab}
\end{table}

%%%%%%%%%%%%%%%%%%%%%%%%%%%%%%%%%%%%%%%%%%%%%%%%%%%%%%%%%
\section{A new model and future developments}
\label{newModel}

Using SLAPP, \href{https://terna.github.io/SLAPP/}{https://terna.github.io/SLAPP/} a second model is under development, with a ratio of 1:100 to the Piedmont population, so 43,500 agents. It will contain the same items as the current one, plus transportation and aggregation places: happy hours, nightlife, sports, stadiums, discotheques, etc. We will also consider networks as family networks, professional networks, high-contact individual networks \cite{manzo2020}. Finally, we will take into consideration the socioeconomic conditions of the individuals.

As seen, the S.I.s.a.R. model is a tool for comparative analyses, not for forecasting, mainly due to the enormous standard deviation values intrinsic to the problem. 

The model is highly parametric, and more it will be, precisely in the comparative perspective. It also represents a small step in using artificial intelligence tools and the inverse generative perspective \cite{inverseGen} in agent-based models.

%%%%%%%%%%%%%%%%%%%%%%%%%%%%%%%%%%%%%%
%%%%%%%%%%%%%%%%%%%%%%%%%%%%%%%%%%%%%%
%%%%%%%%%%%%%%%%%%%%%%%%%%%%%%%%%%%%%%
\section{Appendix 1---Parameter values}
\label{app1}

We report here the values of parameters of Fig. \ref{outline}, with their short names used in program scripts, in round brackets. Look at Section \ref{par} for the definition. Day numbering is related to actual dates via the Table \ref{dates}. Day 1 is February \nth{4}, 2020.

The values adopted in the experiments reported in this work are the following.

\begin{enumerate}[label=\roman*]

\item \label{pp1} The values of \emph{probabilityOfGettingInfection} (\verb|prob|) are: 0.05 (starting phase); 0.02 at day 49 (adoption of non-pharmaceutical measures); 0.035 at day 149 (some relaxation in compliance); 0.02 at day 266 (again, compliance to rules). 

\item \label{pD} The value of \emph{D\%} is -50 in all the runs.

\item \label{pp2} \emph{intrinsicSusceptibility} is set discussing Eq. \ref{intrinsic} in Section \ref{par}.

\item \label{pp3} The values of \emph{\%PeopleAnyTypeNotSymptomaticLeavingHome} (\verb|%PeopleAny|) are:
at (day) 20, 90; at 28, 80; at 31, 0; at 106, 80; at 110, 95; at112, 85; at 117, 95; at 121, 90; at 259, 90; at 266, 80; at 277, 50;
at 302, 70; at 320, 90; at 325, 50; at 329, 80; at 332, 50; at 336, 80; at 337, 50; at 339, 80.

\item \label{pp4} The values of \emph{\%PeopleNotFragileNotSymptomaticLeavingHome} (\verb|%PeopleNot|) are:
at (day) 31, 80; at 35, 70; at 36, 65; at 38, 15; at 42, 25; at 84, 30; at 106, 0; at 302, 90; at 325, 50; at 332, 50; 
at 337, 50; at 339, 100; at 349, 90.

\item \label{pp5} The values of \emph{\%openFactoriesWhenLimitationsOn} (\verb|%Fac|) are:
at (day) 38, value4 0; at 49, 20; at 84, 70; at 106, 100; at 266, 90; at 277, 70; at 302, 80; at 320, 90; at 325, 30;
at 329, 90; at 332, 30; 336, 90; at 337, 30; at 339, 100.

\item \label{pp6} \emph{stopFragileWorkers} (\verb|sFW|): by default, 0; in one of the experiments we used \verb|sFW| with set to 1 (on) at day 245 and to 0 (off) at day 275.

\item \label{pp7} The values of \emph{activateSchools} (\verb|aSch|) are:
at (day) 1, on; at 17, off; at 225, on; at 325, off; at 339, on;

the values of \emph{\%Students} (\verb|%St|) are: 
at (day) 0, 100; at 277, 50; at 339, 50; at 350, 50 (repeated values are not relevant for the model, but for the use of the programmer-author).

\item \label{pp8} The value of \emph{radiusOfInfection} (\verb|radius|) is 0.2; in the model, space is missing of a scale, but forcing the area to be in the scale of a region as Piedmont, 0.2 is equivalent to 20 meters; we have to better calibrate this measure with movements and probabilities; this is a critical step in future developments of the model.

\item \label{pp9} The values of \emph{asymptomaticRegularInfected\%} and \emph{asymptomaticFragileInfected\%} are 95 and 20.

\end{enumerate}

\begin{table}[t]
\begin{center}
\begin{footnotesize}
\begin{tabular}{rrrrrrrrrrr}
\toprule
 Day & Date &~~~~~~~~& Day & Date &~~~~~~~~& Day & Date &~~~~~~~~& Day & Date \\
 \midrule
 25 & 28- 2-2020 & & 200 & 21- 8-2020 & & 375 & 12- 2-2021 & & 550 & 6- 8-2021 \\
 50 & 24- 3-2020 & & 225 & 15- 9-2020 & & 400 & 9- 3-2021 & & 575 & 31- 8-2021 \\
 75 & 18- 4-2020 & & 250 & 10-10-2020 & & 425 & 3- 4-2021 & & 600 & 25- 9-2021 \\
100 & 13- 5-2020 & & 275 & 4-11-2020 & & 450 & 28- 4-2021 & & 625 & 20-10-2021 \\
125 & 7- 6-2020 & & 300 & 29-11-2020 & & 475 & 23- 5-2021 & & 650 & 14-11-2021 \\
150 & 2- 7-2020 & & 325 & 24-12-2020 & & 500 & 17- 6-2021 & & 675 & 9-12-2021 \\
175 & 27- 7-2020 & & 350 & 18- 1-2021 & & 525 & 12- 7-2021 & & 700 & 3- 1-2022 \\
\bottomrule
\end{tabular}
\end{footnotesize}
\caption{The days of the simulation and their equivalent dates in the calendar}
\label{dates}
\end{center}
\end{table}

%%%%%%%%%%%%%%%%%%%%%%%%%%%%%%%%%%%%%%
%%%%%%%%%%%%%%%%%%%%%%%%%%%%%%%%%%%%%%
%%%%%%%%%%%%%%%%%%%%%%%%%%%%%%%%%%%%%%
\section{Appendix 2---A gallery of contagion sequences}
\label{app2}

The gallery of contagion sequences, reported in Table \ref{gallery}, shows the vast variety of situations generated by our agent-based simulations. What is significant is the variety of the situations.

\begin{itemize}
\item [1a] An outbreak without containment measures, with a unique wave, but very heavy: contagions are in nursing homes (orange), workplaces (brown), homes (cyan), hospitals (pink).
\item [1b] This is the previous epidemic without containment measures, considering the first 200 infections, with the main contribution of nursing homes (orange) and workplaces (brown).
\item [1c] Another outbreak, always without containment measures: nursing homes (orange) as a starter.

\item [2a] The [1c] epidemic, without containment measures, first 200 infections: nursing homes (orange) as a starter; around day 70, a unique contagion at home makes the epidemic continue.
\item [2b] Another case without containment measures showing the initial action of contagions in workplaces (brown) and homes (cyan).
\item [2c] Here we see the first 200 infections showing that the initial profound effects of contagions in workplaces (brown) and homes are due, in the beginning, to fragile persons, also asymptomatic,

\item [3a] An outbreak with containment measures, where we see another influential contribution of workplaces (brown) and homes (cyan) to the epidemic diffusion.
\item [3b] Here the first 200 infections: after day 100, we observe many significant cases of fragile workers diffusing the infection.
\item [3c] In this outbreak, with containment measures, the infections arise from workplaces (brown), nursing homes (orange), and homes (cyan), but also hospitals (pink).

\item [4a] Here we explore the first 200 infections of [3c]: in the beginning, workplaces (brown), hospitals (pink), nursing homes (orange), and homes (cyan) are interweaving.
\item [4b] An outbreak with containment measures where the effect of the contagions in workplaces (brown), nursing homes (orange), and homes (cyan) is evident.
\item [4c] In the first 200 infections of [4b], workplaces (brown) and nursing homes (orange) are strictly interweaving.

\item [5a] An outbreak with containment measures where the effect of nursing homes (orange) is prevalent.
\item [5b] An outbreak with containment measures with a highly significant effect from workplaces (brown).
\item [5c] Stopping fragile workers at day 20 in the previous case, we obtain a beneficial effect, but home contagions (cyan) keep alive the pandemic, which explodes again in workplaces (brown).

\item [6a] Exploring the first 200 infections of the case [5c], we have evidence of the event around day 110 with the new phase due to a unique asymptomatic worker.
\item [6b] Finally, the same epidemic stopping fragile workers and any fragility at day 15 case and isolating nursing homes.
\item [6c] An outbreak with containment measures spontaneously stopping in a short period.
\end{itemize}

 \begin{table}
 \centering
 \begin{tabular}{cccc}
 \toprule
 Nr. & a & b & c \\
 \midrule
1 & \includegraphics[width=12em]{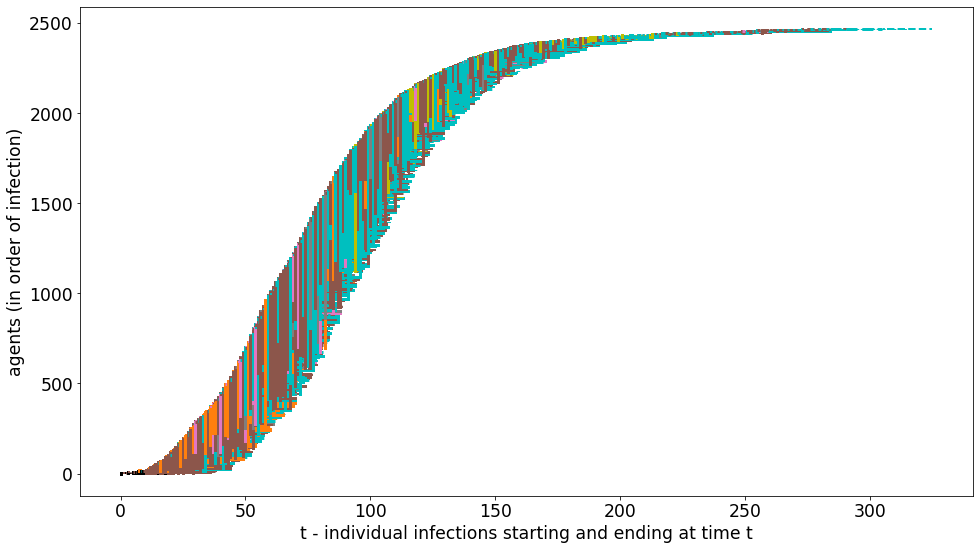} & % no control case 10932215 10933923 in SIsaR_0.9.4.1tmp experiments 2 seeds no control-table_5000.csv, file noControl_10932215_10933923.csv
 \includegraphics[width=12em]{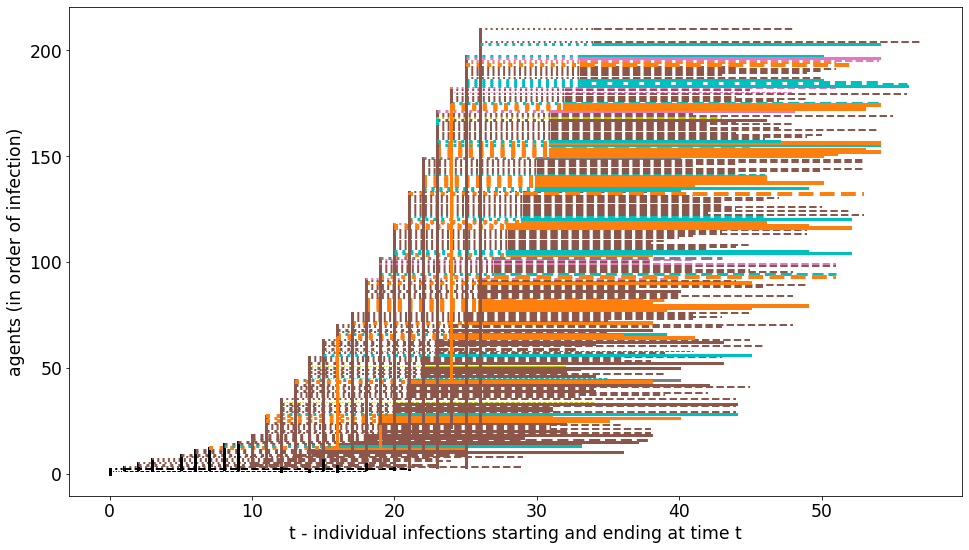} & % no control case 10932215_10933923 in SIsaR_0.9.4.1tmp experiments 2 seeds no control-table_5000.csv, file noControl_10932215_10933923.csv
 \includegraphics[width=12em]{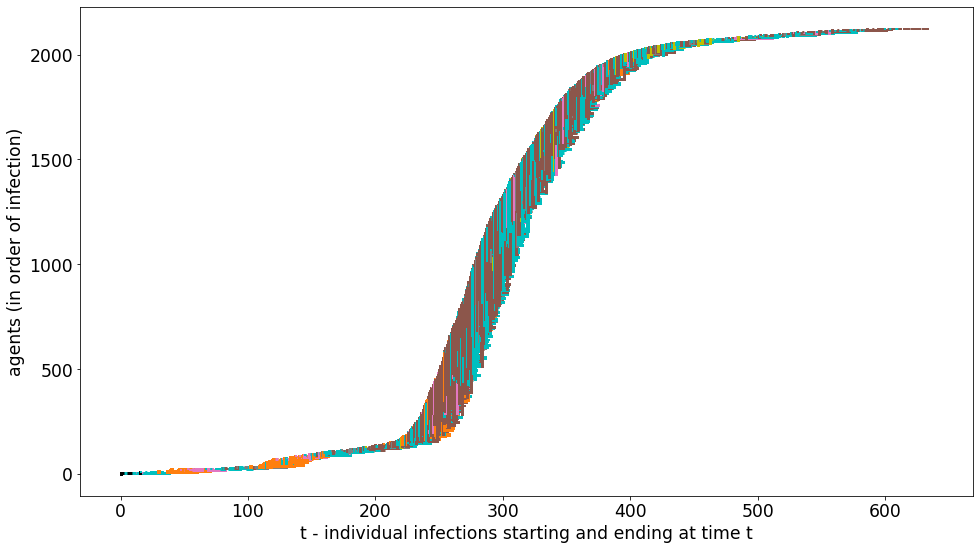} % no control case 12518034 12520144 in SIsaR_0.9.4.1tmp experiments 2 seeds no control-table_5000.csv, file noControl_12518034_12520144.csv
\\
2 & \includegraphics[width=12em]{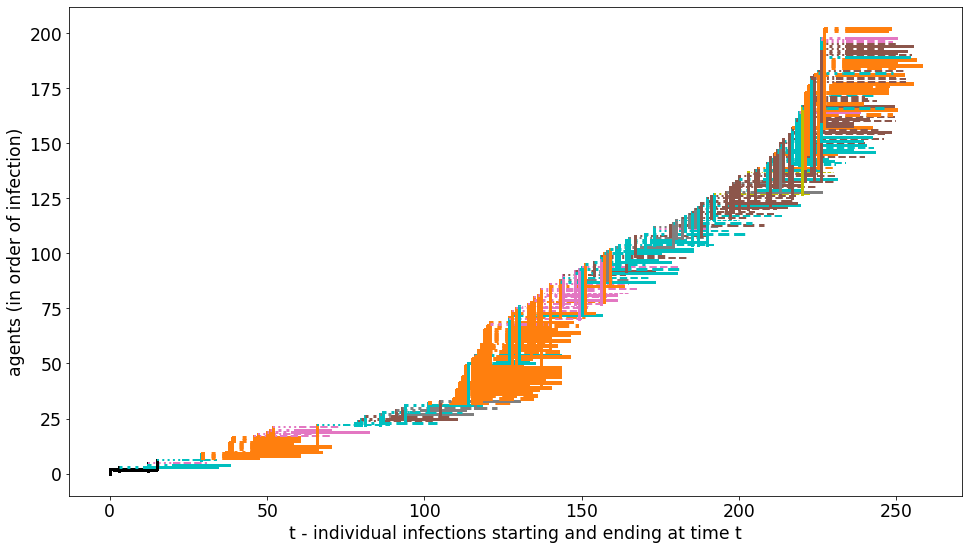} & % no control case 12518034 12520144 in SIsaR_0.9.4.1tmp experiments 2 seeds no control-table_5000.csv, filee noControl_12518034_12520144.csv
 \includegraphics[width=12em]{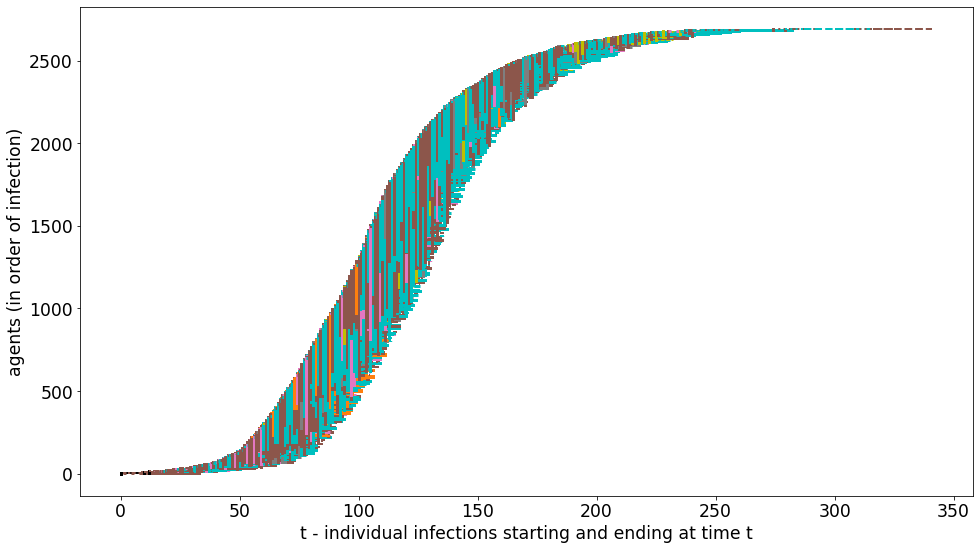} & % no control case 113790741 13792228 in SIsaR_0.9.4.1tmp experiments 2 seeds no control-table_5000.csv, file noControl_13790741 13792228.csv
 \includegraphics[width=12em]{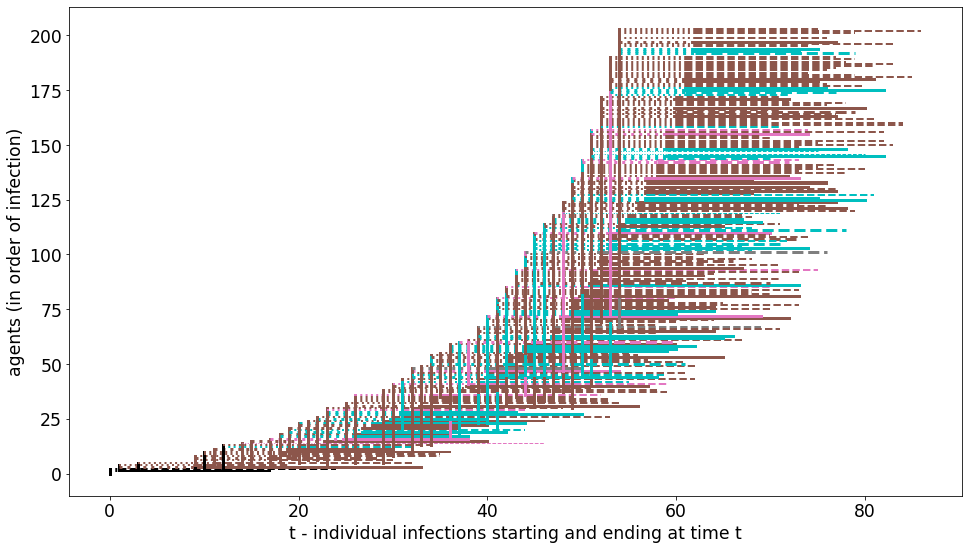} % no control case 113790741 13792228 in SIsaR_0.9.4.1tmp experiments 2 seeds no control-table_5000.csv, file noControl_13790741 13792228.csv
\\
3 & \includegraphics[width=12em]{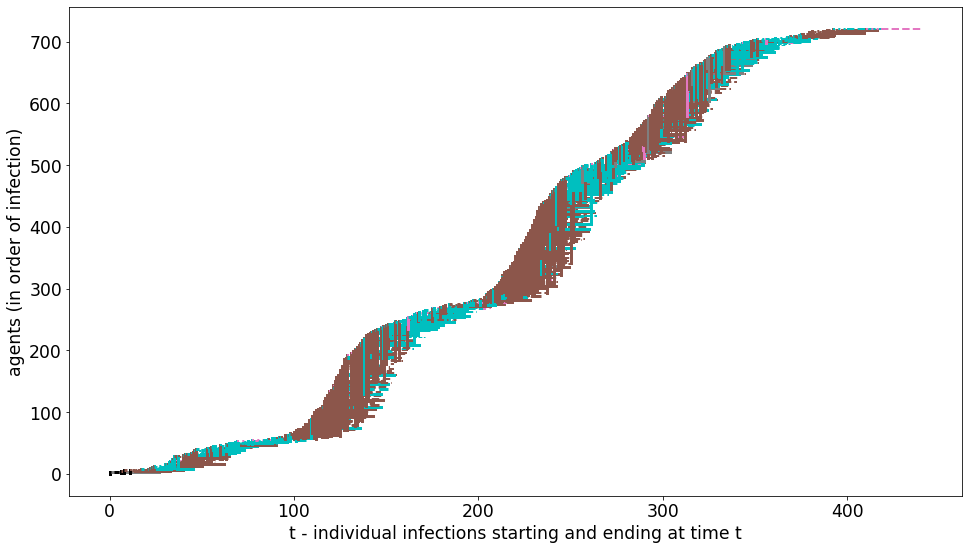} & % with control case 268411 270578 in SIsaR_0.9.4.1 experiments 2 seeds with control-table_10000.csv, file withControl_268411_270578.csv
 \includegraphics[width=12em]{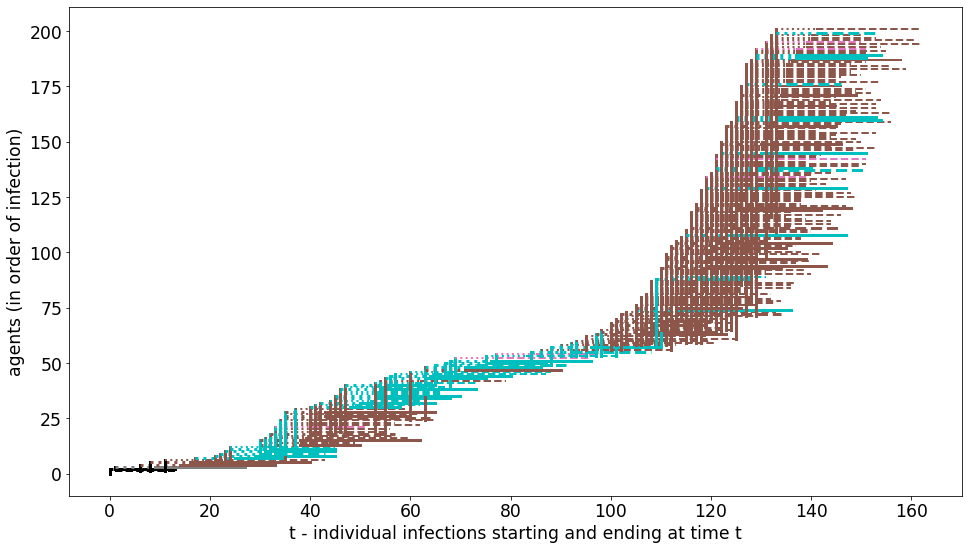} & % with control case 268411 270578 in SIsaR_0.9.4.1 experiments 2 seeds with control-table_10000.csv, file withControl_268411_270578.csv
 \includegraphics[width=12em]{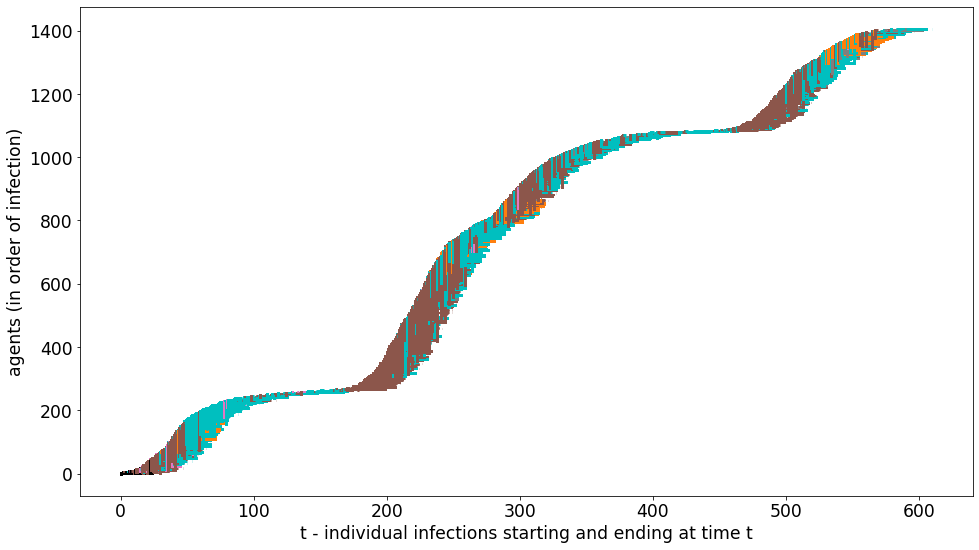} % with control case 2722110 2723664 in SIsaR_0.9.4.1 experiments 2 seeds with control-table_10000.csv, file withControl_2722110_2723664.csv
\\
4 & \includegraphics[width=12em]{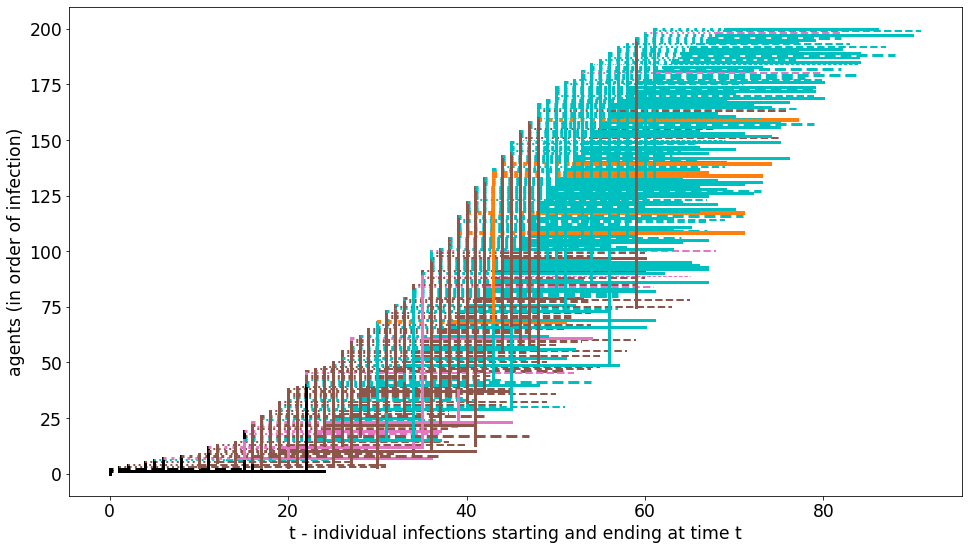} & % with control case 2722110 2723664 in SIsaR_0.9.4.1 experiments 2 seeds with control-table_10000.csv, file withControl_2722110_2723664.csv
 \includegraphics[width=12em]{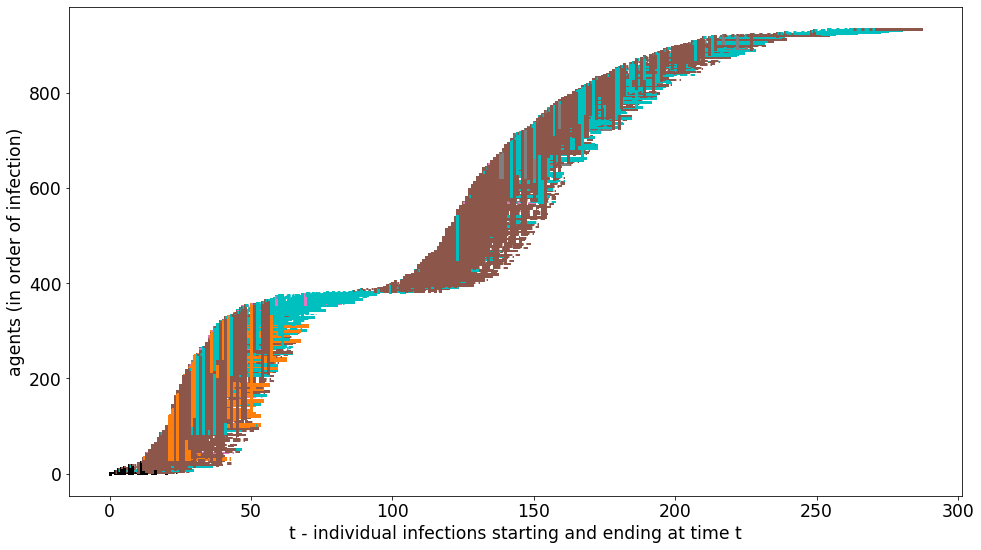} & % with control case 473323 474697 in SIsaR_0.9.4.1 experiments 2 seeds with control-table_10000.csv, file withControl_473323_474697.csv
 \includegraphics[width=12em]{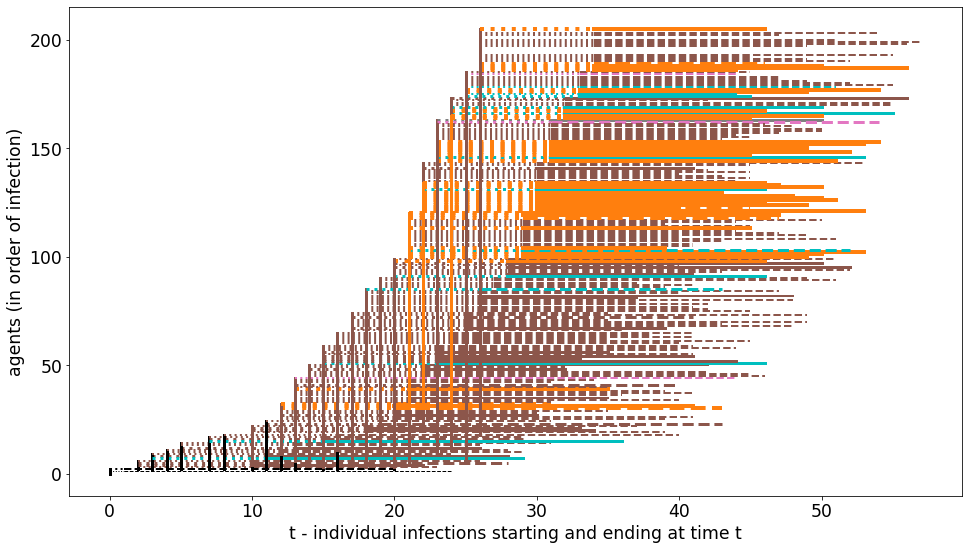} % with control case 473323 474697 in SIsaR_0.9.4.1 experiments 2 seeds with control-table_10000.csv, file withControl_473323_474697.csv
\\
5 & \includegraphics[width=12em]{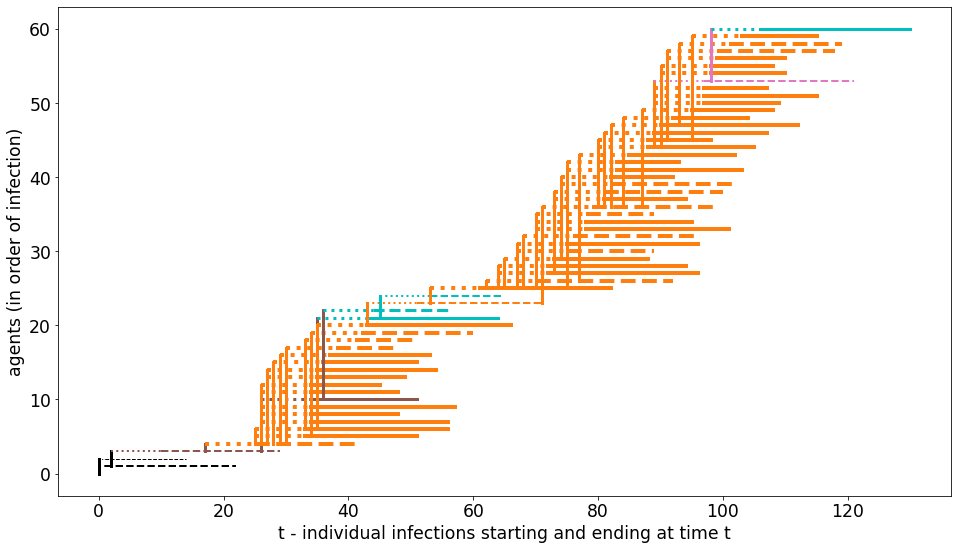} & %using SIsaR 9.4.2 as is (basic control) with 123456_22313, file ex123456_22313.csv
 \includegraphics[width=12em]{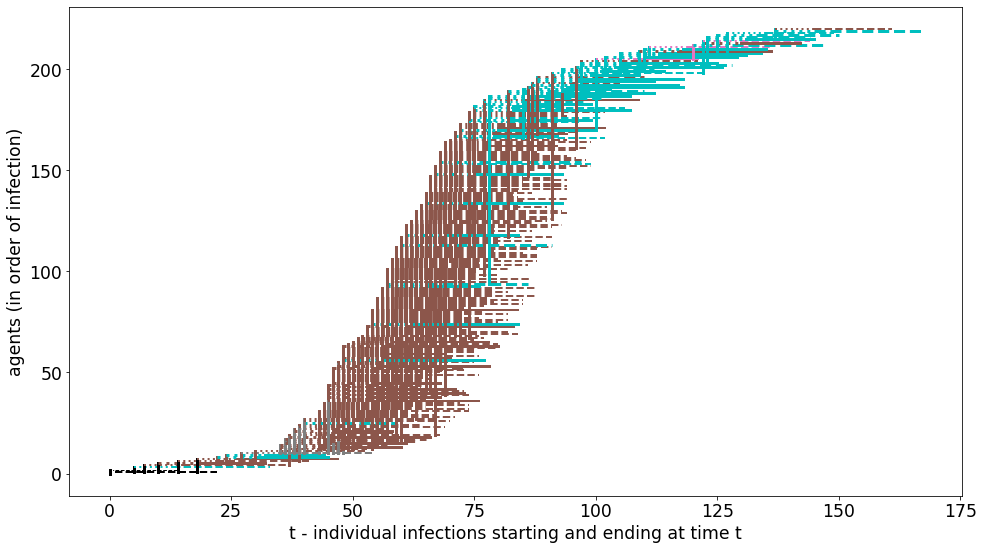} & %using SIsaR 9.4.2 as is (basic control) with 123456_22314
 \includegraphics[width=12em]{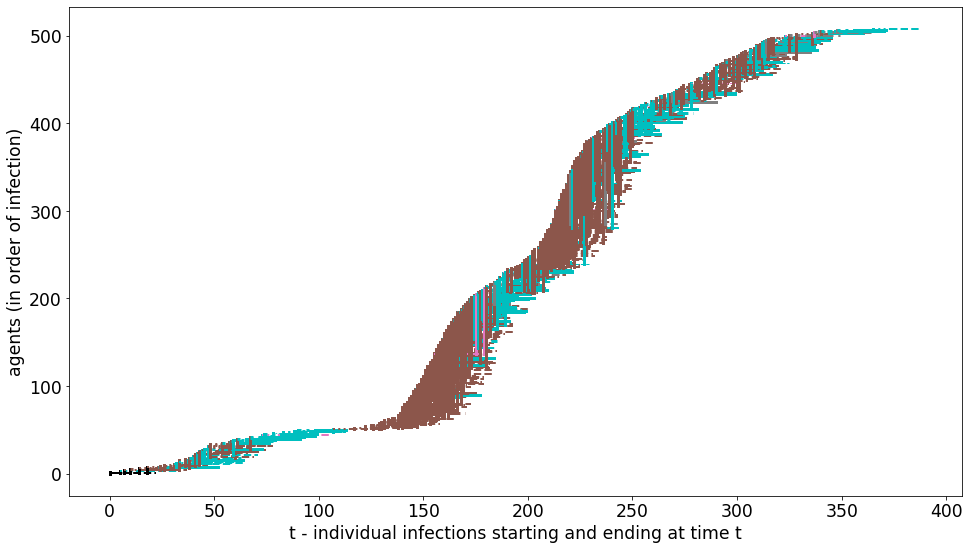} %using SIsaR 9.4.2 as is (basic control + 20 sFW 1,i.e., stop Fragile Workers) with 123456_22314
\\
6 & \includegraphics[width=12em]{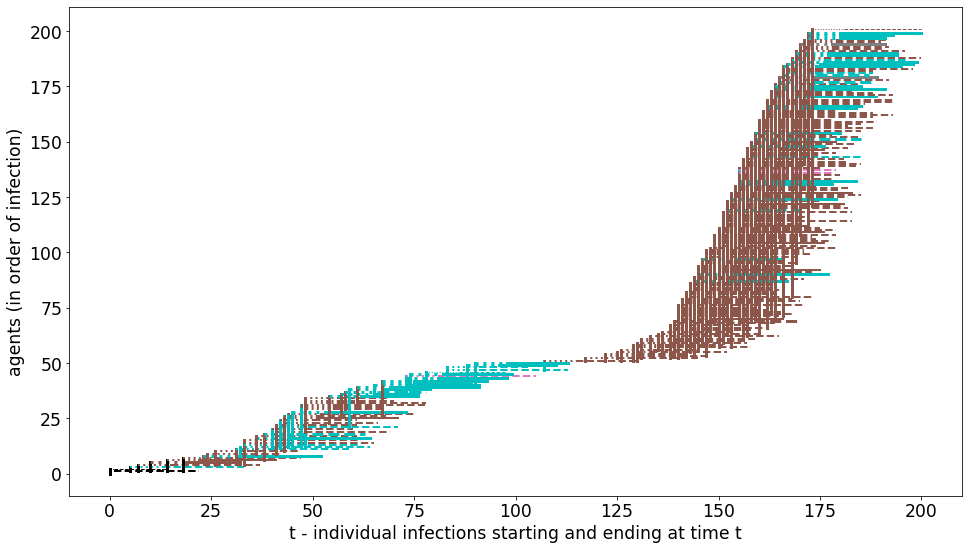} & %using SIsaR 9.4.2 as is (basic control + 20 sFW 1,i.e., stop Fragile Workers) with 123456_22314
 \includegraphics[width=12em]{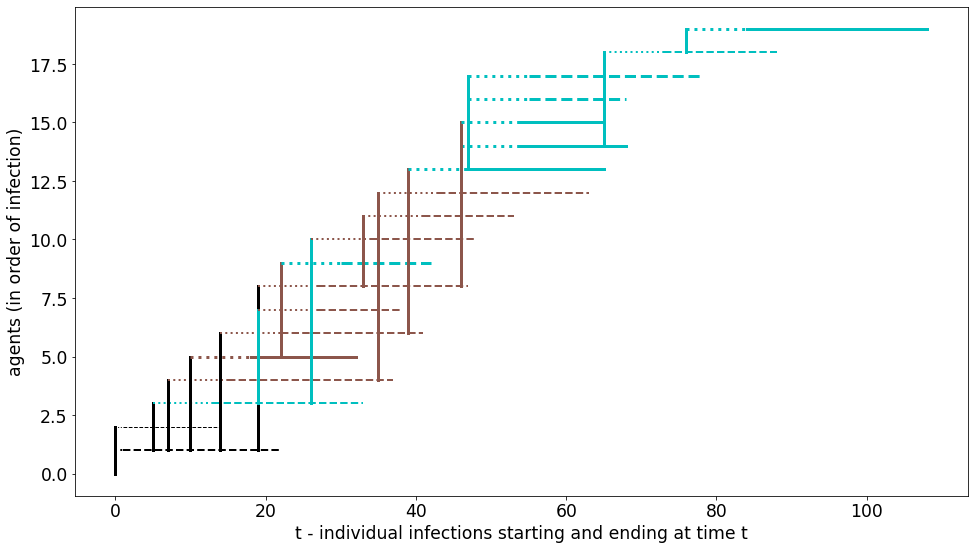} & %using SIsaR 9.4.2 as is (basic control + $$$) with 123456_22314
 \includegraphics[width=12em]{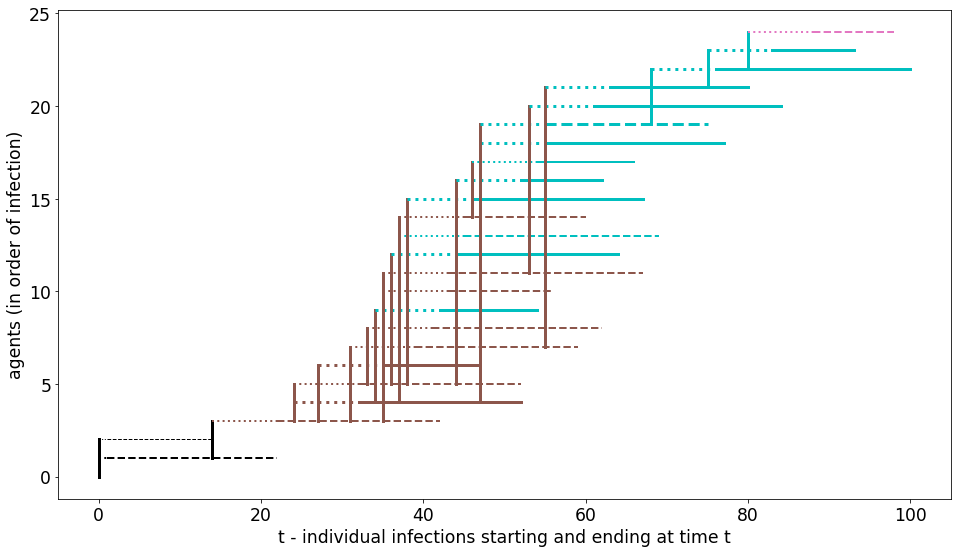} %using SIsaR 9.4.2 as is (basic control) with 123456_22315, file ex123456_22315.csv123456_22314
\\
 \bottomrule
 \end{tabular}
 \caption{Gallery of sequences, symptomatic and asymptomatic agents}
 \label{gallery}
 \end{table}

\section*{Acknowledgements}

Many thanks to Simone Landini, Nizar Mansour, Matteo Morini, Fabio Pammolli, Enrico Scalas, and Federico Tedeschi for their highly valuable discussions, insights, and critics. The usual disclaimer applies.

To run our model, we tremendously benefit from the use of HPC facilities provided by: Sarah de Nigris and Matteo Morini; \href{http://tomorrowdata.io}{http://tomorrowdata.io}; the HP4AI (\href{https://hpc4ai.unito.it/}{https://hpc4ai.unito.it/}) and C3S centers at University of Torino.

\bibliographystyle{spphys}
\bibliography{bibliografiaGenerale}

\end{document}